\documentclass[prd,aps,reprint,onecolumn,groupedaddress,tightenlines,nofootinbib,eqsecnum,preprintnumbers,longbibliography,12pt]{revtex4-2}
\usepackage{epsfig,amsfonts,mathrsfs,amsmath,amssymb,graphicx,color,slashed,multirow}
\usepackage{amsmath,latexsym,amssymb,graphicx,slashed,hyperref,color,enumerate,url,etoolbox,cancel,gensymb,physics}
\hypersetup{colorlinks,citecolor= nicegreen,linkcolor= nicered}
\usepackage[usenames,dvipsnames]{xcolor}
\definecolor{nicered}{rgb}{0.7,0.1,0.1}
\definecolor{nicegreen}{rgb}{0.1,0.5,0.1}
\definecolor{niceblue}{rgb}{0.1,0.2,0.6}

\usepackage{cancel}
\usepackage{bbm}
\usepackage{subfig}
\usepackage{physics}
\usepackage{dsfont}
\usepackage{cancel}
\usepackage{tikz-feynman}
\usepackage{float}
\usepackage{bm}
\tikzset{arrowlabel/.style = {midway, above}}

\begin{document}

\title{Analytic Approximations for Fermionic Preheating}

\author{Heather E.\ Logan}
\email{logan@physics.carleton.ca}
\author{Daniel Stolarski}
\email{stolar@physics.carleton.ca}
\author{Fazlul Yasin}
\email{FAZLULYASIN@cmail.carleton.ca }

\affiliation{Ottawa-Carleton Institute for Physics, Carleton University,
1125 Colonel By Drive, Ottawa, Ontario K1S 5B6, Canada}


\begin{abstract}
Non-thermal fermions can be produced non-perturbatively in the early universe during coherent oscillations of a scalar field. We explore fermion production in $\lambda\phi^{4}$ inflation through this mechanism and analyze the momentum spectrum of the fermions produced, which depends on a coupling parameter $q$. For $q \gtrsim 0.01$, the main contribution to the total number density comes from an approximately half-filled Fermi sphere as a result of non-adiabaticity. For $q\lesssim 0.01$, we find that the major contributions instead come from resonance peaks at higher momentum values. We find a simple relation to predict the momentum values corresponding to resonance peaks for any $q$. We observe power-law behaviour for the total number density of fermions as a function of $q$ and find numerical approximations for the integrals describing the total number density. We find that it is proportional to $q^{1/2}$ for $q\lesssim 0.01$ and proportional to $q^{3/4}$ for $q\gtrsim 10$. If fermions produced by this mechanism make up the entirety of dark matter, we estimate lower bounds on their mass. 
\end{abstract}


\maketitle

\tableofcontents 

\section{Introduction}\label{sec: 1}
Particles can be produced non-perturbatively during coherent oscillations of a scalar field before it decays. If the scalar field is the inflaton, the mechanism is called preheating \cite{Kofman:1994rk,Dolgov:1989us,Traschen:1990sw,Shtanov:1994ce}. We explore a case of the production of fermions through this mechanism, resulting in the production of particles out of thermal equilibrium \cite{Greene:1998nh}. The production of particles  during the oscillatory phase of the inflaton field was first described by perturbative decays of the inflaton field through a process termed reheating \cite{Dolgov:1982th}. However, it was found that particle production can occur explosively before the perturbative decay of the inflaton through the process of preheating \cite{Kofman:1994rk,Kofman:1997yn}.\footnote{
For a recent discussion of fermion production after inflation, see \cite{Bhusal:2025oqg}.
}

The description of reheating is through individual inflaton quanta decaying into produced particles, whereas preheating is production from an oscillating inflaton field. The mechanism of preheating is described by an oscillator-like differential equation (called the `mode equation') and the number density per mode of the fermions produced is described by a model-independent equation. The method of calculating it involves techniques, such as Bogoliubov transformations \cite{Bogolyubov:1958kj}, that are well defined in the context of particle production from time-dependent background fields \cite{Mostepanenko:1974im} and are also used in gravitational particle production \cite{Kolb:2023ydq} and the production of particles from an oscillating axion field \cite{Adshead:2015kza,Adshead:2018oaa}.

Particle production through preheating in the early universe affects the expansion history of the universe. It leads to different kinetic distributions and number densities of particles in the early universe than what is expected from perturbative reheating, which may lead to a different value of the reheating temperature ($T_{rh}$) \cite{Kofman:1997yn}. Unlike reheating, preheating also allows the production of particles with a mass greater than the mass of the inflaton \cite{Bassett:2005xm}. Massive particles produced during preheating may play an important role in baryogenesis \cite{Kolb:1996jt} and leptogenesis \cite{Giudice:1999fb} scenarios to explain the observed baryon asymmetry of the universe. Significant gravitational waves may also be generated as a consequence of the preheating mechanism \cite{Caprini:2018mtu}. Preheating can also lead to the production of particles out of thermal equilibrium that may act as a viable candidate for dark matter \cite{Chung:1998ua,Carena:2021bqm}. For a general review of preheating scenarios and their consequences, see~\cite{Allahverdi:2010xz,Amin:2014eta}.

In the momentum spectrum of fermions produced through preheating, there is a high occupation of particles per mode at low momentum (which we call the `bulk region') as a consequence of the non-adiabaticity condition for fermion mass oscillations \cite{Kofman:1997yn}. The condition essentially implies that the energy cost of particle production becomes zero when the effective mass of the fermions arising from the coupling with the inflaton becomes zero. For larger momentum values outside the bulk region, high occupation of particles per mode occurs only at specific discrete momentum values, giving rise to resonance peaks in the momentum spectrum of produced fermions.

In this paper, we analyze particle production assuming a $\lambda \phi^{4}$ quartic potential for the inflaton \cite{Linde:1983gd}, following from \cite{Greene:1998nh,Greene:2000ew,Greene:2002uot}. Even though $\lambda \phi^{4}$ inflation is strongly disfavoured due to constraints on slow-roll parameters \cite{Planck:2018jri}, we focus on describing only the oscillations of the inflaton field about its minimum by a $\lambda \phi^{4}$ potential. We analyze fermion production during the oscillatory phase of the inflaton, which is decoupled from the dynamics and constraints of the slow-roll phase. 
We do so under the assumption that the bare mass of the fermions $m_{\psi}$ is small compared to the scale of inflaton oscillations. As done before in \cite{Greene:1998nh,Greene:2000ew,Greene:2002uot}, the number density and the momentum spectrum can be obtained through numerical solutions of the mode equation and are dependent on a coupling parameter $q=h^{2}/\lambda$, which represents the strength of the coupling between the inflaton and the fermions. Here $h$ is the Yukawa coupling between the inflaton and fermions, and $\lambda$ is the quartic coupling for the inflaton. Fermion production through preheating can also occur with other inflation potentials, such as a hybrid inflation potential \cite{Garcia-Bellido:2000woy}.\footnote{For any case of particle production from a coherently oscillating field with a scalar coupling to fermions, the description of the mechanism will be similar to $\lambda\phi^{4}$ preheating.} Back-reaction effects can also be relevant for such a production mechanism \cite{Giudice:1999fb}, but we will assume that $q$ is sufficiently small that they can be neglected.

In \cite{Greene:1998nh}, it was found that the momentum values of the resonance peaks in the momentum spectrum can be predicted by a simple relation for $q \ll 1$. Although the mechanism is non-perturbative, we show this relation to be a consequence of energy conservation in a perturbative approximation for the description of resonant fermion production. We also generalize the idea to a semi-analytic relation that predicts the momentum values corresponding to resonance peaks in the momentum spectrum for any $q$, without the need to numerically solve the mode equation.

In \cite{Greene:1998nh,Greene:2000ew,Garcia-Bellido:2000woy}, it was shown that the momentum spectrum of the fermions produced during preheating is not degenerate.  For large $q$, the major contribution to the total number density of fermions comes from the bulk region, which was shown earlier in \cite{Giudice:1999fb}. However, for small $q$ ($\lesssim 0.01$), we find that the major contributions actually come from resonance peaks instead of the bulk region. We observe power-law behaviour for the total number density of fermions as a function of $q$ and find numerical approximations for the integrals describing the total number density, accounting for contributions from the full momentum distribution. We find good approximations in two regimes: the total number density is proportional to $q^{1/2}$ for $q\lesssim 0.01$ and proportional to $q^{3/4}$ for $q\gtrsim 10$.
We also give a detailed review on the derivation of the number density per mode of fermions and show how different descriptions in the literature (from \cite{Greene:1998nh}, \cite{Greene:2002uot} and \cite{Garcia-Bellido:2000woy}) are equivalent.

Fermions produced during preheating can be a promising dark matter candidate \cite{Carena:2021bqm,Bezrukov:2008ut,Garcia:2021iag,Klaric:2022qly,Kolb:1998ki}. Fermionic dark matter can also be produced non-perturbatively from a coupling to a coherently oscillating field that is not the inflaton \cite{Bjaelde:2010vt,Choi:2020nan,Choi:2020tqp}. Even if the fermions are produced with zero temperature, fermion degeneracy can make the fermions relativistic if the Fermi momentum is greater than their mass. One can then place cosmological constraints, for example from structure formation, on these fermions that will impose a lower bound on their mass $m_{\psi}$.  We adapt the bounds from \cite{Carena:2021bqm} to our model for approximately degenerate fermions and find that for small $q$ ($\lesssim 0.01$), the bound on $m_{\psi}$ is mildly strengthened. For larger $q$, we obtain the same bound on $m_{\psi}$ as in \cite{Carena:2021bqm}.

The organization of this paper is as follows. In Section \ref{sec: 2}, we describe fermionic preheating, the mechanism through which particles are produced during the oscillation the inflaton, and give expressions for the number density of the fermions produced. In Section \ref{sec: 3}, we find a simple semi-analytic relation describing the momentum values that correspond to resonance peaks in the momentum distribution of the fermions. In Section \ref{sec: 4}, we calculate the contributions of various components of the momentum distribution of fermions to their total number density and find approximations  for the contributions of both the bulk region and the resonance peaks to the total number density of fermions. In Section \ref{sec: 5}, we discuss the possibility of fermions produced by the aforementioned mechanism making up the observed dark matter density and state our conclusions. Various technical details and derivations are given in the appendices. 

\section{Fermionic Preheating}\label{sec: 2}

Fermionic particles can be produced non-perturbatively and out of thermal equilibrium through a coupling with a coherently oscillating scalar field. If the scalar driving this fermion production is the inflaton, then the process is called Fermionic Preheating \cite{Greene:1998nh,Giudice:1999fb,Greene:2000ew}. We choose to study the mechanism of production of fermions through preheating by considering inflation to be described by a simple $\lambda \phi^{4}$ quartic potential \cite{Greene:2002uot}.
\subsection{Inflaton Oscillations}\label{sec:Inflation}

The evolution of the inflaton ($\phi$) field is described by the following Klein-Gordon equation:
\begin{equation}\label{eq:Inf. KG eq_t}
\phi''(t)+3 H(t) \phi'(t) + \frac{\partial V}{\partial \phi}=0,
\end{equation}
where $'$ indicates derivatives with respect to time $t$. Here, $V$ represents the inflaton potential and $H(t)\equiv a'(t)/a(t)$ is the Hubble parameter (and $a$ is the scale factor), which is given by the Friedmann equation \cite{Kolb:1990vq}:
\begin{equation}\label{eq:Inf. Fm. eq}
H^{2}=\left(\frac{a'(t)}{a(t)}\right)^{2}=\frac{1}{3 M_{Pl}^{2}}\left(\frac{\phi'(t)^{2}}{2}+V(\phi)\right),
\end{equation}
where $M_{Pl}$ is the reduced Planck mass (i.e.~$M_{Pl}=1/\sqrt{8\pi G_{N}}\approx 2.44\times10^{18}$ GeV). For our paper, we take $V(\phi)=\lambda\phi^{4}/4$.

Inflation occurs as a `slow-roll' phase followed by inflaton oscillations. Particle production occurs only during the oscillatory phase of the inflaton evolution and not when it is undergoing slow-roll. The oscillations of the inflaton field have an attractor solution, which means that irrespective of the initial amplitude, the inflaton will evolve towards the same pattern of oscillations. Hence, it is possible to describe the oscillations of the inflaton field using a function that is independent of the initial amplitude. In order to do this, first the Klein-Gordon equation for the inflaton from Eq.~(\ref{eq:Inf. KG eq_t}) is written in terms of conformal variables, $\varphi=a\phi$, $\eta=\int dt/a(t)$, to remove the damping term:
\begin{equation}\label{eq:Inf. KG eq_eta}
\phi''(t)+3 H(t) \phi'(t) + \lambda\phi(t)^{3}=0
\rightarrow
\varphi''+\lambda \varphi^{3}\approx0,
\end{equation}
where $'$ now represents derivatives with respect to the conformal time $\eta$. We have neglected the $-a''\varphi/a$ term from the equation in conformal variables, as we can approximate that the scale factor $a$ grows linearly with $\eta$ during the oscillatory phase, which implies $a''(\eta)\approx0$. The conformal variables can then be rescaled using the initial amplitude $\varphi_{0}$:
\begin{equation}\label{eq:Inf. var. transf.}
f(\tau)=\varphi(\tau)/\varphi_{0},
\quad
\tau=\sqrt{\lambda}\varphi_{0}\eta.
\end{equation}
Using $\dot{}$ to represent derivatives with respect to $\tau$, we can now describe the inflaton oscillations in dimensionless variables by:
\begin{equation}\label{eq:Inf. f eq}
\ddot{f}+f^{3}=0.
\end{equation}
The solution of this equation is a Jacobi Elliptic Cosine\footnote{We have used the notation from Wolfram Mathematica \cite{reference.wolfram_2025_jacobicn} for the Jacobi Elliptic Cosine, while some references like \cite{Greene:2002uot} use the notation $cn(u,k)$ where $k=\sqrt{m}$.} of the form \cite{Greene:2002uot}:
\begin{equation}\label{eq:Inf. f=cn}
f(\tau)=cn\left(\tau-\tau_{1},m\right);
\quad
m=1/2.
\end{equation}
We set $\tau_{1}$ to $0$ to ensure that $f(\tau)$ is symmetric around $\tau=0$. This is important for the derivation of equations needed \cite{Mostepanenko:1974im} to study the late-time behaviour of the produced fermions, and we elaborate on this in Appendix \ref{AppE}.
The general $cn(u,m)$ function can be represented as a series expansion \cite{abramowitz+stegun}. For the case of $m=1/2$, it is written as \cite{Greene:2002uot}:
\begin{equation}\label{eq:Inf. cn(u,1/2)}
cn(u,1/2)=\frac{8 \pi \sqrt{2}}{T} \sum_{n=1}^{\infty} \frac{e^{-\pi(n-1 / 2)}}{1+e^{-2\pi(n-1 / 2)}} \cos \left(\frac{2 \pi(2 n-1) u}{T}\right).
\end{equation}
The Jacobi Elliptic function is periodic with its period for $m=1/2$ given by an elliptic integral as:
\begin{equation}\label{eq:Inf TimePeriod}
T=4\int_0^{\pi / 2} \frac{d \theta}{\left(1-\frac{1}{2}\sin ^2 \theta\right)^{1 / 2}}\approx 7.4163.
\end{equation}
The oscillatory phase of the inflaton dynamics can thus be described by:
\begin{equation}\label{eq:Inf. phi4 solution_t oscillation}
\phi_{Osc}(t)=\left(\frac{3 M_{Pl}^{2}}{\lambda}\right)^{1/4}\frac{cn\left((48\lambda M_{Pl}^{2})^{1/4}t^{1/2}\right)}{t^{1/2}}.
\end{equation}
\subsection{Production and Number Density of Fermions}\label{sec:NumDen}
Fermions ($\psi$) are assumed to have a $h\phi \overline{\psi}\psi$ Yukawa coupling to the inflaton field, along with a mass term $m_{\psi}\overline{\psi}\psi$. As in the previous section, we take the expansion of the universe into account by first using the following conformal transformations of the relevant quantities:
\begin{equation}\label{eq:NumDen t to eta transf.}
\psi \rightarrow \Psi=a^{3/2}\psi,
\quad
\phi \rightarrow \varphi=a\phi,
\quad
t\rightarrow\eta=\int\frac{dt}{a(t)}.
\end{equation}
Then, the fermion field satisfies the Dirac equation: 
\begin{equation}\label{eq:NumDen Dirac eq_eta}
\left(i\gamma^{\mu}\partial_{\mu}-h\varphi(\eta)-m_{\psi}a(\eta)\right)\Psi = 0,
\end{equation}
where $\partial_{\mu}$ is in terms of conformal variables. The solution for this equation can be found by considering an auxiliary field $X(\mathbf{x},\eta)$, separated into components (with comoving momentum $\mathbf{k}$ and position $\mathbf{x}$) as follows:
\begin{equation}\label{eq:NumDen X(x,eta)}
X(\mathbf{x},\eta)=e^{(i\mathbf{k}\cdot\mathbf{x})}X_{k}(\eta)R_{\pm}(\mathbf{k}),
\quad
\text{or:}
\quad
X(\mathbf{x},\eta)=e^{(-i\mathbf{k}\cdot\mathbf{x})}X_{k}(\eta)\overline{R}_{\pm}(\mathbf{k}),
\end{equation}
where: $R_{\pm}$, $\overline{R}_{\pm}$ are eigenvectors of $\gamma^{0}$ and of the helicity operator $\mathbf{\hat{k}}\cdot\mathbf{\Sigma}$ (see Appendix \ref{AppA}). The relevant ansatz to solve the Dirac equation is then written as follows:
\begin{equation}\label{eq:NumDem psi ansatz eta}
\Psi=\left(i\gamma^{\mu}\partial_{\mu}+h\varphi(\eta)+m_{\psi}a(\eta)\right)X(\mathbf{x},\eta).
\end{equation}
Plugging this ansatz in Eq.~(\ref{eq:NumDen Dirac eq_eta}) and using Eq.~\eqref{eq:NumDen X(x,eta)} yields a differential equation:
\begin{equation}\label{eq:NumDen X_k mode eq_eta}
X_{k}''(\eta)+\left[
k^{2}+
\left(h\varphi(\eta)+m_{\psi}a(\eta)\right)^{2}-i
\left(h\varphi'(\eta)+m_{\psi}a'(\eta)\right)
\right]
X_{k}(\eta) = 0,
\end{equation}
where $'$ represents derivatives with respect to the conformal time $\eta$. This is an oscillator-like differential equation and we can represent its `frequency' as:
\begin{equation}\label{eq: Omega_k(eta)}
\Omega_{k}^{2}(\eta)=k^{2}+(h\varphi(\eta)+m_{\psi}a(\eta))^{2}.
\end{equation}
This `mode equation' of Eq.~\eqref{eq:NumDen X_k mode eq_eta} is independent of the inflaton potential. In fact, it will be valid in the study of fermion production through a Yukawa coupling to any coherently oscillating scalar field.

In order to study fermion production, the field can also be written as:
\begin{equation}\label{eq:NumDen Psi(eta,x) operators}
\Psi(\eta,\mathbf{x})=\sum_{s=\pm}\int\frac{d^{3}k}{(2\pi)^{3}}\left(\hat{a}_{k,s}u_{k,s}(\eta)e^{+i\mathbf{k} \cdot \mathbf{x}}+\hat{b}^{\dagger}_{k,s}v_{k,s}(\eta)e^{-i\mathbf{k} \cdot \mathbf{x}}\right),
\end{equation}
where $\hat{a}^{\dagger}_{k,s}, \hat{a}_{k,s}$ and $\hat{b}^{\dagger}_{k,s}, \hat{b}_{k,s}$ are fermion creation and annihilation operators for particles and antiparticles and $u_{k,s}(\eta)$ and $v_{k,s}(\eta)$ are respectively the positive- and negative-frequency eigenspinors of the Dirac equation, Eq.~(\ref{eq:NumDen Dirac eq_eta}) (see Appendix \ref{AppA}).

The Hamiltonian for the system is written in conformal variables as follows \cite{Greene:2002uot}:
\begin{equation}\label{eq:NumDen H_D}
H_{D}=\int d^{3}x\left(i\Psi^{\dagger}\partial_{\eta}\Psi\right).
\end{equation}
$H_{D}$ is diagonal in the creation and annihilation operators at an initial time $\eta=\eta_{0}$ but not diagonal at later times. This breaking of the time-translational symmetry of the Hamiltonian is a signature of particle creation. The number density per mode of the produced fermions i.e.~$n_{k}(\eta)$, can then be calculated after diagonalising $H_{D}$ at time $\eta$ using a Bogoliubov transformation. A Bogoliubov transformation \cite{Bogolyubov:1958kj} is a rotation of the creation and annihilation operators that preserves anti-commutators for fermions (or commutators for bosons) and diagonalises the Hamiltonian in terms of the new operators.

First, we use Eq.~(\ref{eq:NumDen Psi(eta,x) operators}) to evaluate the Hamiltonian expression of Eq.~\eqref{eq:NumDen H_D} in terms of the fermion creation and annihilation operators (see Appendix \ref{AppA} for some details used in the derivation):
\begin{eqnarray}\label{eq:NumDen H_D operators}
H_{D}(\eta)&=&\int \frac{d^{3}k}{(2\pi)^{3}}
\left[E(\eta)\left(\hat{a}_{k,+}^{\dagger}\hat{a}_{k,+}-\hat{b}_{k,+}\hat{b}_{k,+}^{\dagger}+\hat{a}_{k,-}^{\dagger}\hat{a}_{k,-}-\hat{b}_{k,-}\hat{b}_{k,-}^{\dagger}\right)\right. \nonumber \\
&& \left. +F(\eta)
\left(\hat{a}_{k,+}^{\dagger}\hat{b}_{-k,-}^{\dagger}+\hat{a}_{k,-}^{\dagger}\hat{b}_{-k,+}^{\dagger}\right)
+F^{*}(\eta)
\left(\hat{b}_{-k,+}\hat{a}_{k,-}+\hat{b}_{-k,-}\hat{a}_{k,+}\right) 
\right].
\end{eqnarray}
Here we have defined:
\begin{equation}\label{eq:NumDen E(eta),F(eta)}
\begin{split}
&E(\eta)=\left(h\varphi(\eta)+m_{\psi}a(\eta)\right)\left(|X'_{k}(\eta)|^{2}+\Omega_{k}^{2}(\eta)|X_{k}(\eta)|^{2}\right)+2\Omega_{k}^{2}(\eta)\text{Im}(X_{k}(\eta)X'^{*}_{k}(\eta)),\\
&F(\eta)=k(X'^{*}_{k}(\eta))^{2}+k\Omega_{k}^{2}(\eta)(X^{*}_{k}(\eta))^{2}.
\end{split}
\end{equation}

For any two solutions $f(\eta)$ and $g(\eta)$ of the mode equation in Eq.~\eqref{eq:NumDen X_k mode eq_eta}, including the case where $f(\eta) = g(\eta)$, the following equation will hold \cite{Mostepanenko:1974im}:
\begin{equation}\label{eq:NumDen f, g identity}
f'(\eta)g'^{*}(\eta)+\Omega_{k}^{2}(\eta)f(\eta)g^{*}(\eta)+i(m_{\psi}a(\eta)+h\varphi(\eta))(f'(\eta)g^{*}(\eta)-f(\eta)g'^{*}(\eta))=C,
\end{equation}
where $C$ is a constant. This can be easily shown by taking the derivative of both sides of Eq.~\eqref{eq:NumDen f, g identity} with respect to $\eta$ and then substituting the mode equation from Eq.~\eqref{eq:NumDen X_k mode eq_eta}.
We can then use the particular case of Eq.~(\ref{eq:NumDen f, g identity}) for which $f(\eta)=g(\eta)\equiv X_{k}(\eta)$, in order to show that (Appendix \ref{AppB}):
\begin{equation}\label{eq:NumDen E^2+F^2, with C}
E(\eta)^2+|F(\eta)|^2=C^2\Omega_{k}^2(\eta).
\end{equation}
This is a generalisation of a similar equation in \cite{Dolgov:1989us}, where $C$ is taken to be $1$. However, $C$ actually depends on the initial conditions of the mode Eq.~(\ref{eq:NumDen X_k mode eq_eta}).

The initial conditions should ensure that $H_{D}$ is diagonal at $\eta=\eta_{0}$. One such choice of initial conditions is as follows \cite{Garcia-Bellido:2000woy}:
\begin{equation}\label{eq:NumDen X_k intial cond.}
\begin{split}
&X_{k}(\eta_{0})=\left\{2\Omega_{k}(\eta_{0})\left[\Omega_{k}(\eta_{0})+(h\varphi(\eta_{0})+m_{\psi}a(\eta_{0}))\right]\right\}^{-1/2},\\
&X'_{k}(\eta_{0})=-i\Omega_{k}(\eta_{0})X_{k}(\eta_{0}).
\end{split}
\end{equation}
This choice of $X'_{k}(\eta_{0})$ ensures that $H_{D}(\eta_{0})$ is diagonal in the creation and annihilation operators because $F(\eta_{0})=0$. We have chosen $X_{k}(\eta_{0})$ so that $E(\eta_{0})=\Omega_{k}(\eta_{0})$ and therefore $C=1$. We will see that this choice also ensures consistency with the initial condition for the number density per mode, i.e.~$n_{k}(\eta_{0})=0$, from the expression we will derive for $n_{k}(\eta)$. We can find the expression for the number density of fermions after applying a Bogoliubov transformation and diagonalising the transformed $H_{D}(\eta)$ by setting the non-diagonal coefficients to $0$. The relevant Bogoliubov transformation is as follows \cite{Dolgov:1989us}:
\begin{equation}\label{eq:NumDen Bogoliubov transf.}
\begin{split}
&\hat{a}_{k,+}=\alpha_{1}^{*}\hat{c}_{k,+}-\beta_{1} \hat{d}_{k,+}^{\dagger},
\quad
\hat{a}_{k,-}=\alpha_{2}^{*}\hat{c}_{k,-}-\beta_{2} \hat{d}_{k,-}^{\dagger},\\
&\hat{b}_{-k,+}=\alpha_{2}^{*}\hat{d}_{k,-}+\beta_{2} \hat{c}_{k,-}^{\dagger},
\quad
\hat{b}_{-k,-}=\alpha_{1}^{*}\hat{d}_{k,+}+\beta_{1} \hat{c}_{k,+}^{\dagger}.
\end{split}
\end{equation}
Preserving the canonical anti-commutation relations $\{\hat{a}_{k,s},\hat{a}_{k,s}^{\dagger}\}=1$ and the same for the other three sets of operators implies that the Bogoliubov coefficients $\alpha_{j},\beta_{j}$ have the following relation:
\begin{equation}\label{eq:NumDen Bogoliubov transf. properties}
|\alpha_{j}|^{2}+|\beta_{j}|^{2}=1, \qquad \qquad j=1,2.
\end{equation}
We can assume that the operators with $s=+$ and comoving momentum $\mathbf{k}$ are equivalent to other operators with $s=-$ and comoving momentum $-\mathbf{k}$, which implies: $\hat{c}_{k,\pm}\equiv \hat{c}_{-k,\mp}, \quad \hat{d}_{k,\pm}\equiv \hat{d}_{-k,\mp}$. This is a consequence of a similar correspondence between the $R_{\pm}$ and $\overline{R}_{\pm}$ eigenvectors (see Eq.~\eqref{eq: R pm relations}). We use the properties of the Bogoliubov coefficients and the non-diagonal coefficients to first see: $\alpha_{1}=\alpha_{2}=\alpha, \quad \beta_{1}=\beta_{2}=\beta$. Using these, we rewrite Eq.~(\ref{eq:NumDen H_D operators}) in terms of the time-dependent Bogoliubov coefficients and the new operators:
\begin{equation}\label{eq:NumDen H_D Bogoliubov coeff.}
\begin{split}
&H_{D}(\eta)=\int \frac{d^{3}k}{(2\pi)^{3}}\sum_{s=\pm}\left\{
\hat{c}_{k,s}^{\dagger}\hat{c}_{k,s}\left(E(|\alpha|^{2}-|\beta|^{2})+F\alpha\beta^{*}+F^{*}\alpha^{*}\beta\right)\right. \\
-&\hat{d}_{k,s}\hat{d}_{k,s}^{\dagger}\left(E(|\alpha|^{2}-|\beta|^{2})+F\alpha\beta^{*}+F^{*}\alpha^{*}\beta\right)\\
+&\left. \hat{c}_{k,s}^{\dagger}\hat{d}_{k,s}^{\dagger}\left(F\alpha^{2}-F^{*}\beta^{2}-2E\alpha\beta\right)
+\hat{d}_{k,s}\hat{c}_{k,s}\left(F^{*}(\alpha^{*})^{2}-F(\beta^{*})^{2}-2E\alpha^{*}\beta^{*}\right)\right\}.
\end{split}
\end{equation}
If we apply the number density operator to the ground state of the Hamiltonian and use the Bogoliubov transformation from Eq.~\eqref{eq:NumDen Bogoliubov transf.}, we can see that the number density per $k$ mode of particles is given by $n_{k}(\eta)=|\beta|^{2}$. We shall henceforth call $n_{k}(\eta)$ the number density per mode and show how to compute it
in later sections. We can evaluate $|\beta|^{2}$ from Eq.~\eqref{eq:NumDen H_D Bogoliubov coeff.} (Appendix \ref{AppC}) and get: 
\begin{equation}\label{eq:NumDen n_k(eta), with C}
n_{k}(\eta)=|\beta|^{2}=\frac{C\Omega_{k}(\eta)-E(\eta)}{2C\Omega_{k}(\eta)}.
\end{equation}
We have used Eq.~(\ref{eq:NumDen E^2+F^2, with C}) here to simplify the expression. Clearly, for the choice of initial conditions Eq.~(\ref{eq:NumDen X_k intial cond.}), the resulting values of $C=1$ and $E(\eta_{0})=\Omega_{k}(\eta_{0})$ ensure that $n_{k}(\eta_{0})=0$, i.e. no particles are produced before oscillations start, exactly what we expect.

Similarly to the mode equation, this general expression for $n_{k}(\eta)$ will also be valid for any scalar field potential. The form for $\Omega_{k}(\eta)$ and $E(\eta)$ will also remain exactly the same, with the only difference being the way the scalar oscillations $\varphi(\eta)$ are described. The $n_{k}$ expressions in the literature (\cite{Greene:1998nh}, \cite{Greene:2002uot}, \cite{Garcia-Bellido:2000woy}) look different from our expression, but they represent the same number density under the choice of initial conditions of the mode equation, which we show in Appendix~\ref{AppD}.
\subsection{Number Density of Fermions in $\lambda\phi^{4}$ inflation}\label{sec:Lambda phi4}
Now, we shall use the general framework above and apply it to the case of $\lambda\phi^{4}$ inflation with an inflaton quartic coupling $\lambda$ and Yukawa coupling $h$, as in Eq.~\eqref{eq:Inf. KG eq_eta} and Eq.~\eqref{eq:NumDen Dirac eq_eta} respectively.
All conformal variables can first be transformed into dimensionless variables using these couplings and the initial amplitude of the inflaton $\varphi_{0}$, as follows:
\begin{equation}\label{eq:NumDen eta to tau transf.}
\begin{split}
&\kappa^{2}=\frac{k^{2}}{\lambda \varphi_{0}^{2}},
\quad
q=\frac{h^{2}}{\lambda}, 
\quad
\tau=(\sqrt{\lambda}\varphi_{0})\eta,\\ 
X_{\kappa}=&(\sqrt{\lambda}\varphi_{0})X_{k},
\quad
f(\tau)=\frac{\varphi(\tau)}{\varphi_{0}}, 
\quad
\Omega_{\kappa}(\tau)=\frac{\Omega_{k}(\eta)}{\sqrt{\lambda}\varphi_{0}}.   
\end{split}
\end{equation}
We take $f(\tau)$ to be the Jacobi Elliptic Cosine from Eq.~\eqref{eq:Inf. f=cn}, thereby neglecting back-reaction \cite{Giudice:1999fb}. We show that this assumption is justified in Appendix~\ref{AppF}.
We consider fermions to be light compared to the amplitude of inflaton oscillations and therefore can assume the bare mass of the fermion $m_{\psi}=0$. Then, using $\dot{}$ to represent derivatives with respect to $\tau$, the mode equation can be written using dimensionless variables:
\begin{equation}\label{eq:NumDen X_kappa mode eq.}
\ddot{X_{\kappa}}(\tau)+\left(\kappa^{2}+q f^{2}(\tau)-i\sqrt{q}\dot{f}(\tau)\right)X_{\kappa}(\tau)=0.
\end{equation}
And we can write:
\begin{equation}\label{eq: Omega_k(tau)}
\Omega_{\kappa}(\tau)=\sqrt{\kappa^{2}+q f^{2}(\tau)}.
\end{equation}
The parameter $q$ is called the `coupling parameter' for this mode equation, while the parameter $\kappa$ characterises the various fermion momentum modes. We choose the same initial conditions of the mode equation, which were described earlier in Eq.~\eqref{eq:NumDen X_k intial cond.}, and they can be represented in these new variables (with $\tau=0$ representing the beginning of oscillations):
\begin{equation}\label{eq:NumDen X_kappa initial cond.}
\begin{split}
&X_{\kappa}(0)=\left[2\Omega_{\kappa}(0)\left[\Omega_{\kappa}(0)+\sqrt{q}f(0)\right]\right]^{-1/2},\\
&\dot{X}_{\kappa}(0)=-i\Omega_{\kappa}(0)X_{\kappa}(0).
\end{split}
\end{equation}
These initial conditions ensure $C=1$ again and $E_{\kappa}(0)=\Omega_{\kappa}(0)$; hence the relevant number density expression can be written in the dimensionless variables:
\begin{equation}\label{eq:NumDen n_kappa(tau), C=1}
n_{\kappa}(\tau)=\frac{\Omega_{\kappa}(\tau)- E_{\kappa}(\tau)}{2\Omega_{\kappa}(\tau)},
\end{equation}
where:
\begin{equation}\label{eq:NumDen E_kappa(tau)}
E_{\kappa}(\tau)=\sqrt{q}f(\tau)(|\dot{X}_{\kappa}|^{2}+\Omega_{\kappa}^{2}(\tau)|X_{\kappa}|^{2})+2\Omega_{\kappa}^{2}(\tau)\text{Im}(X_{\kappa}\dot{X}_{\kappa}^{*}).
\end{equation}

\begin{figure}[H]
    \centering
    \includegraphics[width=\textwidth]{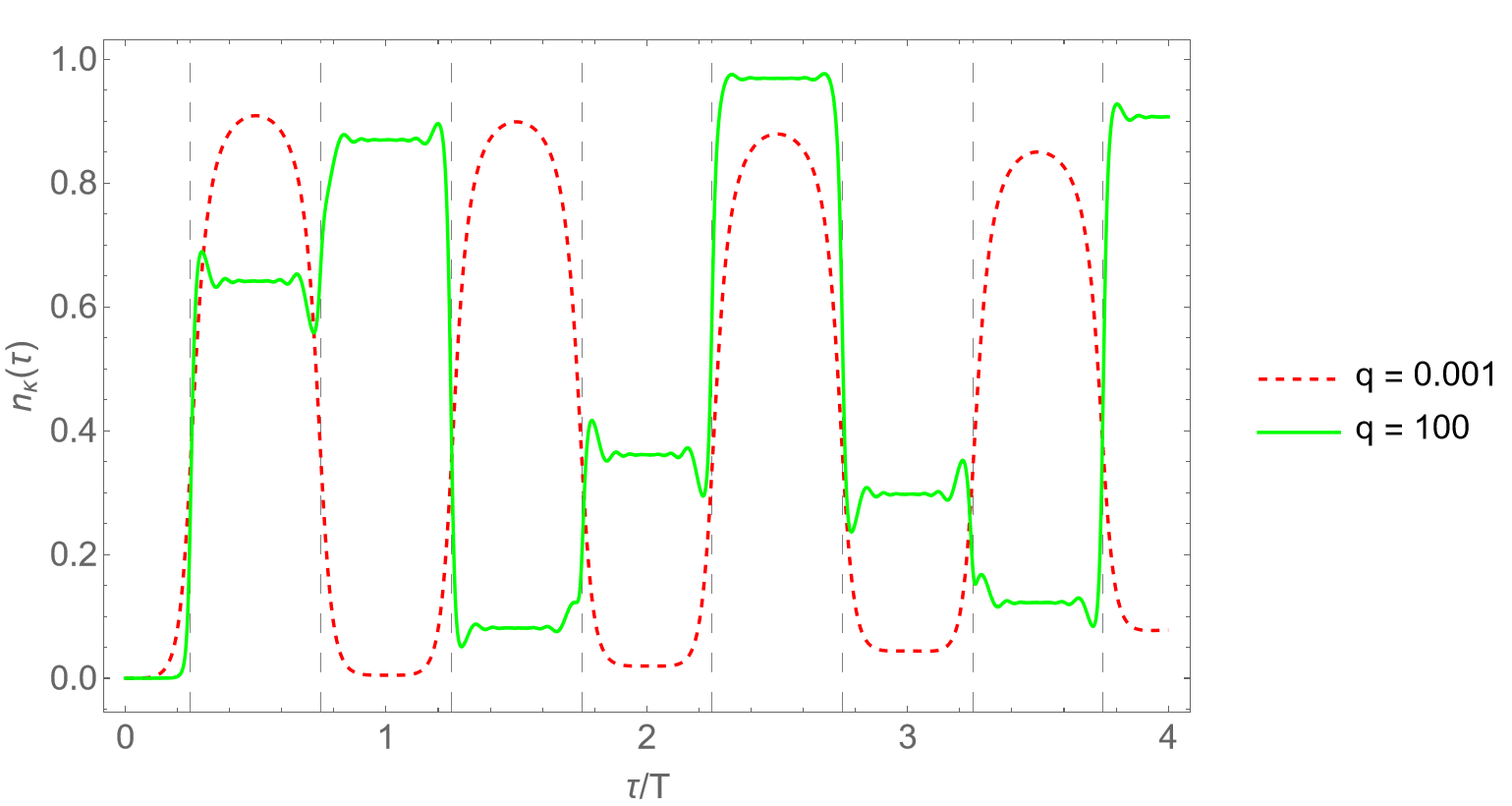}
	\caption{$n_{\kappa}(\tau)$ from Eq.~\eqref{eq:NumDen n_kappa(tau), C=1} as a function of time, using the period $T$ of $f(\tau)$ for two different $q$ values. The dashed vertical lines correspond to $\tau=(2n+1)T/4,n=0,1,2,\dots$, the instants of time when $f(\tau)=0$. We choose $\kappa=0.01$ for $q=0.001$ and $\kappa=1$ for $q=100$, which we explain in Section~\ref{sec:LateTime}.}\label{fig: nk small and large q kap in bulk}
\end{figure}

For small values of the coupling parameter $q$, the number density as a function of dimensionless conformal time $\tau$ follows a smooth pattern. However, for larger values of $q$, the variation of the number density with time has very rapid changes as a consequence of non-adiabaticity, with jumps in the pattern whenever $f(\tau)=0$. These patterns are shown in Fig.~\ref{fig: nk small and large q kap in bulk}.

\begin{figure}[H]
	\centering
	\includegraphics[width=\textwidth]{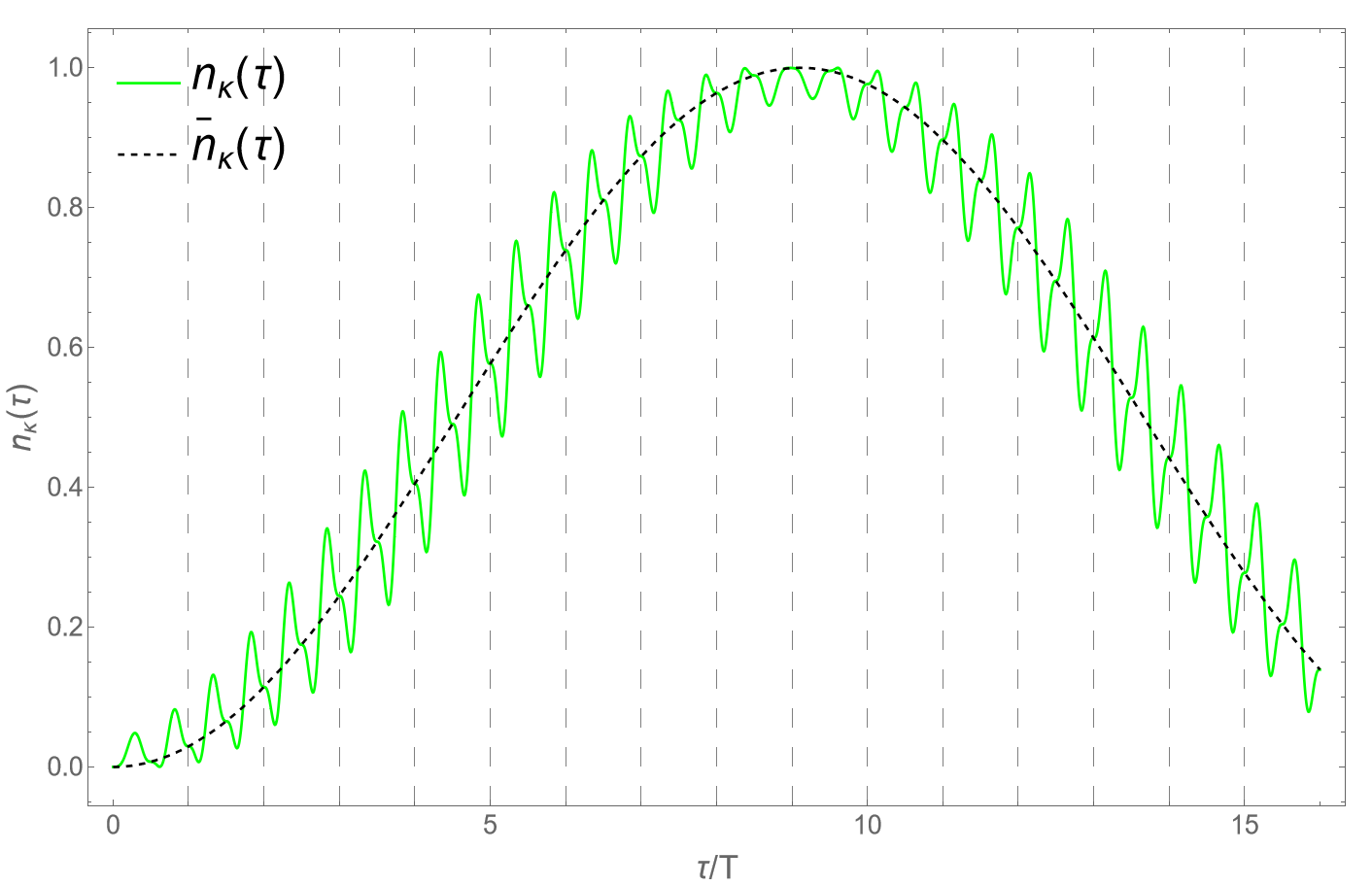}
	\caption{Numerically evaluated functions $\overline{n}_{\kappa}(\tau)$ and $n_{\kappa}(\tau)$ as a function to $\tau/T$, where $T$ is the period of the inflaton solution $f(\tau)$. We choose $q=1$ and $\kappa=1.05236$, which corresponds to one of the resonance peaks as shown in Fig.~\ref{fig: nbarkap for q = 0.01, 1, 100}. $\overline{n}_{\kappa}(\tau)$ and $n_{\kappa}(\tau)$ are guaranteed to be identical at integer multiples of $T$ (and also match at some other points). Dashed vertical lines represent when $\tau$ is an integer multiple of $T$.}
    \label{fig: n_kappa and bar n_kappa}
\end{figure}

The spiky pattern for larger $q$ values can be approximated by a  smooth function $\overline{n}_{\kappa}(\tau)$ that matches the value of $n_{\kappa}(\tau)$ when $\tau$ is an integer multiple of the period $T$. This $\overline{n}_{\kappa}(\tau)$ is given by \cite{Mostepanenko:1974im,Greene:2002uot}: 
\begin{equation}\label{eq:NumDen bar n_kappa(tau)}
\overline{n}_{\kappa}(\tau)=F_{\kappa}\sin^{2}\left(\nu_{\kappa}\tau\right),
\end{equation}
where:
\begin{equation}\label{eq:NumDen nu_kappa and F_kappa}
\begin{split}
&\nu_{\kappa}=\frac{1}{T}\cos^{-1}\left(|\text{Re}\left(X^{1}_{\kappa}(T)\right)|\right),\\
&F_{\kappa}=\frac{1}{\sin^{2}(\nu_{\kappa}T)}\cdot\frac{\kappa^{2}}{\Omega_{\kappa}^{2}}\left(\text{Im}\left(X^{1}_{\kappa}(T)\right)\right)^{2}.
\end{split}
\end{equation}
$X^{1}_{\kappa}(\tau)$ used here is the numerical solution of the same differential Eq.~\eqref{eq:NumDen X_kappa mode eq.}, but with the initial conditions:
\begin{equation}
\label{eq:NumDen X^1_kappa inital cond.}
X^{1}_{\kappa}(0)=1,
\quad
\dot{X}^{1}_{\kappa}(0)=0.
\end{equation}
The details of why $n_\kappa(\tau)$ matches $\overline{n}_{\kappa}(\tau)$ at certain times is discussed in \cite{Mostepanenko:1974im}, and we describe the procedure in our notation in Appendix~\ref{AppE}. As an example, we plot $n_{\kappa}(\tau)$ from Eq.~\eqref{eq:NumDen n_kappa(tau), C=1} and $\overline{n}_{\kappa}(\tau)$ from Eq.~\eqref{eq:NumDen bar n_kappa(tau)} for $q=1$ in Fig. \ref{fig: n_kappa and bar n_kappa}. We have chosen a particular $\kappa$ that we explain in Section~\ref{sec: 3}.

Instead of $n_{\kappa}(\tau)$ and $\overline{n}_{\kappa}(\tau)$ as functions of time, if we plot them as functions of $\kappa$ for a fixed value of time that is an integral multiple of $T$, $n_{\kappa}(nT)$ and $\overline{n}_{\kappa}(nT)$ as functions of $\kappa$ will match for any given $q$. We use $\overline{n}_{\kappa}(\tau)$ to approximate the long-period behaviour of $n_{\kappa}(\tau)$ because $\overline{n}_{\kappa}(\tau)$ is numerically much simpler to calculate. As they have the same pattern as functions of $\kappa$ at points of time where they match, we can use the $\overline{n}_{\kappa}(\tau)$ approximation to study the pattern of filling of the $\kappa$ modes for the number density of produced fermions. 
\subsection{Late-Time Filling of Fermion Modes}\label{sec:LateTime}
Now, we can study the filling of the $\kappa$ modes by seeing the variation of $\overline{n}_{\kappa}(\tau)$ with $\kappa$ for fixed values of $q$ and at fixed instants of time. We do so for four values of the coupling parameter: $q=0.001$, $q=0.01$, $q=1$, and $q=100$, and at two particular instants of time: $\tau=5T$ and $20T$. The plots for these cases are shown in Fig.~\ref{fig: nbarkap for q = 0.01, 1, 100}. We also plot the $F_{\kappa}$ envelopes (see Eq.~\eqref{eq:NumDen nu_kappa and F_kappa}) for the $q$ values on their respective graphs.

From the plots in Fig.~\ref{fig: nbarkap for q = 0.01, 1, 100}, we can see that the lower $\kappa$ modes of the $F_{\kappa}$ envelopes are well populated. We call this low $\kappa$ regime `bulk region.' The higher $\kappa$ modes exhibit resonance peaks where $F_\kappa$ is only large near discrete values of $\kappa$, and these resonances are a consequence of constructive interference between the inflaton oscillations and the fermion modes. We describe how $\kappa$ modes corresponding to resonance peaks can be predicted using an analytic or semi-analytic relation, without having to solve the mode equation in Eq.~\eqref{eq:NumDen X_kappa mode eq.}, in Section \ref{sec: 3}. 
In Fig.~\ref{fig: nk small and large q kap in bulk}, we choose $\kappa=0.01$ and $\kappa=1$ for $q=0.001$ and $q=100$, respectively, because those values of $\kappa$ are in the bulk region as can be seen in Fig.~\ref{fig: nbarkap for q = 0.01, 1, 100}. 

From Fig.~\ref{fig: nbarkap for q = 0.01, 1, 100}, we see that the occupancy of each mode oscillates with $\kappa$ within the envelope $F_\kappa$, and that the frequency of the oscilations in $\kappa$ space increase with time. This means that at late time, modes that are nearby one another in $\kappa$ will decohere. Hence, if the inflaton oscillations and subsequent particle production occur for a long time $\tau \gg T$, we can average the $\sin^{2}\left(\nu_{\kappa}\tau\right)$ term in Eq.~\eqref{eq:NumDen bar n_kappa(tau)} to $1/2$, and we can approximate $\overline{n}_{\kappa}$ simply from the $F_{\kappa}$ envelope, as: 
\begin{equation}\label{eq:LateTime nbar tau avg}
\bar n_{\kappa}(\tau) \xrightarrow{\tau\gg T}  \langle F_{\kappa} \sin^2(\nu_k \tau) \rangle = F_{\kappa}/2.
\end{equation}
We conclude from this that we can approximate the filling of the $\kappa$ modes of the produced fermions by only the $F_{\kappa}$ envelope. As such, we study the $\kappa$ modes of the $F_{\kappa}$ envelope in terms of its bulk region and resonance peaks, which are a consequence of the behaviour of the oscillator-like equation from Eq.~\eqref{eq:NumDen X_kappa mode eq.}. The exact patterns of how momentum modes are filled vary with $q$, depending on whether $q$ is less than $\mathcal{O}(0.1)$ or greater. Although $F_{\kappa}$ can be numerically evaluated, it becomes difficult to numerically resolve resonance peaks at higher $\kappa$ values, in particular for smaller values of $q$, as the peak widths are very narrow. Thus, it becomes important to use analytical approximations to estimate the contributions of different $\kappa$ regions to the number density.

To obtain an expression for the total number density of fermions, we first see that we can use the conformal variables and the density of states in $\mathbf{k}$  space to write the total comoving number density of fermions by integrating over the number density per mode as:
\begin{equation}\label{eq:NumDen total n, conformal}
n_{\psi+\overline{\psi}}=\frac{2\times2\times4\pi}{(2\pi)^{3}}\int dk k^{2} n_{k}=\frac{2}{\pi^{2}}\int dk k^{2} n_{k}.
\end{equation}
The additional factors of $2$ come from counting particles and anti-particles, and taking into account spin-up and spin-down fermions. The $4\pi$ comes from considering a sphere in $\mathbf{k}$ space and $(2\pi)^{3}$ comes from the expression for the density of states. We approximate $n_{\kappa}$ by $\overline{n}_{\kappa}$ from Eq.~\eqref{eq:NumDen bar n_kappa(tau)}, and use $\overline{n}_{\kappa}(\tau)\xrightarrow{\tau\gg T} F_{\kappa}/2$. Hence, if we convert the integral above into dimensionless variables using $k=\kappa\sqrt{\lambda}\varphi_{0}$ (from Eq.~\eqref{eq:NumDen eta to tau transf.}), we can write the total comoving number density as:
\begin{equation}\label{eq:NumDen total n, dimless}
n_{\psi+\overline{\psi}}=\frac{2\lambda^{3/2}\varphi_{0}^{3}}{\pi^{2}}\int d\kappa \kappa^{2} \overline{n}_{\kappa}\approx
\frac{\lambda^{3/2}\varphi_{0}^{3}}{\pi^{2}}\int d\kappa \kappa^{2}F_{\kappa}.
\end{equation}
The total physical number density is obtained from the total comoving number density by dividing by $a^{3}$ (where $a$ is the scale factor) and hence:
\begin{equation}\label{eq:NumDen total n, physical, with a}
n=\frac{n_{\psi+\overline{\psi}}}{a^{3}}=\frac{2\lambda^{3/2}\varphi_{0}^{3}}{\pi^{2}a^{3}}\int d\kappa \kappa^{2} \overline{n}_{\kappa}\approx
\frac{\lambda^{3/2}\varphi_{0}^{3}}{\pi^{2}a^{3}}\int d\kappa \kappa^{2}F_{\kappa}.
\end{equation}

\begin{figure}[H]
    \subfloat
    {%
    \includegraphics[width=0.5\textwidth,height=0.2\textheight]{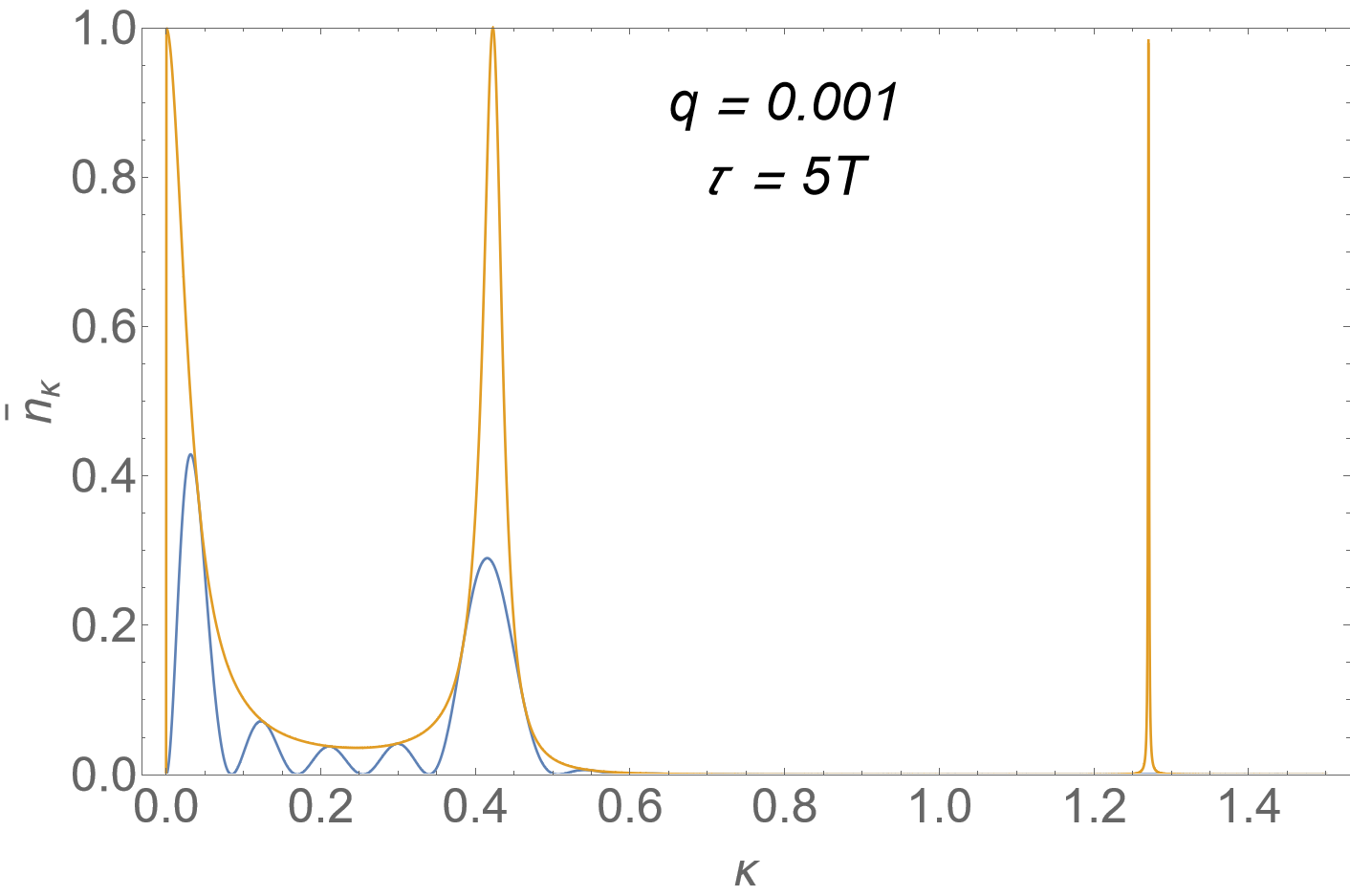}
    }
    \hfill
    \subfloat
    {%
    \includegraphics[width=0.5\textwidth,height=0.2\textheight]{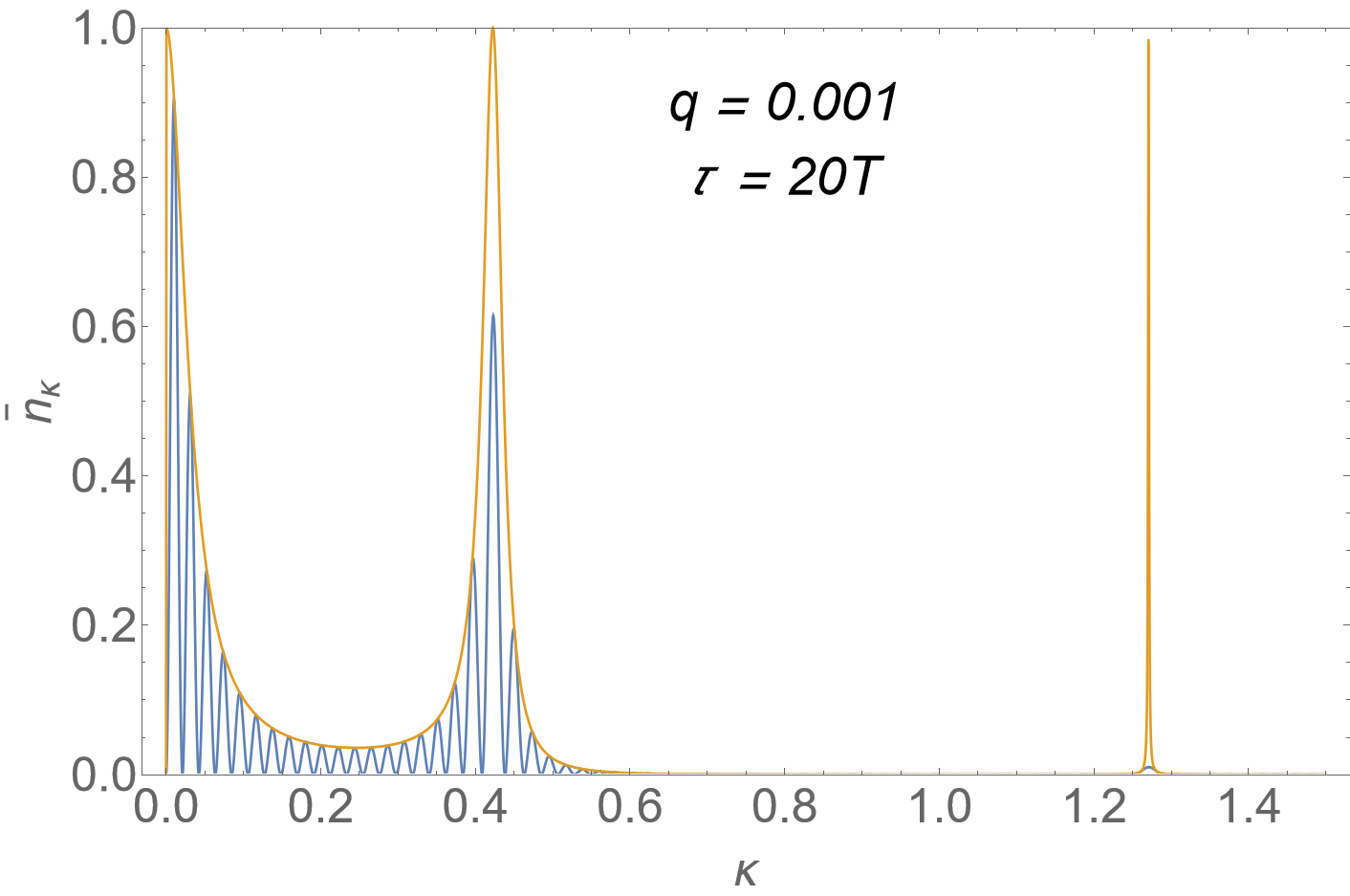}
    }
    \\
    \subfloat
    {%
    \includegraphics[width=0.5\textwidth,height=0.2\textheight]{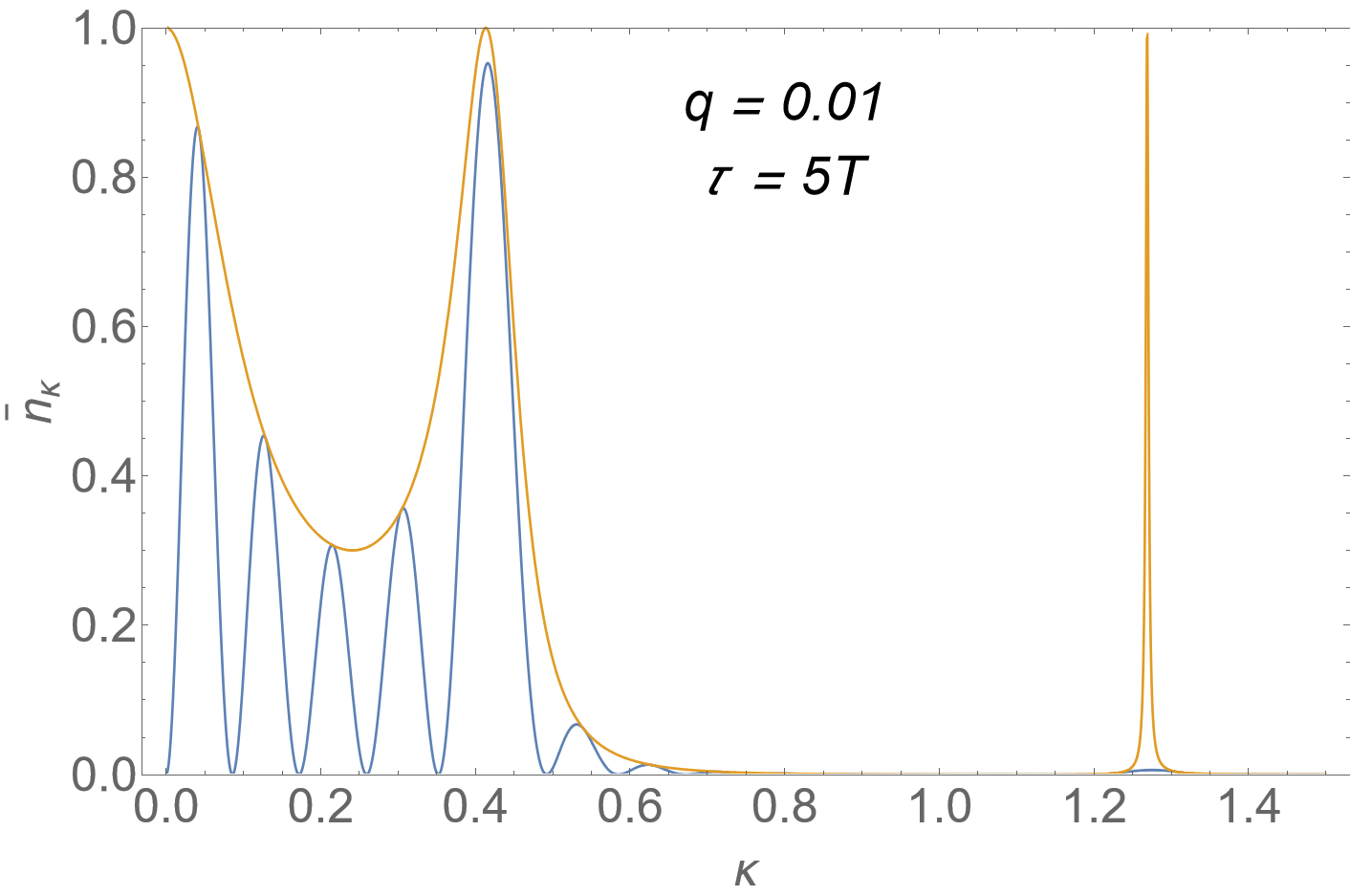}
    }
    \hfill
    \subfloat
    {%
    \includegraphics[width=0.5\textwidth,height=0.2\textheight]{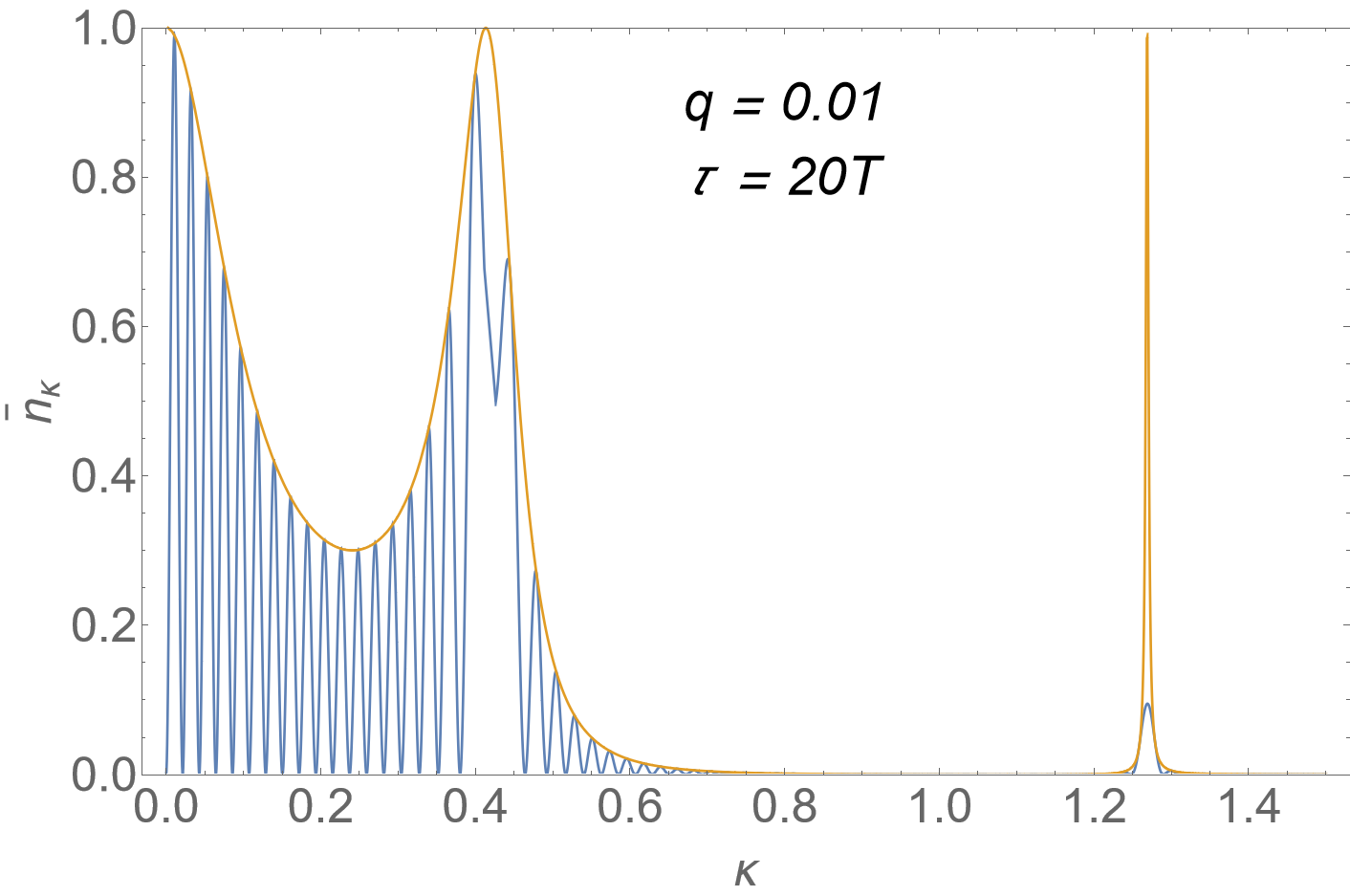}
    }
    \\
    \subfloat
    {%
    \includegraphics[width=0.5\textwidth,height=0.2\textheight]{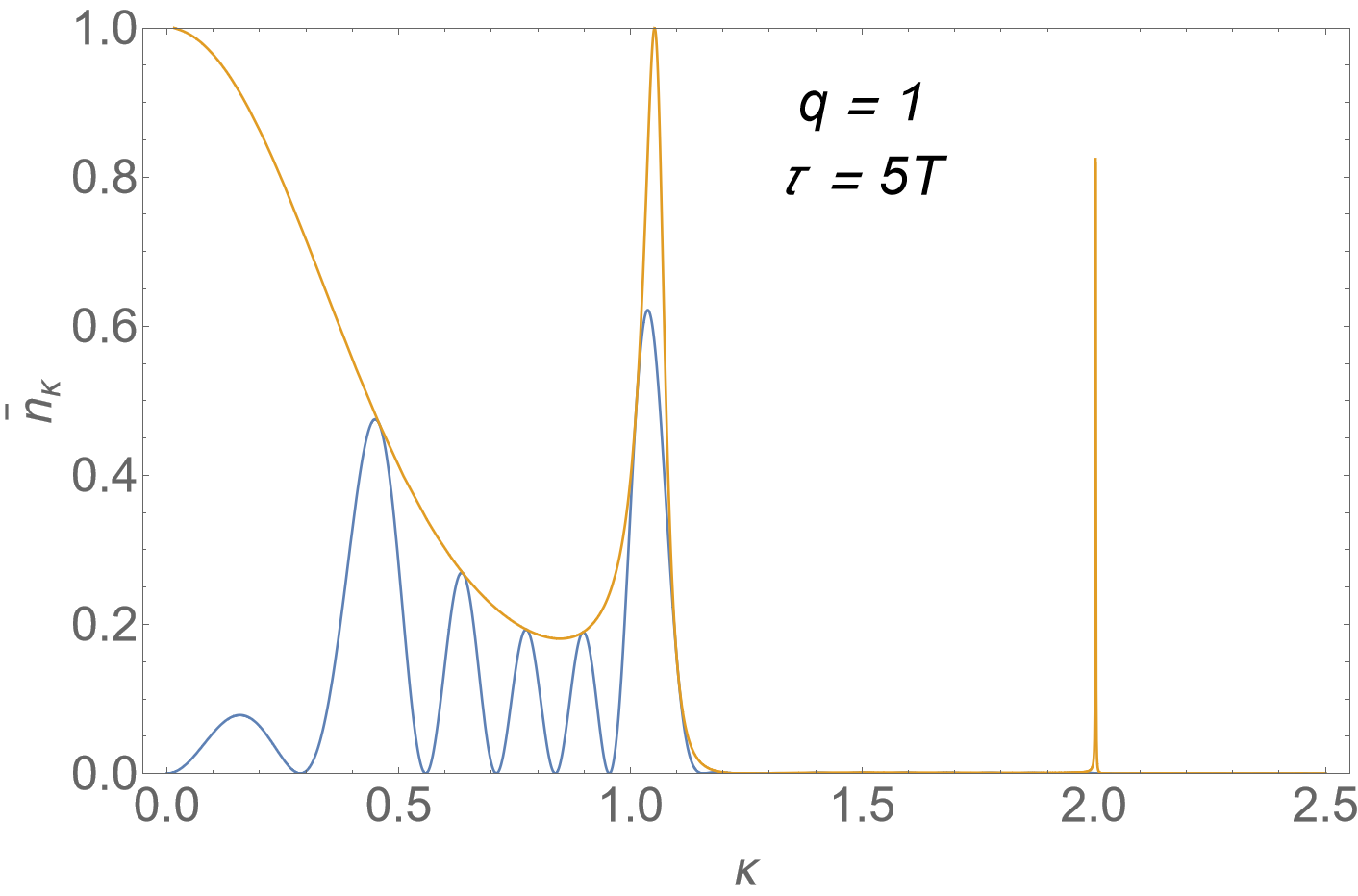}
    }
    \hfill
    \subfloat
    {%
    \includegraphics[width=0.5\textwidth,height=0.2\textheight]{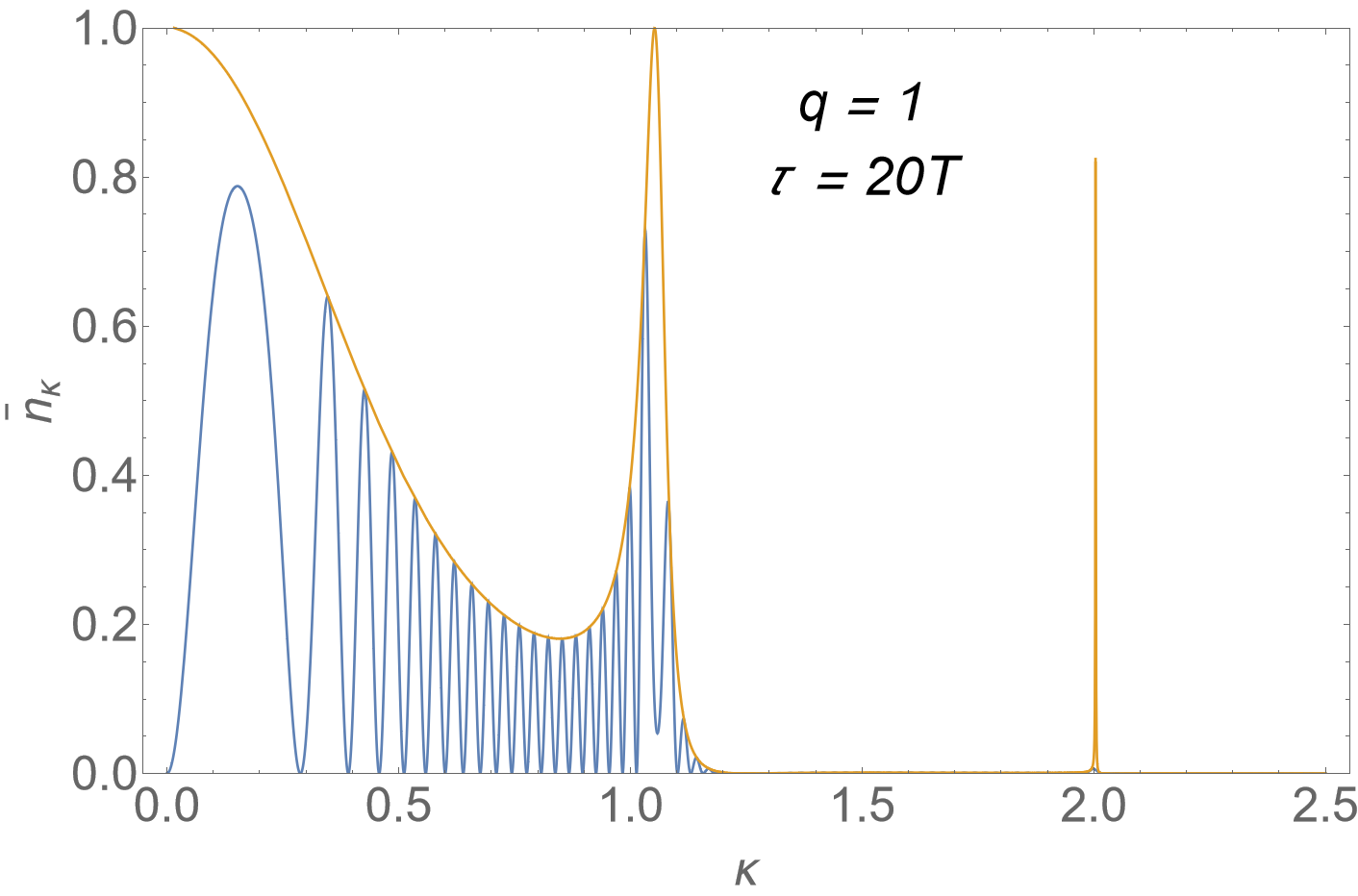}
    }
    \\
    \subfloat
    {%
    \includegraphics[width=0.5\textwidth,height=0.2\textheight]{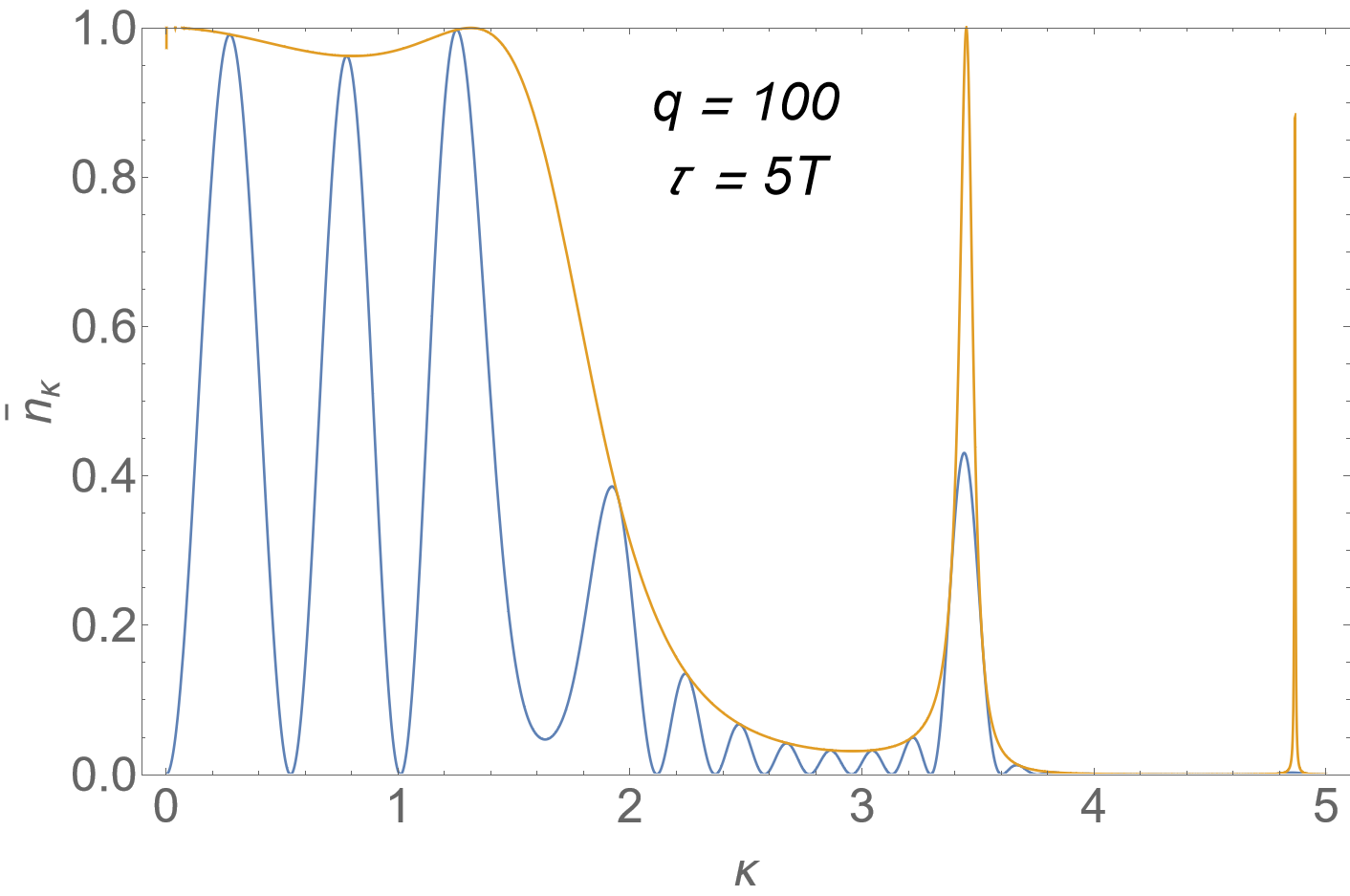}
    }
    \hfill
    \subfloat
    {%
    \includegraphics[width=0.5\textwidth,height=0.2\textheight]{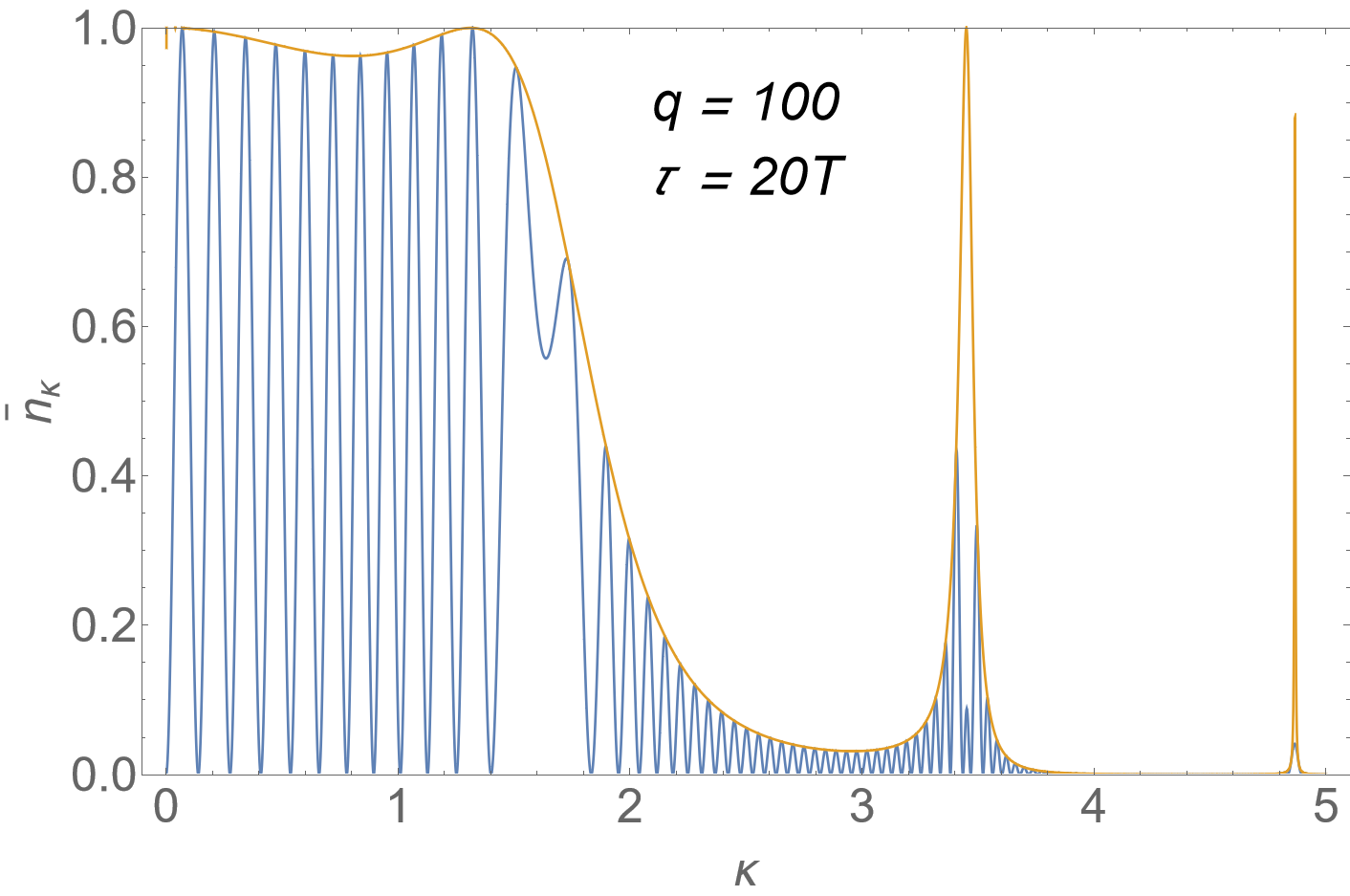}
    }
    \caption{$\overline{n}_{\kappa}(\tau)$ (in blue) for $q=0.001$ (top row), $q=0.01$ (second row), $q=1$ (third row) and $q=100$ (bottom row) with its $F_{\kappa}$ envelope (in orange) at $\tau=5T$ on the left and $\tau=20T$ on the right. 
    }\label{fig: nbarkap for q = 0.01, 1, 100}
\end{figure}


\section{Predicting the Locations of Resonance Peaks}\label{sec: 3}
In this section, we describe how the $\kappa$ values of the resonance peaks of the momentum spectrum can be found analytically or semi-analytically, without the need to solve a differential equation. We also provide physical explanations for the relations that help predict the $\kappa$ values of the resonance peaks. 

\begin{figure}[H]
	\centering
	\includegraphics[width=\textwidth]{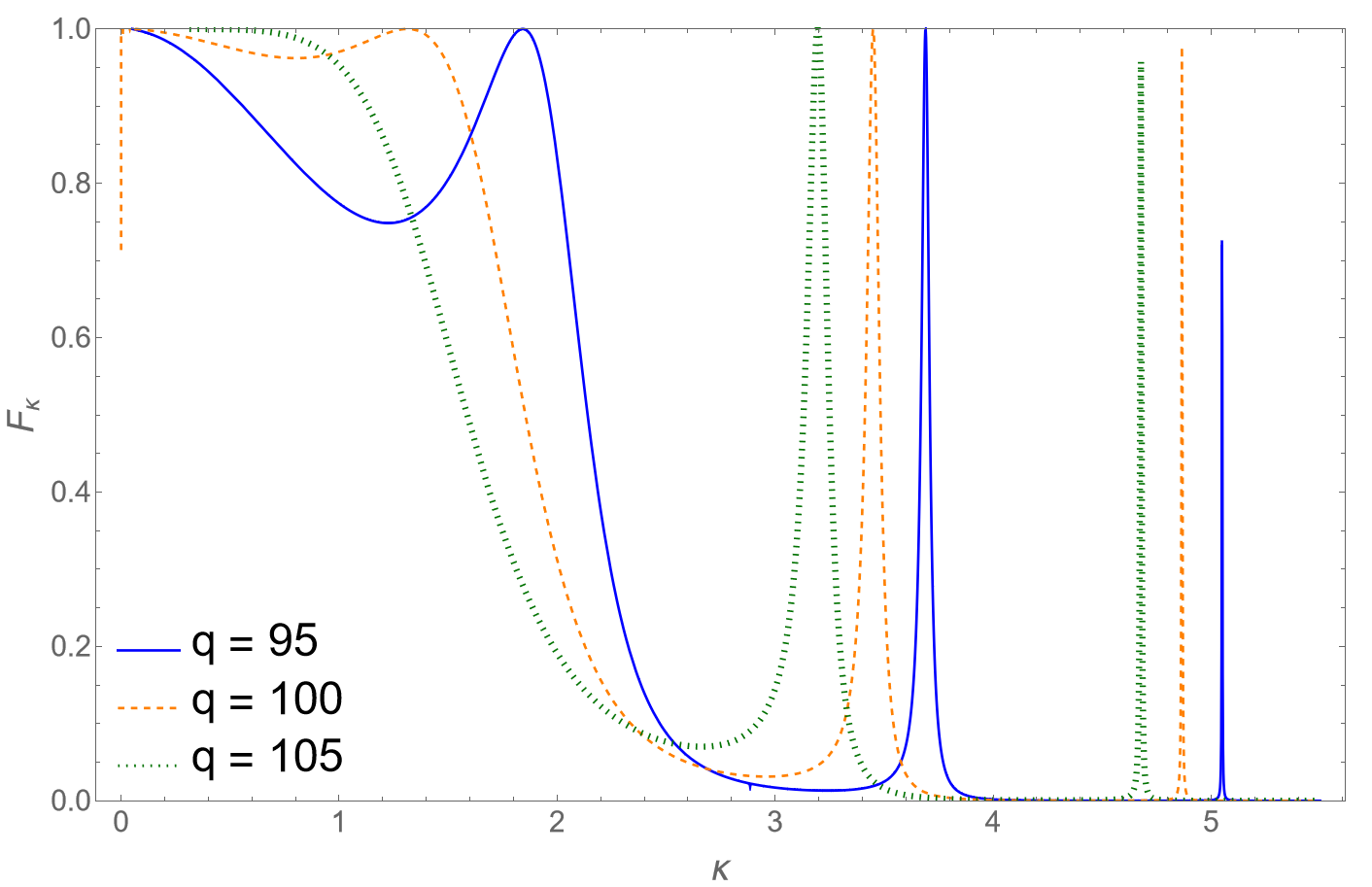}
	\caption{$F_{\kappa}$ as a function of $\kappa$ for three different values of $q$, showing how the envelope evolves as $q$ increases. The $\kappa$ value of each resonance peak decreases with increasing $q$.}\label{fig: Fkap evolution with q}
\end{figure}

The envelope $F_{\kappa}$ is defined in Eq.~\eqref{eq:NumDen nu_kappa and F_kappa} and computed by numerically solving the mode equation of Eq.~\eqref{eq:NumDen X_kappa mode eq.} with the initial conditions of Eq.~\eqref{eq:NumDen X^1_kappa inital cond.}. For small values of the coupling parameter $q$ ($\lesssim 0.01$), the bulk region located at small $\kappa$ is narrow, while the resonance peaks (the spikes at larger values of $\kappa$) are independent of $q$. The bulk region and the resonance peaks are well separated for small $q$ values. For large $q$ ($\gtrsim 1$), the resonance peak positions decrease in $\kappa$ with increasing $q$ and we can see this evolution in Fig.~\ref{fig: Fkap evolution with q}. For larger $q$ values, the bulk region of the $F_{\kappa}$ envelope can merge with a resonance peak for some values of $q$, and we can also see this effect in Fig.~\ref{fig: Fkap evolution with q}.

\begin{figure}[H]
	\centering
	\includegraphics[width=0.7\textwidth]{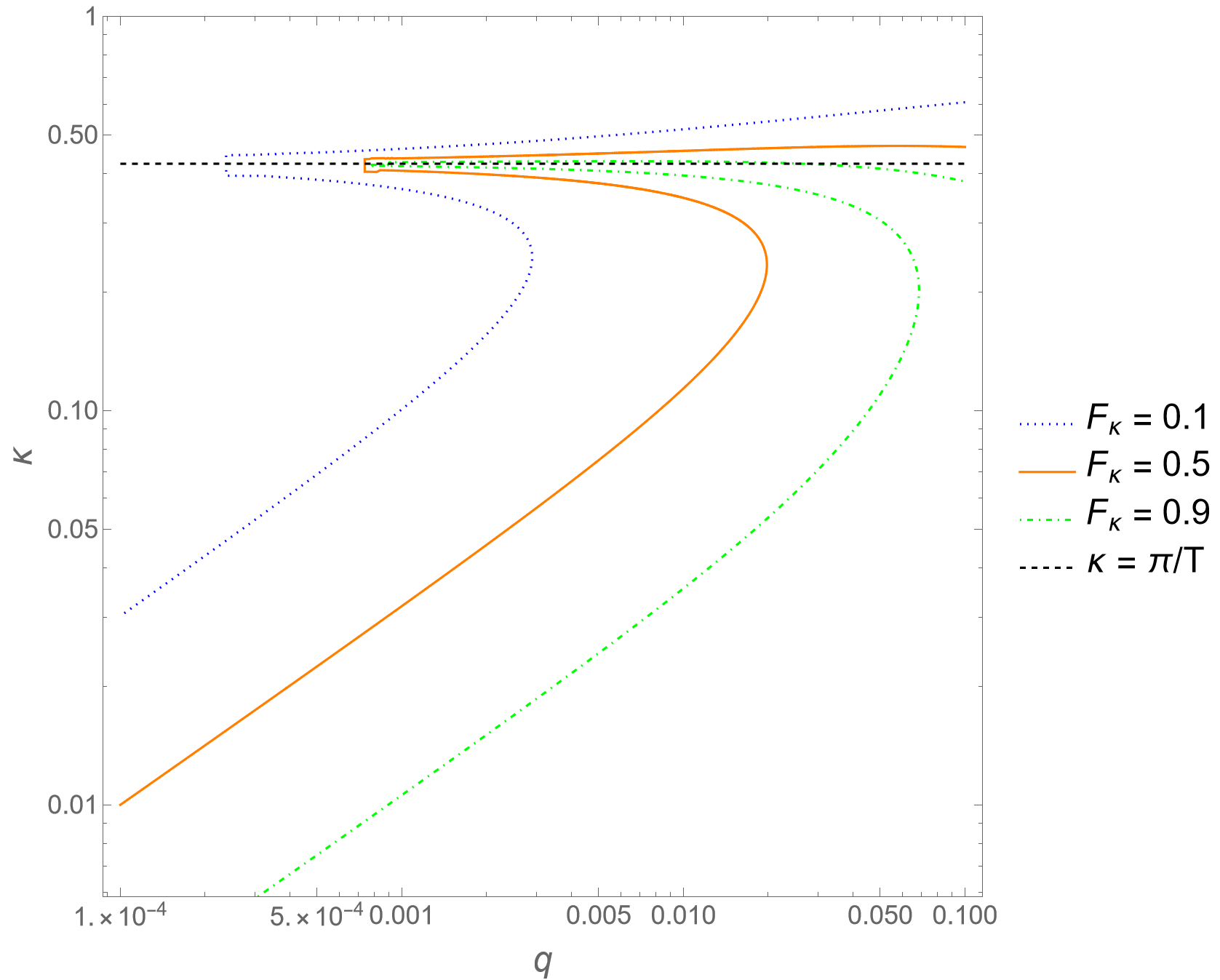}
	\caption{$F_{\kappa}$ contours in the $(q,\kappa)$ plane, with both axes scaled logarithmically. $F_k$ gets large at $\kappa \approx \pi/T$ (black dashed line), the resonance peak corresponding to $l=0$ from Eq.~\eqref{eq:ResPeakPos peak position small q}.}\label{fig: Fkap contours_q vs kap_bulkwidth_small q}
\end{figure}

In \cite{Greene:1998nh}, it was found that for $q\ll 1$, the resonance peaks are located at odd multiples of $\pi/T$, which can be written as:
\begin{equation}\label{eq:ResPeakPos peak position small q}
\kappa=\frac{\pi}{T}(2l+1)
\text{, where }
l=0,1,2,3,\dots
\text{ (small }q)
\end{equation}
We numerically verify this relation for $l=0$ in Fig.~\ref{fig: Fkap contours_q vs kap_bulkwidth_small q}. 
For small values of $q$, it is difficult to verify Eq.~\eqref{eq:ResPeakPos peak position small q} explicitly for peaks at higher values of $l$ (and thus larger $\kappa$) as those peaks become very narrow and difficult to resolve numerically.

We now give a physics argument for why Eq.~\eqref{eq:ResPeakPos peak position small q} holds, based on a semi-perturbative description of the process $n\varphi\rightarrow\Psi\overline{\Psi}$. Effectively, the relation is a consequence of energy conservation in this process. To show this, we define the time average of $\Omega_{\kappa}(\tau)$ from Eq.~\eqref{eq: Omega_k(tau)} over a single period of inflaton oscillations as follows:
\begin{equation}\label{eq:ResPeakPos Omkapavg}
\overline{\Omega}_{\kappa}(q) \equiv \frac{1}{T}\int_{0}^{T}d\tau\Omega_{\kappa}(\tau),
\end{equation}
where $\Omega_{\kappa}(\tau)$ and $\overline{\Omega}_{\kappa}(q)$ are  functions of both $q$ and $\kappa$. 
$\overline{\Omega}_{\kappa}(q) T$ represents the phase accumulated in the $\Psi$ and $\overline{\Psi}$ wavefunctions over one time period of the inflaton oscillations (refer to Eq.~(3.16) of \cite{Greene:2002uot}). Then, the integrals over internal spacetime points that appear in the amplitude for the process $n\varphi\rightarrow\Psi\overline{\Psi}$ yield a term of the form:
\begin{equation}\label{eq:ResPeakPos integral amp}
\int d^4x f(\tau)^n e^{i(\kappa^{\mu}_{\Psi}+\kappa^{\mu}_{\bar \Psi}) x_{\mu}} \sim
\delta^{(3)}\left(\bm{\kappa}_{\Psi}+\bm{\kappa}_{\overline{\Psi}}\right)
\int d\tau e^{-in\omega_{\varphi}\tau}e^{2i\Omega_{\kappa}(\tau) \tau},
\end{equation}
where we keep the term involving the fundamental frequency of $f(\tau)$ (see Eq.~\eqref{eq:Inf. cn(u,1/2)}), and $\omega_{\varphi}=2\pi/T$, with $T$ being the period of inflaton oscillations.\footnote{Higher harmonics in $f(\tau)$ from Eq.~\eqref{eq:Inf. cn(u,1/2)} contribute only odd multiples of $\omega_{\varphi}$; this holds for the solution of any scalar potential symmetric under $\varphi \to -\varphi$ and will be important to preserve our conclusion below that only odd values of $n$ contribute.} The $\varphi$ field is a classical background field with trivial spatial dependence which implies that it does not carry any 3-momentum. The $\delta^{(3)}$ function then ensures momentum conservation, so that the fermion 3-momentum is equal in magnitude and opposite in direction to that of the anti-fermion, i.e.~$\bm{\kappa}_{\Psi}=-\bm{\kappa}_{\overline{\Psi}}$, and $\kappa_{\Psi}=\kappa_{\overline{\Psi}}=\kappa$. 

If $\Omega_{\kappa}(\tau)$ were a constant, the $\tau$ integral in Eq.~\eqref{eq:ResPeakPos integral amp} would yield the usual energy-conserving $\delta$ function.  Instead, the periodic variation of $\Omega_{\kappa}(\tau)$ causes the phase to oscillate more and more rapidly as $\tau$ becomes large, while the accumulated phase for the fermion-antifermion pair approaches $2 \overline{\Omega}_{\kappa}(q) \tau$ to higher and higher fractional precision.  Replacing $\Omega_{\kappa}(\tau)$ with its time-average then gives the energy-conservation condition:
\begin{equation}\label{eq:ResPeakPos Omkapavg with n varphi}
n\omega_{\varphi}=2\overline{\Omega}_{\kappa}(q) \implies
\overline{\Omega}_{\kappa}(q) = \frac{\pi}{T} n.
\end{equation}
At small $q$, $\overline{\Omega}_{\kappa}(q) \approx\kappa$ so this reduces to Eq.~\eqref{eq:ResPeakPos peak position small q} if $n$ is odd.

\begin{figure}[H]
    \centering
    \begin{tikzpicture}[baseline=(b.base)]
    \begin{feynman}
        \vertex (a) {$\varphi$};
        \vertex [right=of a] (b);
        \vertex [above right=of b] (f1) {$\Psi$};
        \vertex [below right=of b] (f2) {$\overline{\Psi}$};
        \diagram*{
        (a) -- [scalar] (b) -- [fermion] (f1),
        (b) -- [anti fermion] (f2)
        };
    \end{feynman}    
    \end{tikzpicture}\quad
    \begin{tikzpicture}[baseline=(b.base)]
    \begin{feynman}
        \vertex (a) ;
        \vertex [right=of a] (b);
        \vertex [above right=of b] (f1) {$\Psi$};
        \vertex [below right=of b] (f2) {$\overline{\Psi}$};
        \vertex [above left=of a] (a1) {$\varphi$};
        \vertex [left=of a] (a2) {$\varphi$};
        \vertex [below left=of a] (a3) {$\varphi$};
        \diagram*{
        (a1) -- [scalar] (a) -- [scalar] (b) -- [fermion] (f1),
        (b) -- [anti fermion] (f2),
        (a2) -- [scalar] (a), (a3) -- [scalar] (a)
        };
    \end{feynman}    
    \end{tikzpicture}\quad
    \begin{tikzpicture}[baseline=(b.base)]
    \begin{feynman}
        \vertex (a) ;
        \vertex [right=of a] (b);
        \vertex [above right=of b] (f1) {$\Psi$};
        \vertex [below right=of b] (f2) {$\overline{\Psi}$};
        \vertex [above left=of a] (a1) {$\varphi$};
        \vertex [left=of a] (a2) {$\varphi$};
        \vertex [below left=of a] (a3);
        \vertex [left=of a3] (a3I) {$\varphi$}; 
        \vertex [below left=of a3] (a3II) {$\varphi$};
        \vertex [below=of a3] (a3III) {$\varphi$}; 
        \diagram*{
        (a1) -- [scalar] (a) -- [scalar] (b) -- [fermion] (f1),
        (b) -- [anti fermion] (f2),
        (a2) -- [scalar] (a), (a3) -- [scalar] (a),
        (a3I) -- [scalar] (a3), (a3II) -- [scalar] (a3), 
        (a3III) -- [scalar] (a3)
        };
    \end{feynman}    
    \end{tikzpicture}
    \caption{Feynman diagrams describing the perturbative process $n\varphi\rightarrow\Psi\overline{\Psi}$, for small $q$. Shown are the diagrams for $n=1$ (left), $n=3$ (middle), and a representative diagram for $n=5$ (right).}
    \label{fig: feyndiags small q ResPeaks}
\end{figure}
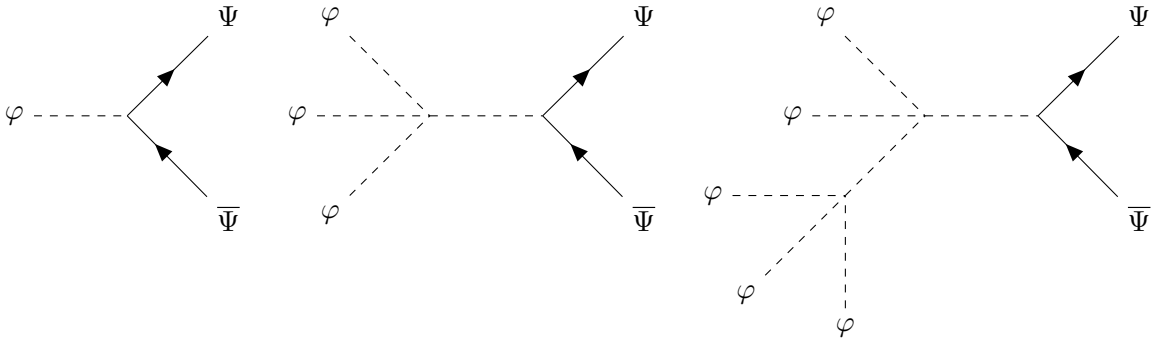

We can qualitatively explain the resonance structure of Eq.~\eqref{eq:ResPeakPos Omkapavg with n varphi} from the perturbative amplitude of $n\varphi\rightarrow\Psi\overline{\Psi}$ using Feynman diagrams.  
For small values of $q=h^{2}/\lambda$ (from Eq.~\eqref{eq:NumDen eta to tau transf.}), the inflaton quartic coupling $\lambda$ is much larger than the square of the Yukawa coupling $h$.  Then the process is dominated by Feynman diagrams of the form shown in Fig.~\ref{fig: feyndiags small q ResPeaks}.  Clearly, for each subsequent addition of a $\varphi^{4}$ vertex to the $n=1$ case (left diagram of Fig.~\ref{fig: feyndiags small q ResPeaks}), the number of external $\varphi$ increases by $2$. This implies that $n$ must be odd, i.e.~$n=2l+1$, where $l=0,1,2,\dots$, and we have proved Eq.~\eqref{eq:ResPeakPos peak position small q} for small $q$.  Note that this argument will hold for any scalar potential symmetric under $\varphi \to -\varphi$.
 
\begin{figure}[H]
\centering
\begin{tikzpicture}
\begin{feynman}
    \vertex (phi1) at (-2,2) {$\varphi_{1}$};
    \vertex (phi2) at (-2,1) {$\varphi_{2}$};
    \vertex (phi3) at (-2,-1) {$\varphi_{n-1}$};
    \vertex (phi4) at (-2,-2) {$\varphi_{n}$};

    \vertex (v1) at (0,2);
    \vertex (v2) at (0,1);
    \vertex (vdots) at (-0.5,0) {$\vdots$};
    \vertex (v3) at (0,-1);
    \vertex (v4) at (0,-2);

    \vertex (f1) at (2,2) {$\Psi$};
    \vertex (f2) at (2,-2) {$\overline{\Psi}$};
    
    \diagram*{
        (v4) -- [anti fermion, momentum'={[arrowlabel]$k_{2}$}] (f2),
        (v4) -- [fermion] (v3) -- [fermion] (v2) -- [fermion] (v1) -- [fermion, momentum={[arrowlabel]$k_{1}$}] (f1),
        (phi1) -- [scalar, momentum={[arrowlabel]$p_{1}$}] (v1),
        (phi2) -- [scalar, momentum={[arrowlabel]$p_{2}$}] (v2),
        (phi3) -- [scalar, momentum'={[arrowlabel]$p_{n-1}$}] (v3),
        (phi4) -- [scalar, momentum'={[arrowlabel]$p_{n}$}] (v4),
    };
\end{feynman}
\end{tikzpicture}
\caption{General form of Feynman diagram describing the perturbative process $n\varphi\rightarrow\Psi\overline{\Psi}$, for large $q$.}
\label{fig: feyndiag large q ResPeaks}
\end{figure}
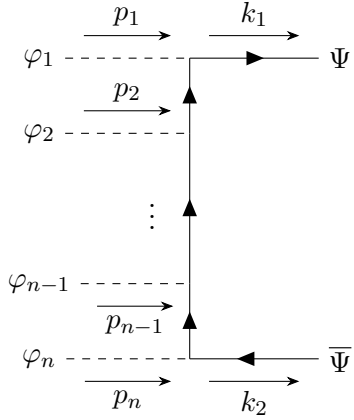

For large values of $q$, the square of the Yukawa coupling is much larger than $\lambda$ and we can neglect the $\varphi^4$ vertices.  Then $n\varphi\rightarrow\Psi\overline{\Psi}$ is described by Feynman diagrams having the general form shown in Fig.~\ref{fig: feyndiag large q ResPeaks}. 

For any even $n$ in Fig.~\ref{fig: feyndiag large q ResPeaks}, i.e.~an even number of external $\varphi$ vertices, the diagram will have an odd number of fermion propagators, in which case the amplitude involves only terms containing an odd number of $\gamma$-matrices in between the external fermion and antifermion spinors.  These can be reduced using anticommutation relations down to a single gamma matrix, which must be dotted into an external fermion momentum because the initial-state scalars are effectively at rest.  For massless fermions, all such terms are then zero by the Dirac equation. Hence, the amplitude of diagrams of the form of Fig.~\ref{fig: feyndiag large q ResPeaks} is non-zero only for odd values of $n$. We thus obtain the same criterion for $n$ as for small $q$, i.e.~$n=2l+1$, where $l=0,1,2,\dots$.

\begin{figure}[H]
	\centering
	\includegraphics[width=0.7\textwidth]{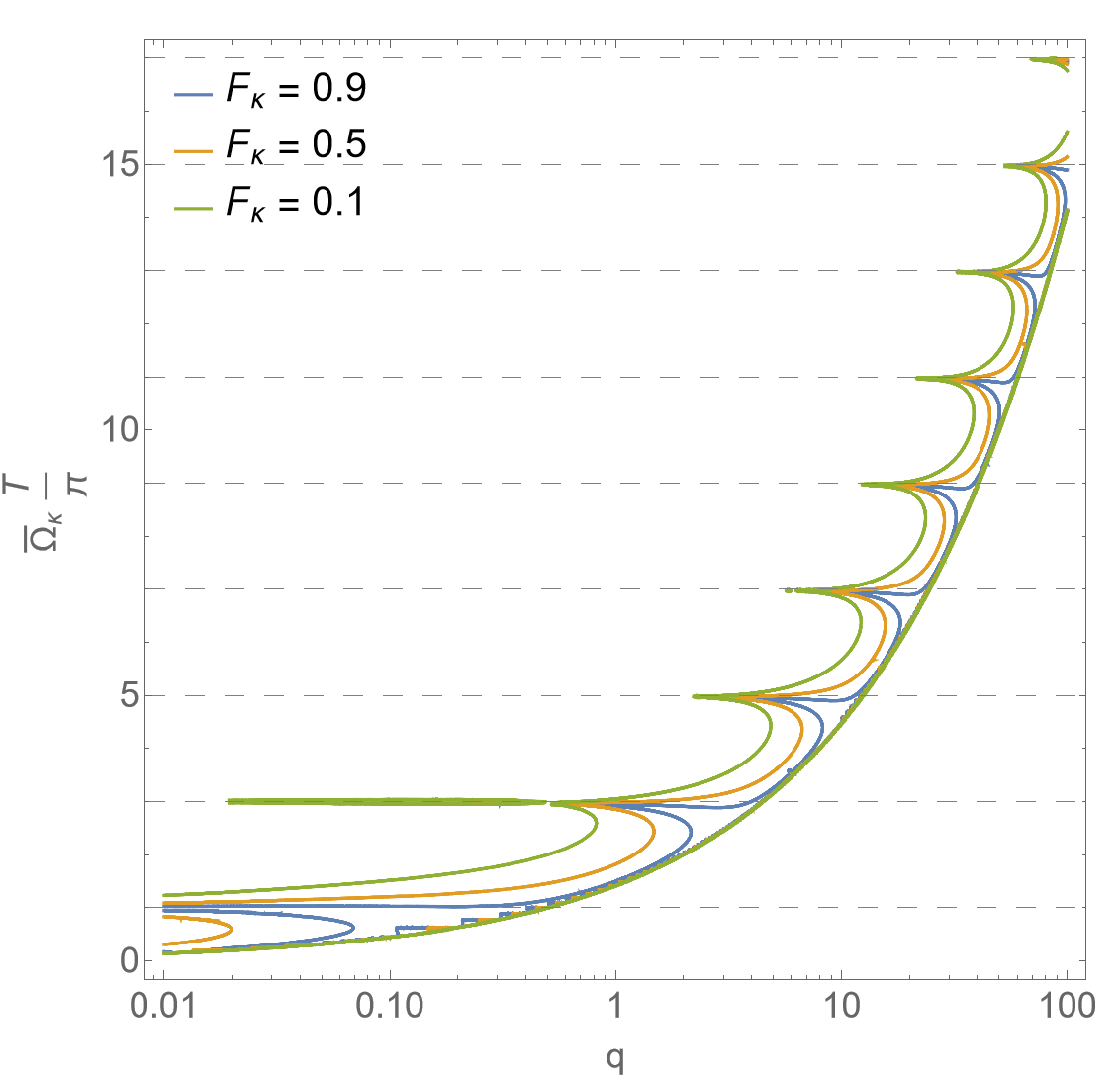}
	\caption{Contours corresponding to $F_{\kappa}=0.9$ (blue), $F_{\kappa}=0.5$ (orange), $F_{\kappa}=0.1$ (green) in the $q$ vs.~$\overline{\Omega}_\kappa T/ \pi$ plane. The horizontal dashed lines are at odd integers. In addition to the spiky pattern of the resonance peaks, there is a smooth curve at the base of the spikes that corresponds to contours for $\kappa=0$ and determines which $l$ peaks are accessible for a particular $q$.}\label{fig: Fkap contours_q vs OmkapavgTbyPi}
\end{figure}

Thus, as a consequence of energy conservation, we can write a general semi-analytic relation that enables us to calculate the $\kappa$ values corresponding to resonance peaks for any $q$:
\begin{equation}\label{eq:ResPeakPos peak position Omkapavg}
\overline{\Omega}_{\kappa}(q)=\frac{\pi}{T}(2l+1)\text{, where }l=0,1,2,3,\dots
\end{equation}
We can numerically verify that Eq.~\eqref{eq:ResPeakPos peak position Omkapavg} holds by plotting the contours of $F_{\kappa}$ in the $(q,\overline{\Omega}_{\kappa}\cdot T/\pi)$ plane instead of the $(q,\kappa)$ plane of Fig.~\ref{fig: Fkap contours_q vs kap_bulkwidth_small q}. We expect resonance peaks whenever $\overline{\Omega}_{\kappa}\cdot T/\pi$ is equal to an odd positive integer and this is exactly what we observe in Fig.~\ref{fig: Fkap contours_q vs OmkapavgTbyPi}. 
We provide further justification for our energy conservation arguments in Appendix~\ref{AppG} by redoing our analysis with a $\phi\phi\overline{\psi}\psi$ interaction rather than a Yukawa coupling.

We can thus identify each resonance peak for all values of $q$ by its corresponding $l$, according to Eq.~\eqref{eq:ResPeakPos peak position Omkapavg}. For a fixed $q$, $\overline{\Omega}_{\kappa}$ cannot be less than its value when $\kappa = 0$. This $\overline{\Omega}_{0}$ (smallest value of $\overline{\Omega}_{\kappa}$ for a fixed $q$) is given by:
\begin{equation}\label{eq:ResPeakPos Omkapavg kap = 0}
\overline{\Omega}_{0}(q) = \langle \Omega_0(\tau) \rangle = q^{1/2} \langle |f(\tau)| \rangle = \frac{\pi}{T}\sqrt{2} \, q^{1/2},
\end{equation}
 where $\langle \cdots \rangle$ denotes a time average over a complete period of inflaton oscillations i.e~$\langle|f|\rangle=(1/T)\int_{0}^{T}d\tau\sqrt{f(\tau)^{2}}=\sqrt{2}\,\pi/T$. This lower bound $\overline{\Omega}_{0}(q)$ is indicated by the lowermost smooth green curve in Fig.~\ref{fig: Fkap contours_q vs OmkapavgTbyPi} and indicates which $l$ values are possible for a fixed $q$.

In order to find the positions of the peaks in $\kappa$ for a fixed value of $q$, we first construct an interpolating function for $\overline{\Omega}_{\kappa}(q)$ for a wide range of values of $q$ and $\kappa$. Then, we can invert Eq.~\eqref{eq:ResPeakPos peak position Omkapavg} for a fixed value of $l$ to get the value of $\kappa$ corresponding to that value of $l$. However, before applying this method, we need to take into account the lowest $l$ value accessible for a particular $q$ due to the lower bound $\overline{\Omega}_{0}(q)$, which can be obtained as follows: 
\begin{equation}\label{eq:ResPeakPos peak position l0}
l_{0}(q)=\left\lceil\frac{1}{2}\left(\overline{\Omega}_{0}(q)\frac{T}{\pi}-1\right)\right\rceil
= \left\lceil\frac{1}{2}\left(\sqrt{2} \, q^{1/2}-1\right)\right\rceil,
\end{equation}
 where $\lceil x\rceil$ represents the ceiling function that gives the nearest integer greater than or equal to $x$. The peak in $F_{\kappa}$ corresponding to this $l_{0}$ is sometimes merged into the bulk region at small $\kappa$, which we can see happening from the evolution of the $F_{\kappa}$ envelope of Fig.~\ref{fig: Fkap evolution with q} with increasing $q$. In general, for any fixed $q$, we can first use Eq.~\eqref{eq:ResPeakPos peak position l0} to find $l_{0}$ and then invert Eq.~\eqref{eq:ResPeakPos peak position Omkapavg} to find the value of $\kappa$ corresponding to the peak $l_{0}$ and all higher peaks.  We used this procedure to find the approximate $\kappa$ value used in Fig.~\ref{fig: n_kappa and bar n_kappa} corresponding to the $l_{0}=1$ peak for $q=1$. Finally, using Eq.~\eqref{eq:ResPeakPos Omkapavg kap = 0}, we are also able to determine the values of $q$ at which the value of $l_{0}$ changes:
\begin{equation}\label{eq:AnalyticF q0(l)}
q_{0}(l)=\left(\frac{\pi(2l+1)}{T\langle|f|\rangle}\right)^{2}
= \frac{(2l+1)^{2}}{2}
\text{, where }l=0,1,2,3,\dots
\end{equation}
 These values will correspond to the discontinuities in the curves in Figs.~\ref{fig:DenCont Large q ratios} and \ref{fig:DenCont Intermediate q ratios}.


\section{Contributions to the Total Number Density}\label{sec: 4}

In this section, we will compute how the bulk region  around $\kappa=0$ and the resonance peaks at higher $\kappa$ contribute to the total number density by numerically evaluating integrals of the form in Eqs.~\eqref{eq:NumDen total n, dimless} and \eqref{eq:NumDen total n, physical, with a}. We will divide up the integral into bulk and resonance peak contributions, show how they behave separately and then find analytical approximations for all integrals. We numerically evaluate the different contributions to the integral:
\begin{equation}\label{eq: total Fk integral I}
I\equiv\int_{0}^{\infty} d\kappa \kappa^{2}F_{\kappa}
\end{equation}
We find approximations (as functions of $q$) for the different contributions and for the total integral itself. In particular, we find good approximations in two regimes: small $q$ ($q\lesssim 0.01$) and large $q$ ($q\gtrsim 10$). We also give numerical results for the intermediate regime, but do not find a good analytical approximation there. 
\subsection{Small $q$}\label{Sec:DenCont Small q}
We first focus on evaluating the integral $I$ over a range of small $q$ values: $10^{-5} \leq q \leq 10^{-2}$. We expect these results to be valid for smaller values of $q$ as well. For all values of $q$ less than $q_{0}(0)=0.5$, $l_{0}= 0$ and, therefore, resonance peaks corresponding to all $l$ values are present for $q < 0.5$. We observe, however, that the resonance peaks corresponding to $l\geq 2$ are very narrow and, as such, the numerical estimations of their contributions to $I$ become unreliable. 
Below we give evidence that that the contribution of the $l=2$ peak is small compared to the total integral.

\begin{figure}[H]
	\centering
	\includegraphics[width=\textwidth]{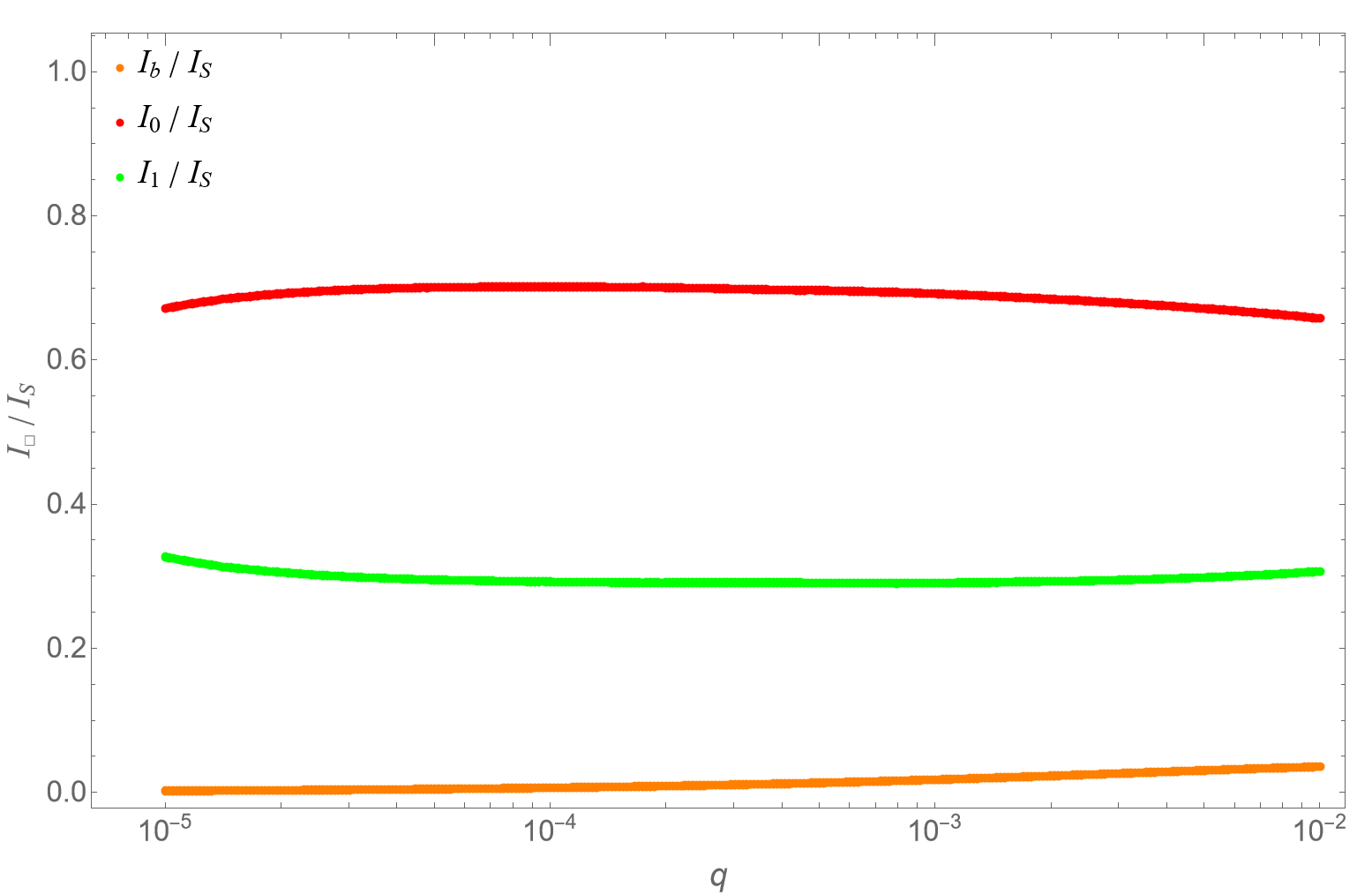}
	\caption{Contributions of different components as fractions of the total integral $I_{S}(q)$ from Eq.~\eqref{eq:DenCont Small q total}\, for $10^{-5} \leq q \leq 10^{-2}$. 
    The contributions defined in Eq.~\eqref{eq:DenCont Small q total parts} are given by the bulk (orange, bottom), the $l=0$ peak  (red, top), and the $l=1 $ peak (green, middle).
    }\label{fig:DenCont Small q ratios}
\end{figure}

To estimate the total integral for small $q$, we integrate the bulk region and the first two resonance peaks corresponding to $l=0$ and $l=1$: 
\begin{equation}\label{eq:DenCont Small q total}
I_{S}(q)=\int_{0}^{(\kappa_{1}+\kappa_{2})/2} d\kappa \kappa^{2}F_{\kappa}.
\end{equation}
The upper limit of the this integral is chosen because it represents the $\kappa$ point midway between the $l=1$ and $l=2$ resonance peaks. Because the peaks are narrow compared to their spacing (see the top row of Fig.~\ref{fig: nbarkap for q = 0.01, 1, 100}), this upper limit of the integral ensures that we are taking into account only the contributions from the bulk and the first two peaks $l=0$ and $l=1$. 
The total integral $I_S$ can be broken up into three parts, with each part describing the contributions from the bulk region and each of the first two resonance peaks, as follows:
\begin{equation}\label{eq:DenCont Small q total parts}
I_{S}(q)
=
\underbrace{\int_{0}^{\kappa_{0}/2} d\kappa \kappa^{2}F_{\kappa}}_\text{Bulk, $I_{b}$}
+
\underbrace{\int_{\kappa_{0}/2}^{(\kappa_{0}+\kappa_{1})/2} d\kappa \kappa^{2}F_{\kappa}}_\text{$l=0$ Peak, $I_{0}$}
+
\underbrace{\int_{(\kappa_{0}+\kappa_{1})/2}^{(\kappa_{1}+\kappa_{2})/2} d\kappa \kappa^{2}F_{\kappa}}_\text{$l=1$ Peak, $I_{1}$}.
\end{equation}
Fig.~\ref{fig:DenCont Small q ratios} shows the value of the three component integrals normalized to the total integral, as a function of $q$. 
We observe in Fig.~\ref{fig:DenCont Small q ratios} that most of the contribution to the fermion density for small $q$ values comes from the resonance peaks $l=0$ and $l=1$. The peak $l=0$ contributes $65-70\%$ of the total integral, the peak $l=1$ contributes $30-35\%$, while the bulk region contributes less than $5\%$ of the total. As such, we focus on finding approximations to the contributions of the resonance peaks. 

\begin{figure}[H]
    \includegraphics[width=\textwidth]{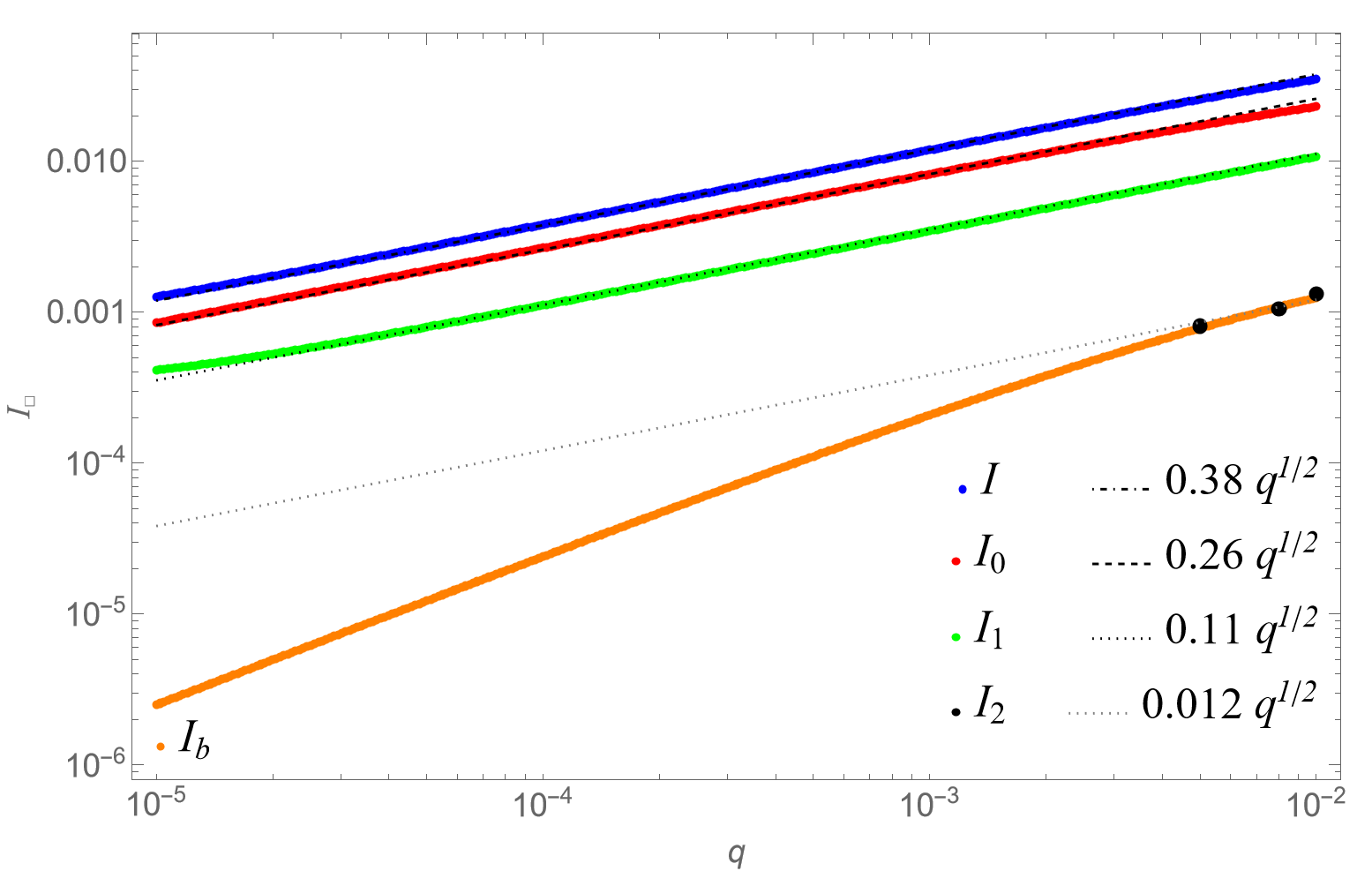}
    \caption{Contributions of different components and the total integral for $10^{-5} \leq q \leq 10^{-2}$. Values of the total integral are given by $I_{S}$ from Eq.~\eqref{eq:DenCont Small q total} (blue, top) with the approximation from Eq.~\eqref{eq:DenCont Small q Total Approx} (black, dot-dashed). Also shown are contributions of the peak $l=0$ (red, second from top) given by $I_{0}$ from Eq.~\eqref{eq:DenCont Small q total parts} with its corresponding approximation from Eq.~\eqref{eq:DenCont Small q l=0 approx} (black, dashed), the peak $l=1$ (green, third from top) given by $I_{1}$ from Eq.~\eqref{eq:DenCont Small q total parts} with its corresponding approximation from Eq.~\eqref{eq:DenCont Small q l=1 approx} (black, dotted), and contributions of the bulk given by $I_{b}$ (orange, bottom) from Eq.~\eqref{eq:DenCont Small q total parts}. We also show contributions of the $l=2$ peak for $q=0.005,0.008,0.01$ given by $I_{2}$ (black circles, bottom right) and its approximation from Eq.~\eqref{eq:DenCont Small q l=2 approx} (gray, dotted) which we extrapolate down to $q=10^{-5}$. The three points for $I_{2}$ are approximately the same size as $I_{b}$, but we expect this to be a coincidence and expect $I_{2}$ to be larger than $I_{b}$ for smaller $q$, as shown by the extrapolation.
    \label{fig: Small q contributions and total subfigs}}
\end{figure}

Plotting $I_0$ vs.~$q$ with both axes logarithmically scaled, we find that  the values follow a $q^{1/2}$ power law. Fitting a straight line to the log-log data using least-squares regression gives \begin{equation}\label{eq:DenCont Small q l=0 approx}
I_{0}(q)=\int_{\kappa_{0}/2}^{(\kappa_{0}+\kappa_{1})/2} d\kappa \kappa^{2}F_{\kappa}
\approx 0.26\times q^{1/2}.
\end{equation}
The numerical estimate as well as the fit line are shown in Fig.~\ref{fig: Small q contributions and total subfigs}. Repeating the procedure for $I_{1}$, we similarly find:
\begin{equation}\label{eq:DenCont Small q l=1 approx}
I_{1}(q)=\int_{(\kappa_{0}+\kappa_{1})/2}^{(\kappa_{1}+\kappa_{2})/2} d\kappa \kappa^{2}F_{\kappa}
\approx 0.11\times q^{1/2}.
\end{equation}
This fit is also shown in Fig.~\ref{fig: Small q contributions and total subfigs}.

Repeating the same procedure for the bulk region, $I_{b}=\int_{0}^{\kappa_{0}/2} d\kappa \kappa^{2}F_{\kappa}$, we find that it follows a power law $\sim q^{0.9}$. Because the bulk is only a small fraction of the total contribution (see Fig.~\ref{fig:DenCont Small q ratios}), the power law dependence of the total integral is still well described by $q^{1/2}$, as shown in Fig.~\ref{fig: Small q contributions and total subfigs}.

We are unable to find a good approximation for the resonance peaks $l=2$ and higher by this method directly, as the peaks are too narrow. The numerical evaluation becomes unreliable due to round-off error as $\text{Re}\left(X^{1}_{\kappa}(T)\right)$ (from Eq.~\eqref{eq:NumDen nu_kappa and F_kappa}) is very close to $-1$. As such, we try to estimate the contribution from the peak $l=2$ by using $Y_{\kappa}=X_{\kappa}+1$ and substituting it for $X_{\kappa}$ in  Eqs.~\eqref{eq:NumDen X_kappa mode eq.}, \eqref{eq:NumDen nu_kappa and F_kappa} and \eqref{eq:NumDen X^1_kappa inital cond.}. We find that this trick improves numerical calculations for a small range of $q$ values between $q=0.005$ and $q=0.01$, since Wolfram Mathematica handles very small values better than small differences between $\mathcal{O}(1)$ values. We then evaluate the integral using the $F_{\kappa}$ from Eq.~\eqref{eq:NumDen nu_kappa and F_kappa} (with $Y_{\kappa}$ now) over a small range of $\kappa$ around $\kappa = 5\pi/T$ (from Eq.~\eqref{eq:ResPeakPos peak position small q}, corresponding to the peak $l=2$ for small $q$) for three values $q=0.005,0.008,0.01$. We find that the log-log data for the three integral values approximately follows a $q^{1/2}$ power law (similar to $I_{0}$ and $I_{1}$) and we find the fit to these data points using least-squares regression as follows:   
\begin{equation}\label{eq:DenCont Small q l=2 approx}
I_{2}(q)=\int_{5\pi/T-0.01}^{5\pi/T+0.01} d\kappa \kappa^{2}F_{\kappa}
\approx 0.012\times q^{1/2}.
\end{equation}
The numerical values of the integral and their corresponding fit are also shown in Fig.~\ref{fig: Small q contributions and total subfigs}. We thus see that $I_2$ is about ten times smaller than $I_1$, justifying that it can be ignored. We note that the three points computed for $I_2$ are approximately the same size as $I_b$, but we expect that this is a coincidence and that at smaller $q$, $I_2$ will be larger than $I_b$ as shown by the extrapolation in Fig.~\ref{fig: Small q contributions and total subfigs}.

We now fit the total integral:
\begin{equation}\label{eq:DenCont Small q Total Approx}
I_{S}(q)=\int_{0}^{(\kappa_{1}+\kappa_{2})/2} d\kappa \kappa^{2}F_{\kappa}
\approx 0.38\times q^{1/2}.
\end{equation}
which is shown as the top lines in Fig.~\ref{fig: Small q contributions and total subfigs}. 
This approximation in Eq.~\eqref{eq:DenCont Small q Total Approx} for $I$ can now be used in the number density formulae from Eqs.~\eqref{eq:NumDen total n, dimless} and \eqref{eq:NumDen total n, physical, with a} to approximate the number density of produced fermions as a function of the coupling parameter $q$ for small values of $q$ ($\lesssim 0.01$). As we have only considered the peaks $l=0$ and $l=1$ for the approximation in Eq.~\eqref{eq:DenCont Small q Total Approx}, we can estimate the error in the approximation by comparing $I_{2}$ from Eq.~\eqref{eq:DenCont Small q l=2 approx} and $I_{S}$ from Eq.~\eqref{eq:DenCont Small q Total Approx}. We find: $I_{2}/I_{S}\approx 0.03$, which implies that dropping the $l=2$ peak gives an error of around $3\%$. We observe from the approximations in Eqs.~\eqref{eq:DenCont Small q l=0 approx}, \eqref{eq:DenCont Small q l=1 approx} and \eqref{eq:DenCont Small q l=2 approx} that the proportionality constant decreases significantly with increasing $l$ from $0$ to $2$. We expect that this pattern continues so that higher $l\,(>2)$ peaks contribute even less to the total number density; we therefore neglect them in our analysis. 
\subsection{Large $q$}\label{Sec:DenCont Large q}
\begin{figure}[H]
	\centering
	\includegraphics[width=0.7\textwidth]{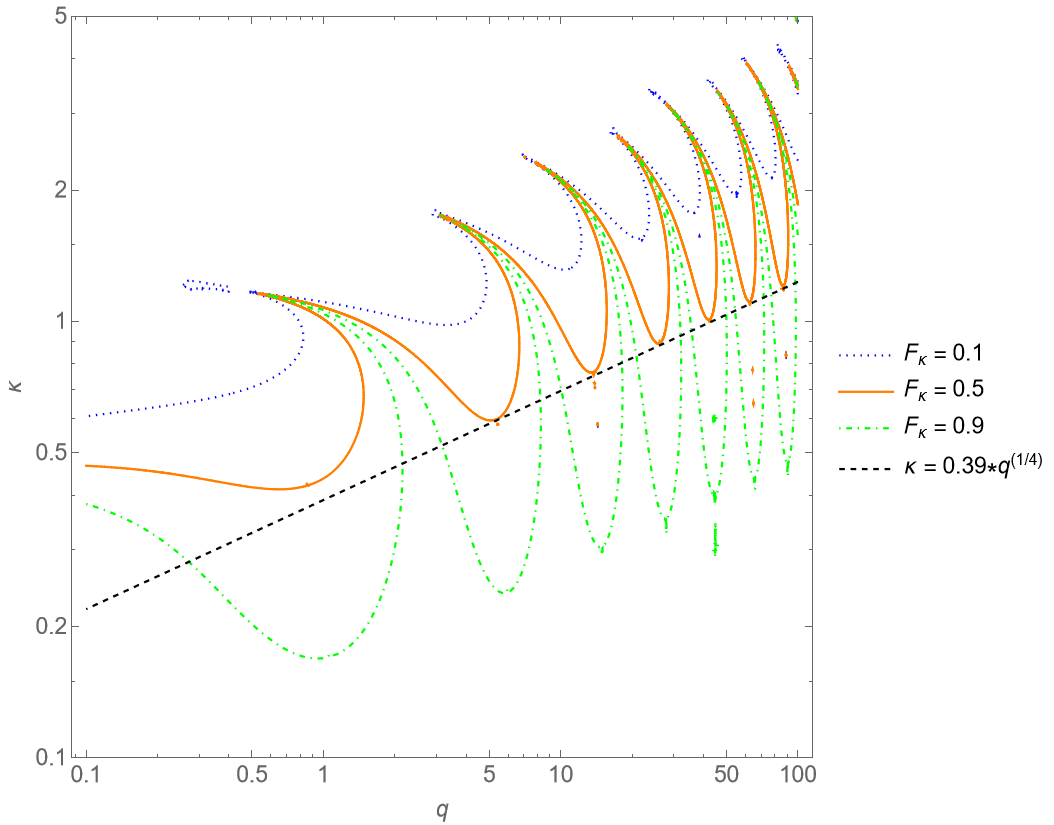}
	\caption{$F_{\kappa}$ contours in the $(q,\kappa)$ plane, with both axes scaled logarithmically. $F_{\kappa}=0.5$ contour follows a pattern which matches the line representing $\kappa=0.39\times q^{1/4}$ for $q\gtrsim 1$.
    }\label{fig: Fkap contours_q vs kap_bulkwidth_large q}
\end{figure}

For larger values of the coupling parameter ($q\gtrsim 10$), we first discuss the boundary of the bulk region and how it could be used to approximate $I$ from Eq.~\eqref{eq: total Fk integral I}. In \cite{Kofman:1997yn}, it was shown that the boundary of the bulk region of $F_{\kappa}$  is proportional to $q^{1/4}$, which was shown to be a consequence of the non-adiabaticity condition for particle production: $|\dot{\Omega}_{\kappa}|\gtrsim\Omega_{\kappa}^{2}$.
The non-adiabaticity condition arises whenever ${\Omega}_{\kappa}$ changes very rapidly and is needed for non-resonant particle production, resulting in the large $F_{\kappa}$ in the bulk region of the momentum distribution. 
For our case, we can first write the non-adiabaticity condition for $\Omega_{\kappa}$ from Eq.~\eqref{eq: Omega_k(tau)} as follows:
\begin{equation}\label{eq:AnalyticF non-adiabatic}
\Biggr|\frac{d\Omega_{\kappa}}{d\tau}\Biggr|\gtrsim\Omega_{\kappa}^{2}
\implies \kappa \lesssim \sqrt{\left(qf\dot{f}\right)^{2/3}-qf^{2}},
\end{equation}
where we have chosen a time when $f$ and $\dot{f}$ are both positive. Particle production for large $q$ occurs whenever the effective fermion mass $\sqrt{q}f(\tau)$ crosses zero.\footnote{As before, we are assuming the bare mass of the fermions is negligible.} We can see this in Fig.~\ref{fig: nk small and large q kap in bulk}, where for large $q$ there is a sudden change in $n_{\kappa}(\tau)$ in the vicinity of times where $f(\tau)=0$. This indicates that the non-adiabaticity condition will be valid in a narrow region of $\tau$ around the points where $f(\tau)=0$. 
Similarly to the procedure in \cite{Kofman:1997yn}, we can find the maximum $\kappa$ that results in particle production by first setting $\kappa$ equal to the right-hand side of the inequality on the right of Eq.~\eqref{eq:AnalyticF non-adiabatic}. Then, if $f_{*}$ is the value of $f$ that corresponds to the maximum value of $\kappa$, it can be found by maximising $\kappa$ with respect to $f$ as follows:
\begin{equation}\label{eq:AnalyticF non-adiabatic f*}
\frac{d\kappa}{df}\Bigr|_{f=f_{*}}=
\frac{d\left(\left(qf\dot{f}\right)^{2/3}-qf^{2}\right)^{1/2}}{df}\Bigr|_{f=f_{*}}=0\implies
f_{*}=\frac{q^{-1/4}\dot{f}_{*}^{1/2}}{3^{3/4}},
\end{equation}
where $\dot{f}_{*}$ is the slope when $f(\tau)=0$ (and at a time when it is positive). As the $\tau$ range for which the non-adiabaticity condition is satisfied is very narrow, we approximate $\dot{f}_{*}$ as a constant equal to the slope of $f$ when it is zero. For $f(\tau)$ from Eq.~\eqref{eq:Inf. f=cn}, $\dot{f}_{*}=1/\sqrt{2}$. We use $f_{*}$ from Eq.~\eqref{eq:AnalyticF non-adiabatic f*} for the maximum $\kappa$ given by the inequality on the right of Eq.~\eqref{eq:AnalyticF non-adiabatic} and see for the corresponding value $\kappa_{*}$:
\begin{equation}\label{eq:AnalyticF kap*}
q f_{*}\dot{f}_{*}=\left(\kappa_{*}^{2}+qf_{*}^{2}\right)^{3/2}
\implies
\kappa_{*}=\left(\frac{2 q^{1/2}\dot{f}_{*}}{3\sqrt{3}}\right)^{1/2}=
\left(\frac{2}{27}\right)^{1/4}q^{1/4}\approx 0.52q^{1/4}.
\end{equation}
This $\kappa_{*}\approx 0.52q^{1/4}$ implies that the boundary of the bulk is $\propto q^{1/4}$.  If the major contribution to the fermion number density comes from the bulk region, we thus expect $I(q)\propto q^{3/4}$; we will show numerically that this is indeed the case. A similar approximation was used in \cite{Greene:1998nh} and \cite{Garcia-Bellido:2000woy}. 
For larger $\kappa$ values, particle production occurs via resonant production as in the small-$q$ case.  Estimating the boundary of the bulk is difficult for large $q$ because for many $q$ values, the lowest peaks are partially merged into the bulk. The points in the contours of $F_{\kappa}$ in the $(q,\kappa)$ plane where the lowest resonance peaks are separated from the bulk correspond to the dips in Fig.~\ref{fig: Fkap contours_q vs kap_bulkwidth_large q} because those are where the bulk is narrowest. Fitting the points at these dips indeed yields a $q^{1/4}$ power law as expected from the non-adiabaticity condition, as we show for the $F_{\kappa}=0.5$ contour in Fig.~\ref{fig: Fkap contours_q vs kap_bulkwidth_large q}.

For many larger values $q$, the bulk region merges with the lowest resonance peak denoted $l_{0}$, given by the function of $q$ in Eq.~\eqref{eq:ResPeakPos peak position l0}. This merging can be seen  in Fig.~\ref{fig: Fkap evolution with q} as the lowest peak moves down in $\kappa$ with increasing $q$. The merging can also be observed in slices of fixed $q$ in Fig.~\ref{fig: Fkap contours_q vs kap_bulkwidth_large q} as for a fixed $q$, we observe that the resonance peak gets wider as it gets closer to the bulk region and ultimately assimilates into the bulk.
Hence, rather than trying to find approximations for the bulk and the first peak of the integral $I$ from Eq.~\eqref{eq: total Fk integral I} (as done for small $q$), we treat the bulk and first peak as a single component. 

\begin{figure}[H]
	\centering
	\includegraphics[width=\textwidth]{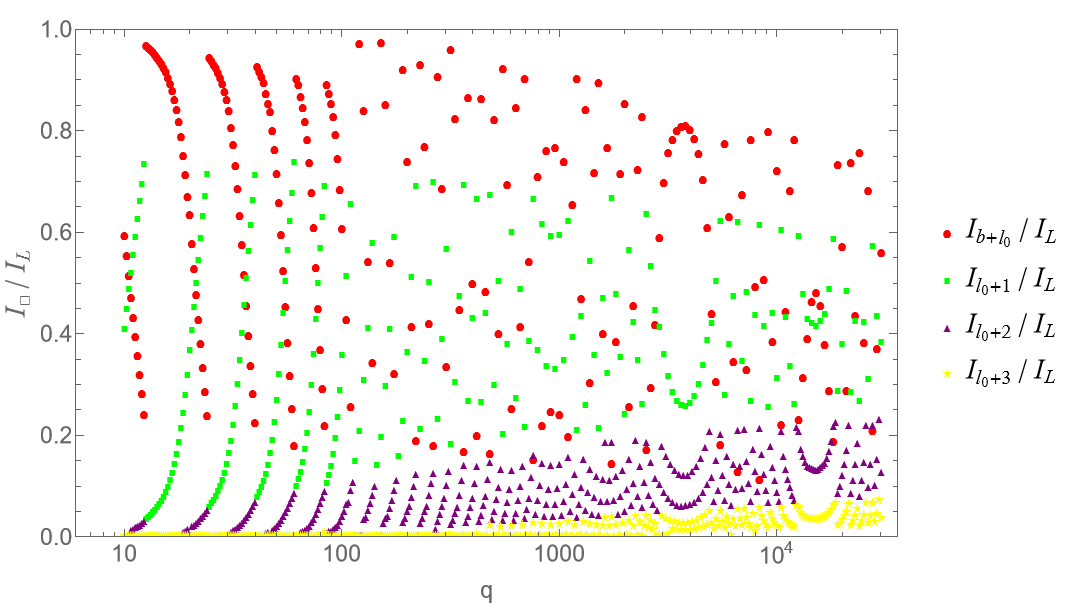}
	\caption{Contributions of different components as fractions of the total integral $I_{L}$ defined in Eq.~\eqref{eq:DenCont Large q Total}. Shown are contributions of a component with the bulk region and peak $l_{0}$ taken together $I_{b+l_{0}}$ (red circles), as well as $I_{l_{0}+1}$ (green squares), $I_{l_{0}+2}$ (purple triangles), $I_{l_{0}+3}$ (yellow stars), with the integrals defined in Eq.~\eqref{eq:DenCont Large q Total Parts}.    
    }\label{fig:DenCont Large q ratios}
\end{figure}

For large $q$ all resonance peaks with  $l \geq l_{0}$ exist, but our numerical analysis indicates that the cumulative sum is well approximated by the bulk and the first six resonance peaks. Therefore, we truncate our computation of the total integral after the sixth peak. 
As before, the positions of the resonances are computed using Eq.~\eqref{eq:ResPeakPos peak position Omkapavg}. Then, similar to Eq.~\eqref{eq:DenCont Small q total}, the total integral for the large $q$ values is evaluated as:
\begin{equation}\label{eq:DenCont Large q Total}
I_{L}(q)=\int_{0}^{(\kappa_{l_{0}+5}+\kappa_{l_{0}+6})/2} d\kappa \kappa^{2}F_{\kappa}.
\end{equation}
We find numerical estimations for different contributions to $I_{L}$ by first assuming that the bulk region and the $l_{0}$ peak form one component and then take into account the other resonance peaks (which are reasonably well separated from the bulk). 
We break up $I_{L}$ into six components with the different components representing contributions from different parts as follows:
\begin{equation}\label{eq:DenCont Large q Total Parts}
\begin{split}
I_{L}(q)
&=
\underbrace{\int_{0}^{(\kappa_{l_{0}}+\kappa_{l_{0}+1})/2} d\kappa \kappa^{2}F_{\kappa}}_\text{Bulk and $l_{0}$ Peak, $I_{b+l_{0}}$}
+
\underbrace{\int_{(\kappa_{l_{0}}+\kappa_{l_{0}+1})/2}^{(\kappa_{l_{0}+1}+\kappa_{l_{0}+2})/2} d\kappa \kappa^{2}F_{\kappa}}_\text{$l_{0}+1$ Peak, $I_{l_{0}+1}$}\\
&+
\underbrace{\int_{(\kappa_{l_{0}+1}+\kappa_{l_{0}+2})/2}^{(\kappa_{l_{0}+2}+\kappa_{l_{0}+3})/2} d\kappa \kappa^{2}F_{\kappa}}_\text{$l_{0}+2$ Peak, $I_{l_{0}+2}$}
+
\underbrace{\int_{(\kappa_{l_{0}+2}+\kappa_{l_{0}+3})/2}^{(\kappa_{l_{0}+3}+\kappa_{l_{0}+4})/2} d\kappa \kappa^{2}F_{\kappa}}_\text{$l_{0}+3$ Peak, $I_{l_{0}+3}$}\\
&+
\underbrace{\int_{(\kappa_{l_{0}+3}+\kappa_{l_{0}+4})/2}^{(\kappa_{l_{0}+4}+\kappa_{l_{0}+5})/2} d\kappa \kappa^{2}F_{\kappa}}_\text{$l_{0}+4$ Peak, $I_{l_{0}+4}$}
+
\underbrace{\int_{(\kappa_{l_{0}+4}+\kappa_{l_{0}+5})/2}^{(\kappa_{l_{0}+5}+\kappa_{l_{0}+6})/2} d\kappa \kappa^{2}F_{\kappa}}_\text{$l_{0}+5$ Peak, $I_{l_{0}+5}$}
\end{split}
\end{equation}
These integrals, normalized to the total, are shown in  Fig.~\ref{fig:DenCont Large q ratios} for various $q$ values in the range $10 < q < 3\cdot 10^4$.
We observe that although the bulk region and the $l_{0}$ peak together (red circles) account for the highest contribution for many values of $q$, the contribution from the $l_{0}+1$ peak (green squares) surpasses that of the bulk+$l_{0}$ component for some values of $q$, implying that it is not only the bulk region that has a relevant contribution. Although it is subdominant, the $l_{0}+2$ peak (purple triangles) also has a significant contribution, as much as 20\% of the total integral, while the $l_{0}+3$ and higher peaks make up a smaller contribution.

\begin{figure}[H]
    \subfloat
    {%
    \includegraphics[width=0.5\textwidth,height=0.2\textheight]{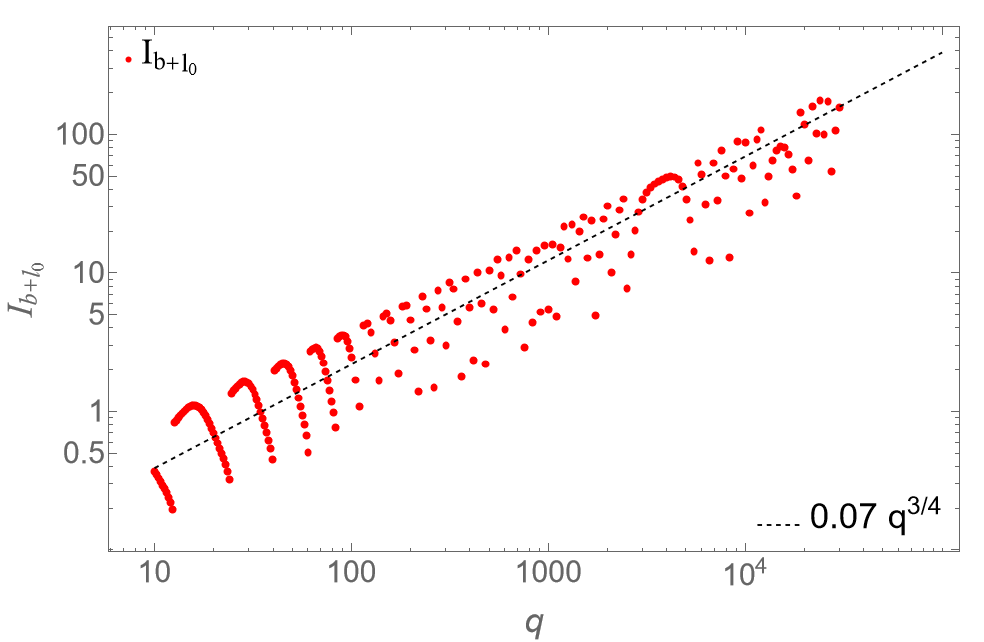}
    }
    \hfill
    \subfloat
    {%
    \includegraphics[width=0.5\textwidth,height=0.2\textheight]{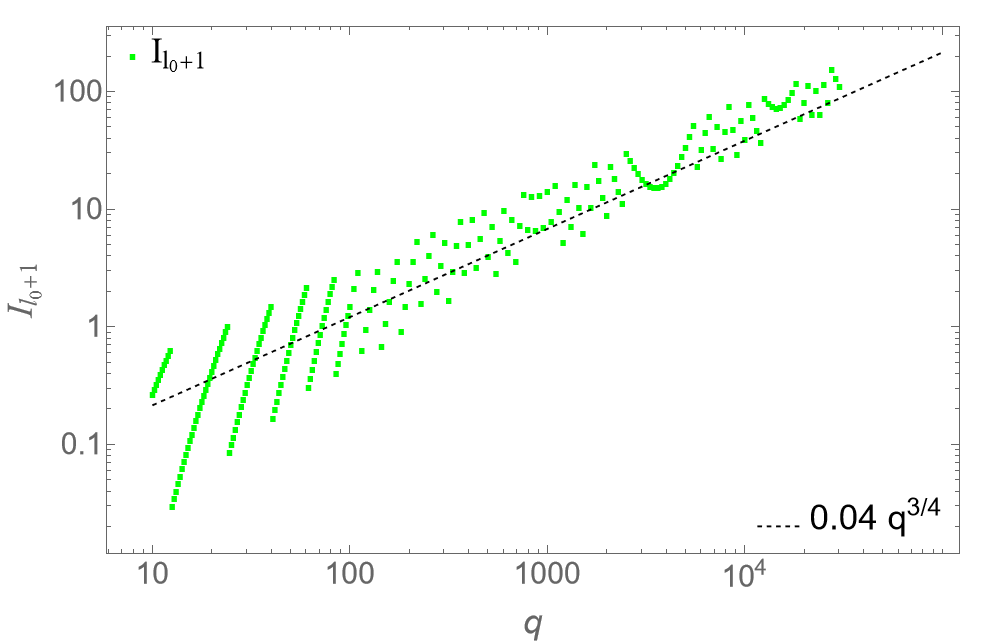}
    }
    \\
    \subfloat
    {%
    \includegraphics[width=0.5\textwidth,height=0.2\textheight]{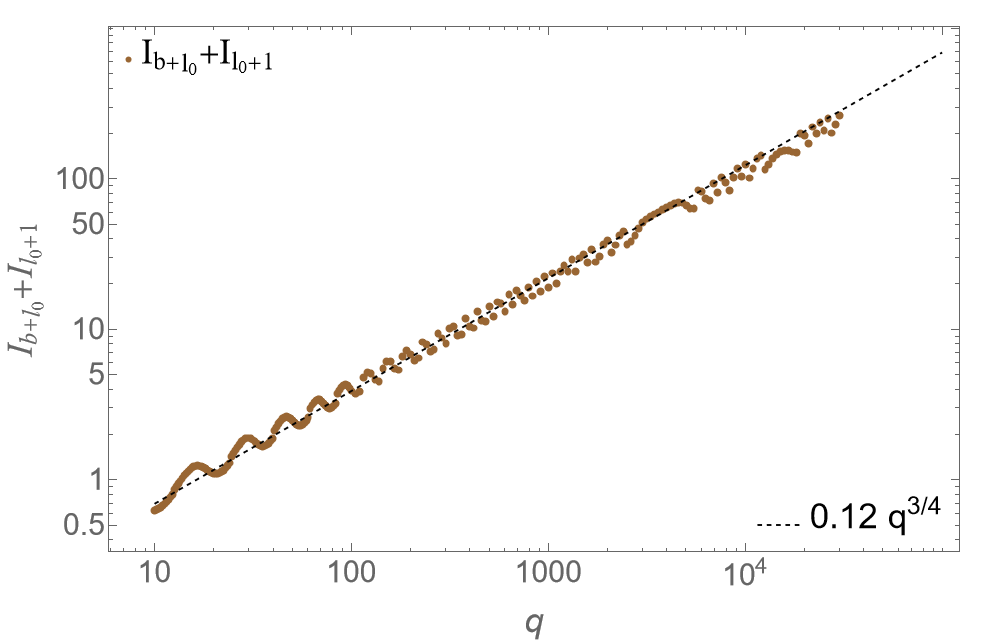}
    }
    \hfill
    \subfloat
    {%
    \includegraphics[width=0.5\textwidth,height=0.2\textheight]{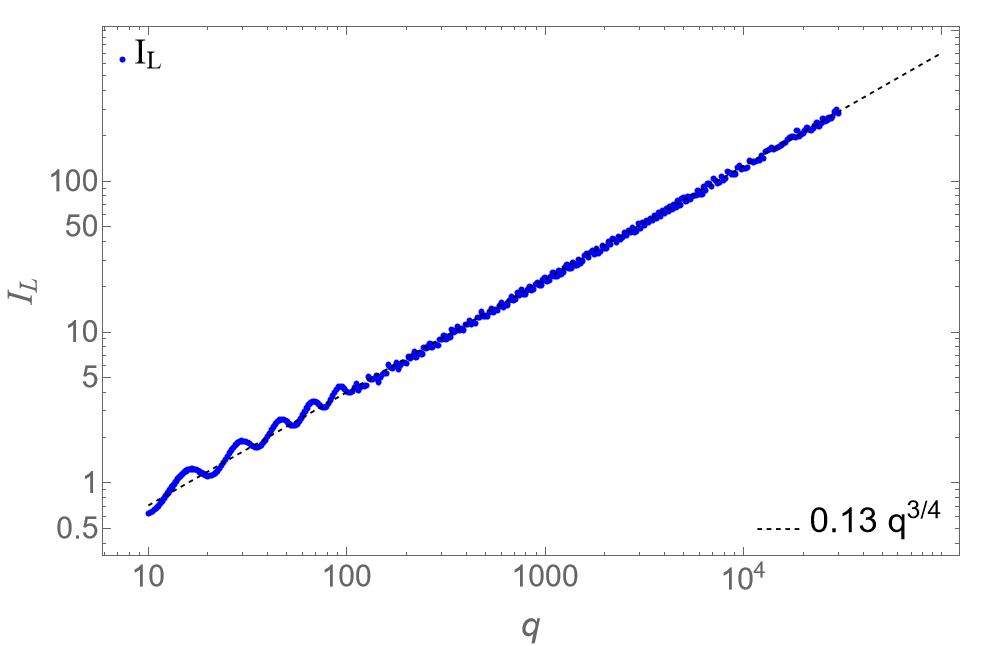}
    }
    \caption{Different integrals as a function of $q$ for $10 < q < 3\cdot 10^4$. 
    \textbf{Top left:} The integral $I_{b+l_{0}}$ (see Eq.~\eqref{eq:DenCont Large q Total Parts}) over the the bulk region and peak $l_{0}$ as well as the approximation from Eq.~\eqref{eq:DenCont Large q bulk+l0 approx}. 
    \textbf{Top right:} $I_{l_{0}+1}$ over the second peak and the approximation from Eq.~\eqref{eq:DenCont Large q l0+1 approx}. 
    \textbf{Bottom left:} $I_{b+l_{0}}+I_{l_{0}+1}$ over the bulk, first peak, and second peak, as well as the fit from Eq.~\eqref{eq:DenCont Large q bulk+l0 + l0+1 approx}. \textbf{Bottom right:} Total integral $I_{L}$ (see Eq.~\eqref{eq:DenCont Large q Total}) with the approximation from Eq.~\eqref{eq:DenCont Large q Total Approx}.   
    }\label{fig: Large q contributions and total subfigs}
\end{figure}

Then, similar to the small $q$ case, we fit the numerical values of the bulk+$l_{0}$ and $l_{0}+1$ integrals as a function of $q$ to a power law. The points are shown in the top two plots of Fig.~\ref{fig: Large q contributions and total subfigs}.
Although the values of the integrals oscillate around the fit as $q$ varies, both contributions are reasonably approximated by a $q^{3/4}$ power law, where the fit is also shown in the top two plots of 
Fig.~\ref{fig: Large q contributions and total subfigs}. The fits are given by:
\begin{equation}\label{eq:DenCont Large q bulk+l0 approx}
I_{b+l_{0}}(q) = \int_{0}^{(\kappa_{l_{0}}+\kappa_{l_{0}+1})/2} d\kappa \kappa^{2}F_{\kappa}
\approx 0.069\times q^{3/4},
\end{equation}
\begin{equation}\label{eq:DenCont Large q l0+1 approx}
I_{l_{0}+1}(q) = \int_{(\kappa_{l_{0}}+\kappa_{l_{0}+1})/2}^{(\kappa_{l_{0}+1}+\kappa_{l_{0}+2})/2} d\kappa \kappa^{2}F_{\kappa}
\approx 0.038\times q^{3/4}.
\end{equation}
When we add up the contributions from the bulk+$l_{0}$ component and the $l_{0}+1$ peak, we find that the scatter diminishes significantly and the data points more clearly follow a $q^{3/4}$ power law, which can be seen in the bottom left plot of Fig.~\ref{fig: Large q contributions and total subfigs}. Hence, an overall approximation for the total contribution from the bulk region, $l_{0}$ peak and the $l_{0}+1$ peak is found from the best fit to the data points as:
\begin{equation}\label{eq:DenCont Large q bulk+l0 + l0+1 approx}
I_{b+l_{0}}+I_{l_{0}+1} = \int_{0}^{(\kappa_{l_{0}+1}+\kappa_{l_{0}+2})/2} d\kappa \kappa^{2}F_{\kappa}
\approx 0.12\times q^{3/4}.
\end{equation}

When we add all the contributions in $I_L$ from Eq.~\eqref{eq:DenCont Large q Total}, we find that the scatter in the calculations reduces even more and the total integral clearly follows a $q^{3/4}$ power law, which is represented in the bottom right plot of Fig.~\ref{fig: Large q contributions and total subfigs}. The proportionality constant is found from the best fit to the data points and hence, the relevant approximation can be written as:
\begin{equation}\label{eq:DenCont Large q Total Approx}
I_{L}(q) = \int_{0}^{(\kappa_{l_{0}+5}+\kappa_{l_{0}+6})/2} d\kappa \kappa^{2}F_{\kappa} \approx 0.13\times q^{3/4}.
\end{equation}
We see that including peaks with $l > l_0+1$ changes the total integral by less than 10\% which shows that the contributions to $I$ from increasing $l$ are rapidly decreasing, implying $I_L$ is a good approximation for $I$.
This fit in Eq.~\eqref{eq:DenCont Large q Total Approx} for $I$ is can be used in the number density formulae from Eqs.~\eqref{eq:NumDen total n, dimless} and \eqref{eq:NumDen total n, physical, with a}, to approximate the total number density of produced fermions as a function of the coupling parameter $q$ for large values of $q\gtrsim 10$.

\subsection{Intermediate $q$}\label{Sec:DenCont Intermediate q}
\begin{figure}[H]
	\centering
	\includegraphics[width=\textwidth]{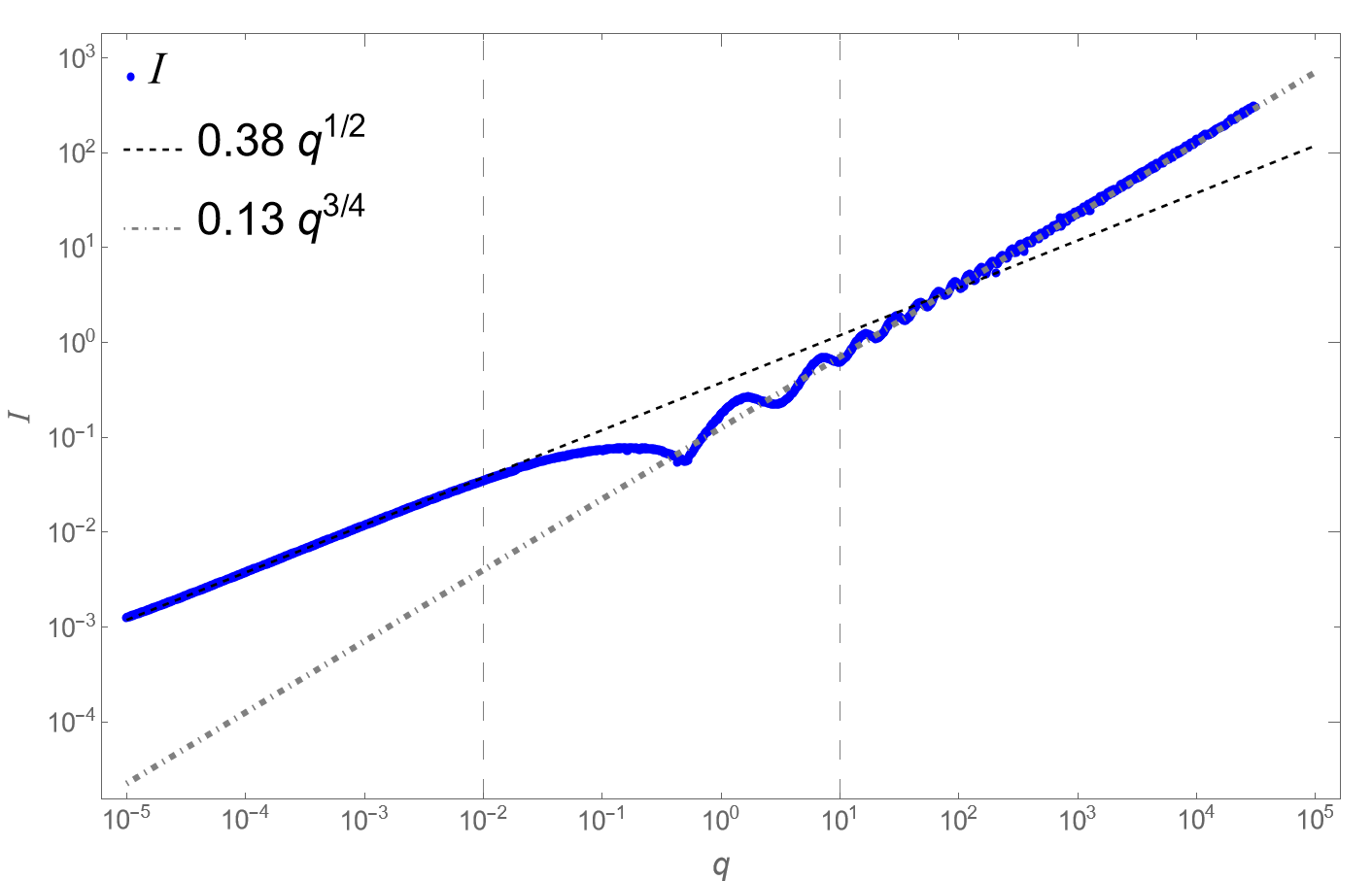}
	\caption{Total integral $I(q)$ for $10^{-5}\leq q \leq 3.0\times 10^{4}$, with the approximations for the small $q$ and large $q$ cases described by Eq.~\eqref{eq:DenCont Small q Total Approx} and Eq.~\eqref{eq:DenCont Large q Total Approx}. The vertical dashed lines divide the small $q$, intermediate $q$ and large $q$ regions.}\label{fig:DenCont All q with fits}
\end{figure}

For the intermediate range of $q$ values between $10^{-2}$ and $10$, we use the same integral limits to evaluate the total integral and the individual contributions as in Eqs.~\eqref{eq:DenCont Large q Total} and \eqref{eq:DenCont Large q Total Parts}. Neither the $q^{1/2}$ approximation from the small $q$ case nor the $q^{3/4}$ approximation from the large $q$ case gives a good approximation for the total integral in this intermediate range of $q$, as can be seen in Fig.~\ref{fig:DenCont All q with fits}. Although there is a major deviation from the $q^{1/2}$ power law for $q\gtrsim 0.01$, we find that for values of $q$ greater than $q_{0}(0) = 0.5$, an oscillating pattern starts to form around the $q^{3/4}$ power law fit which ultimately converges into the expression from Eq.~\eqref{eq:DenCont Large q Total Approx} at larger $q$. The $q^{3/4}$ power law fit seems to be decent for $1<q<10$ but with a maximum relative error $\sim 50\%$. 

\begin{figure}[H]
	\centering
	\includegraphics[width=\textwidth]{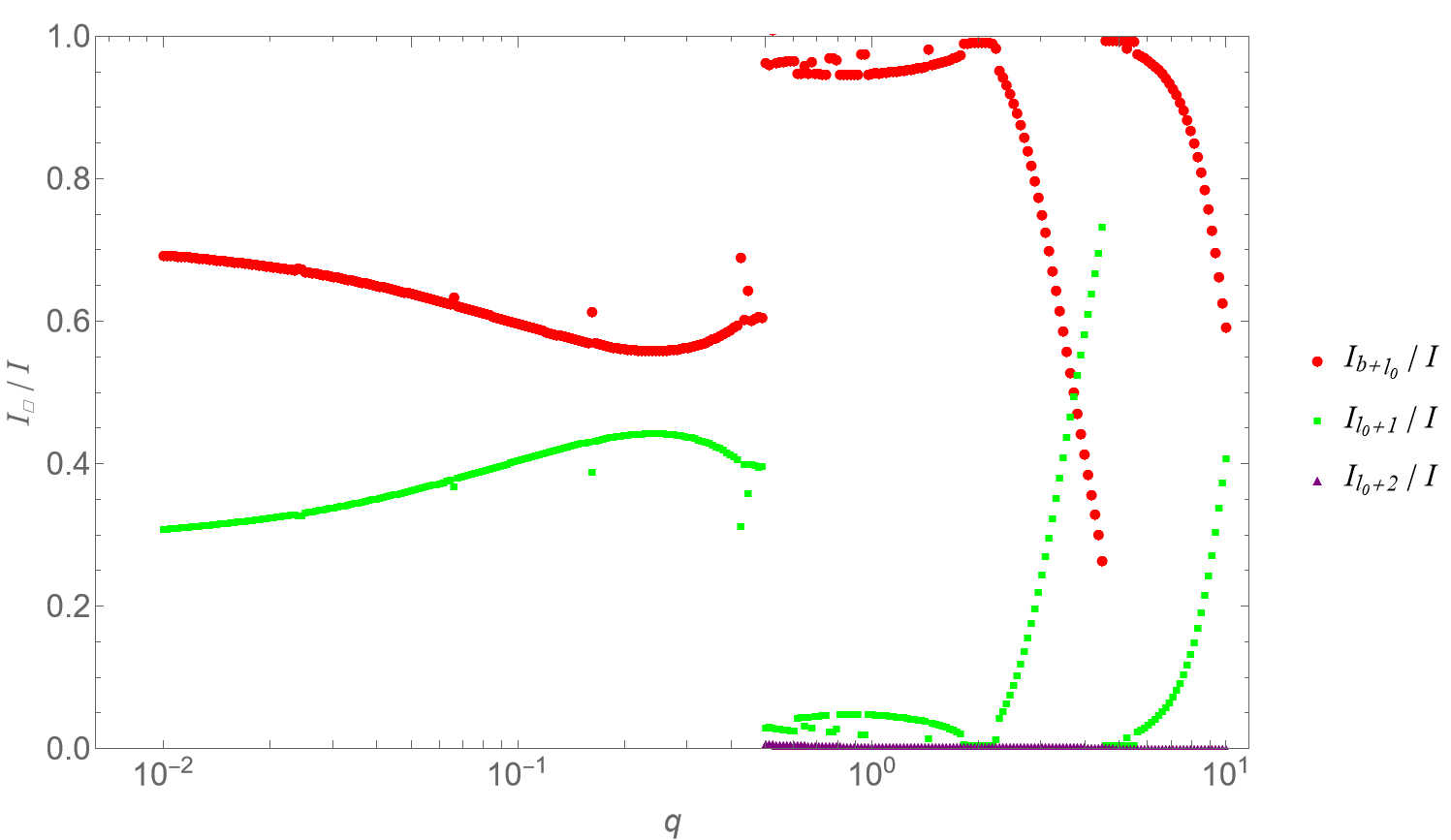}
	\caption{Contributions of different components for $0.01\leq q \leq 10$ as fractions of the total integral. For $q < q_{0}(0) = 0.5$, the normalization is $I_S$ from Eq.~\eqref{eq:DenCont Small q total}, while for $q\geq q_0(0)$, the normalization in $I_L$ from Eq.~\eqref{eq:DenCont Large q Total}. Shown are contributions of the bulk  and  $l_{0}$ peak, $I_{b+l_{0}}$ (red circles), the $l_{0}+1$ peak $I_{l_{0}+1}$ (green squares), and the $l_{0}+2$ peak $I_{l_{0}+2}$ (purple triangles, for $q>q_0(0)$ only).
    }\label{fig:DenCont Intermediate q ratios}
\end{figure}

We also plot fractions of the total integral contributed by the first three components, and this is shown in Fig.~\ref{fig:DenCont Intermediate q ratios}. Similar to the large $q$ case, the major contributions come from the the bulk+$l_{0}$ component and the $l_{0}+1$ peak, which can be seen in a more pronounced manner for this intermediate $q$ range as the contribution of the $l_{0}+2$ peak remains very small. For $q$ less than $q_{0}(0) = 0.5$, the bulk+$l_{0}$ component has the dominant contribution over that of the $l_{0}+1$ peak. Then for values of $q$ just above each transition point $q_{0}(l)$ (from Eq.~\eqref{eq:AnalyticF q0(l)}, with $l=1,2,3,\dots$), the largest contribution again comes from the bulk+$l_{0}$ component. However, this fraction decreases with increasing $q$ and the contribution from the $l_{0}+1$ peak dominates at values of $q$ just below the next transition point $q_{0}(l+1)$. The jumps in Fig.~\ref{fig:DenCont Intermediate q ratios} at $q=0.5$ and $q=4.5$ (corresponding to $q_{0}(0)$ and $q_{0}(1)$ from Eq.~\eqref{eq:AnalyticF q0(l)} respectively) are physical and are a consequence of the $l$ value of the lowest resonance peak changing from $0$ to $1$ and from $1$ to $2$ respectively. We expect such jumps over the whole range of $q$ whenever the $l$ value of the lowest resonance peak changes and we observe this pattern in Fig.~\ref{fig:DenCont Large q ratios} as well. 

\section{Discussion and Conclusions}\label{sec: 5}

We considered non-perturbative production of fermions that are produced out of thermal equilibrium during coherent oscillations of a scalar field. We study the production of such particles through the mechanism of fermionic preheating for $\lambda\phi^{4}$ inflation, under the assumption that the bare mass of the fermions is negligible. Production is described by a differential equation with a time-dependent natural frequency $\Omega_{\kappa}(\tau)\equiv\sqrt{\kappa^{2}+qf^{2}(\tau)}$, which represents an effective relativistic energy, where $\kappa$ is the dimensionless comoving momentum of the fermions, $f(\tau)$ describes inflaton oscillations, and $q=h^{2}/\lambda$ is a coupling parameter with $h$ being the Yukawa coupling between inflaton and fermions, and $\lambda$ being the inflaton quartic coupling. The momentum distribution of the fermions produced is described by an envelope function $F_{\kappa}$, which contains a highly occupied `bulk region' around $\kappa=0$ and resonance peaks for higher $\kappa$.

It was observed in \cite{Greene:1998nh} that for $q\ll1$, resonance peaks of $F_{\kappa}$ are present when $\kappa$ is an odd multiple of $\pi/T$ ($T$ being the time period of inflaton oscillations). We provide a physics explanation for this relation and generalize the idea to find the momentum value of resonance peaks for any $q$. We use the time average of $\Omega_{\kappa}$ to find a simple semi-analytic relation that predicts the $\kappa$ values corresponding to resonance peaks in the momentum distribution of the fermions produced for any $q$ without the need to numerically solve a differential equation. This is described as: $\overline{\Omega}_{\kappa}(q)=\frac{\pi}{T}(2l+1)\text{, }l=0,1,2,3,\dots$, which we use to find and label $\kappa$ corresponding to the resonance peak $l$ as $\kappa_{l}$ for various $q$. However, $l=0$ does not correspond to the first resonance peak for all $q$ values. The $l$ value representing the first resonance peak is called $l_{0}$ and is given by Eq.~\eqref{eq:ResPeakPos peak position l0} as a function of $q$. Because of the physics behind it, we expect this semi-analytic relation describing the positions of resonance peaks in the momentum spectrum to be true for any symmetric inflaton potential.

As shown in Eq.~\eqref{eq:LateTime nbar tau avg}, we use $F_{\kappa}$ to describe the number density per mode of the particles produced ($\bar{n}_{\kappa}$), which implies that for the total number density ($n$):
\begin{equation}\label{eq:Disc total num den}
n \propto N_{f}\int_{0}^{\infty} d\kappa \kappa^{2} \langle\bar{n}_{\kappa}\rangle = \frac{N_{f}}{2}\int_{0}^{\infty} d\kappa \kappa^{2} F_{\kappa}
\equiv \frac{N_{f}}{2}I,
\end{equation}
where $N_{f}$ is the number of chiral fermion species. Hence, we use $F_{\kappa}$ to estimate the total number density of fermions through the approximations we find for the integral $I$.

We observe that the momentum spectrum of the fermions produced is not completely degenerate. We find that for small values of the coupling parameter $q\lesssim 0.01$, the main contributions ($\sim 95\%$) to $I$ are actually from momentum shells at $\kappa$ corresponding to the resonance peaks labeled $l=0$ and $l=1$ (at $\kappa_{0}=\pi/T$ and $\kappa_{1}=3\pi/T$), instead of the bulk region (an approximately filled sphere around $\kappa =0$). We can see this from the plots of the contributions of different components of $F_{\kappa}$ in Figs.~\ref{fig:DenCont Small q ratios} and \ref{fig: Small q contributions and total subfigs}.

For larger $q$, we evaluate and plot the contributions to $I$ of different components of $F_{\kappa}$ in Figs.~\ref{fig:DenCont Large q ratios} and \ref{fig: Large q contributions and total subfigs} using the values of $\kappa_{l}$ for different $q$, and find that $\sim 80$--$90\%$ of $I$ is taken into account with $\kappa$ between $0$ and the peak $l_{0}+1$ (where Eq.~\eqref{eq:ResPeakPos peak position l0} gives $l_{0}$ for any $q$). Overall, we observe contributions from an approximately half-filled sphere with sub-dominant contributions from momentum shells at resonant values of momentum.

Thereafter, we extract novel analytic approximations for the contributions of different components of $F_{\kappa}$ to the number density through $I$.
For $q \lesssim 0.01$, we find that the major contributions, which are from the resonance peaks, follow $q^{1/2}$ power-law approximations such as $I_{0}(q) \approx 0.26 \times q^{1/2}$ for the $l=0$ peak, as shown in Fig.~\ref{fig: Small q contributions and total subfigs}. We therefore find a $q^{1/2}$ power-law approximation for $I$ for small $q$, given by $I_{S}(q)\approx 0.38 \times q^{1/2}$ from Eq.~\eqref{eq:DenCont Small q Total Approx}.  For larger $q$ ($\gtrsim 10$), we find that the contributions of different components follow $q^{3/4}$ power-law approximations, as shown in Fig.~\ref{fig: Large q contributions and total subfigs}. We therefore can approximate the total integral at large $q$ by $I_{L}(q) \approx 0.13 \times q^{3/4}$ from Eq.~\eqref{eq:DenCont Large q Total Approx}. 

The fermions produced by the mechanism described in this work could be a promising dark matter candidate. In Ref.~\cite{Carena:2021bqm}, it was found that for a single flavour of fully-degenerate Majorana fermions $\psi$ to make up all dark matter, their mass $m_{\psi}$ must be greater than about 2~keV as a consequence of the effects on structure formation when their Fermi momentum $p_F$ remains relativistic, i.e. $p_F \gtrsim m_\psi$, to a sufficiently late redshift. We can adapt this limit on $m_\psi$ for our case of Dirac fermions (equivalent to $N_f = 2$ Majorana flavours) produced with the nonthermal momentum distributions predicted by our analysis, assuming that these fermions make up all the dark matter.

We first consider small $q \lesssim 0.01$, for which the fermion momentum spectrum is dominated by the first two resonance peaks. 
To demonstrate the method, we start by considering only the $l=0$ resonance peak and assuming that all the fermions produced are in a single narrow Fermi shell with momentum $\kappa_{0}=\pi/T$.  These fermions become relativistic when $m_\psi \simeq \kappa_0$.  The bound on $m_\psi$ can then be found by rescaling the fermion number density in Ref.~\cite{Carena:2021bqm} by the ratio:
\begin{equation}\label{eq:Disc fraction l=0 only}
\frac{(N_{f}/2)\times I_{0}(q)}{\int_{0}^{p_{F}=\kappa_{0}}p^{2}dp}
=\frac{3N_{f}I_{0}(q)}{2\kappa_{0}^{3}},
\end{equation}
where $I_{0}(q)$ is the integral of the $l = 0$ resonance momentum shell from Eq.~\eqref{eq:DenCont Small q l=0 approx} (note that this first-pass approximation ignores 30--35\% of the fermion number density).
Using the fact that the energy density of dark matter $\rho_{\psi}=nm_{\psi}$ remains fixed, we can then rescale the limit on $m_\psi$ from Ref.~\cite{Carena:2021bqm} as follows: 
\begin{equation}\label{eq:Disc mPsi limit with kap0 only}
m_{\psi} > 2\text{ keV}\left(\frac{2\kappa_{0}^{3}}{3 N_{f}I_{0}(q)}\right)^{1/4}
\approx 2\text{ keV}\left(\frac{0.56}{q^{1/8}}\right).
\end{equation}
For $q=10^{-5}$, we find that the limit is strengthened to $m_{\psi}\gtrsim 5$ keV, but for $q=0.01$, we still find $m_{\psi}\gtrsim 2$ keV. The constraint on $m_{\psi}$ is not improved for $q=0.01$, which we attribute to the fact that the peak $l=0$ for $q=0.01$ is quite broad, as can be seen in Fig.~\ref{fig: nbarkap for q = 0.01, 1, 100}, so that it again approximates a (half-)filled Fermi sphere.

For a better approximation at small $q$, we can instead consider the contributions from the first two momentum shells, at $\kappa_{0}=\pi/T$ and $\kappa_{1}=3\pi/T$, containing $\sim$65\% and $\sim$35\% of the total fermion number density, respectively. Assuming that having only $\sim$35\% of the dark matter be relativistic is sufficient to trigger the constraints from structure formation, we can follow the same procedure, now taking into account the full number density and setting $m_\psi \simeq \kappa_1$.  In this case we find,
\begin{equation}\label{eq:Disc mPsi limit with kap0 and kap1}
m_{\psi} > 2\text{ keV}\left(\frac{2\kappa_{1}^{3}}{3 N_{f}I_{S}(q)}\right)^{1/4}
\approx 2\text{ keV}\left(\frac{1.2}{q^{1/8}}\right).
\end{equation}
The limit on $m_{\psi}$ for the particles to make up all dark matter is then strengthened for $q=10^{-5}$ ($q=0.01$) to $m_{\psi}\gtrsim 10$ keV ($m_{\psi}\gtrsim 4$ keV). We expect the actual bound to be between the ones from \eqref{eq:Disc mPsi limit with kap0 only} and \eqref{eq:Disc mPsi limit with kap0 and kap1}.
 
For larger $q$ values, the momentum spectrum can be reasonably approximated as a half-filled Fermi sphere. Taking into account the extra factor of $N_f = 2$ for Dirac fermions and assuming that these fermions make up all the dark matter, the bound $m_\psi > 2$~keV from Ref.~\cite{Carena:2021bqm} remains unchanged.

Although our results are for $\lambda\phi^{4}$ inflation, they are easily generalizable to $m^{2}\phi^{2}$ inflation \cite{Linde:1983gd}, hybrid inflation \cite{Linde:1993cn}, or coherent axion oscillations \cite{Preskill:1982cy,Abbott:1982af}. We generically expect the total number density of produced fermions for any coherently oscillating scalar field with a symmetric potential to follow the same power laws that we found for $\lambda\phi^{4}$ inflation: proportional to $q^{1/2}$ for small values of $q$, and proportional to $q^{3/4}$ for large values of $q$. We check $I_{S}(q)$ from Eq.~\eqref{eq:DenCont Small q total} for $m^{2}\phi^{2}$ inflation and obtain a nearly identical fit as the one for $\lambda\phi^{4}$ inflation: $I_{S}(q)\approx 0.39\times q^{1/2}$. 


\section*{Acknowledgments}
We thank Ivan L'Heureux for helpful suggestions. We thank the anonymous referee(s) from Physical Review D for suggesting the analysis in Appendix~\ref{AppG}. 
This work was supported in part by the Natural Sciences and Engineering Research Council of Canada (NSERC). This work was performed on the unceded territory of the Algonquin Anishinaabeg nation, and in part at Aspen Center for Physics, which is supported by National Science Foundation grant PHY-2210452.

\appendix
\section{Details of the calculation of $H_D$ in terms of $\hat{a}_{k,s}$, $\hat{b}_{k,s}$ operators}\label{AppA}

Here, we write the steps involved in representing the Hamiltonian $H_{D}(\eta)$ in terms of the creation ($\hat{a}^{\dagger}_{k,s}$, $\hat{b}^{\dagger}_{k,s}$) and annihilation ($\hat{a}_{k,s}$, $\hat{b}_{k,s}$) operators. $H_{D}(\eta_{0})$ is diagonal in these operators and the operators annihilate the ground state of the Hamiltonian. However, $H_{D}(\eta)$ at $\eta > \eta_{0}$ is not diagonal in these operators and that is the expression we are looking for. First, we write the equations that describe all the relevant terms from Eqs.~\eqref{eq:NumDen X(x,eta)} and \eqref{eq:NumDen Psi(eta,x) operators}. The orthonormal basis spinors $R_{\pm}$ and $\overline{R}_{\pm}$ are defined through the eigenequations:
\begin{equation}\label{eq: R eigeneqs.}
\begin{split}
&\gamma^{0} R_{\pm}(\mathbf{k})=+R_{\pm}(\mathbf{k}),
\quad
\gamma^{0} \bar{R}_{\pm}(\mathbf{k})=-\bar{R}_{\pm}(\mathbf{k}),\\
&(\mathbf{\hat{k}} \cdot \mathbf{\Sigma})R_{\pm}(\mathbf{k})=\pm R_{\pm}(\mathbf{k}),
\quad
(\mathbf{\hat{k}} \cdot \mathbf{\Sigma})\bar{R}_{\pm}(\mathbf{k})=\pm \bar{R}_{\pm}(\mathbf{k}).
\end{split}
\end{equation}
For simplicity, the comoving momentum ($\mathbf{k}$) of the particles can be assumed along the $z$-direction i.e.~$\mathbf{k}=(0,0,k)$, and we can use $\mathbf{\Sigma}=\begin{pmatrix}
\bm{\sigma}&0\\
0&\bm{\sigma}\\
\end{pmatrix}$ \cite{Thomson:2013zua}, where $\bm{\sigma}$ are the Pauli matrices. Using this assumption and the eigenequations above, we can show the following relations:
\begin{equation}\label{eq: R pm relations}
R_{\pm}(\pm\mathbf{k})=R_{\mp}(\mp\mathbf{k}), 
\quad
\bar{R}_{\pm}(\pm\mathbf{k})=\bar{R}_{\mp}(\mp\mathbf{k}).
\end{equation}
The ansatz used for $\Psi$ in Eq.~\eqref{eq:NumDem psi ansatz eta} can then be used to define $u_{k,s}(\eta)$, with $'$ here indicating derivatives with respect to $\eta$:
\begin{equation}\label{eq: u spinor}
\begin{split}
&u_{k,\pm}(\eta)e^{+i\mathbf{k} \cdot \mathbf{x}}=\left[i\gamma^{0}\partial_{\eta}+i\gamma^{j}\partial_{j}+(h\varphi(\eta)+m_{\psi}a(\eta))\right]X_{k}(\eta)R_{\pm}(\mathbf{k})e^{+i\mathbf{k} \cdot \mathbf{x}},\\
\implies
&u_{k,\pm}(\eta)=\left[iX'_{k}(\eta)-(\mathbf{\gamma} \cdot \mathbf{k}-(h\varphi(\eta)+m_{\psi}a(\eta))X_{k}(\eta)\right]R_{\pm},
\end{split}
\end{equation}
where $a(\eta)$ is the scale factor. The charge conjugate $v_{k,s}(\eta)$ is defined through the charge conjugation operator $\mathit{C}=i\gamma_{0}\gamma_{2}$ and $\overline{u}_{k,\pm}=u^{\dagger}_{k,\pm}\gamma_{0}$ as:
\begin{equation}\label{eq: v spinor}
\begin{split}
v_{k,\pm}(\eta)&=\mathit{C}\overline{u}^{T}_{k,\pm}(\eta),\\
\implies v_{k,\pm}(\eta)&=\left[-iX'^{*}_{k}(\eta)+(\mathbf{\gamma} \cdot \mathbf{k}+(h\varphi(\eta)+m_{\psi}a(\eta)))X^{*}_{k}(\eta)\right]\overline{R}_{\pm}.
\end{split}
\end{equation}
To evaluate the Hamiltonian $H_{D}$ in terms of the fermion creation and annihilation operators, we first use Eqs.~(\ref{eq: u spinor}) and (\ref{eq: v spinor}) in Eq.~\eqref{eq:NumDen Psi(eta,x) operators} and use the resultant expression in Eq.~\eqref{eq:NumDen H_D}. The definition of the $\delta$-function can then be used to first remove the $d^{3}x$-integral:
\begin{equation}\label{eq: delta-func.}
\int e^{i(\mathbf{k}-\mathbf{K}) \cdot \mathbf{x}}d^{3}x = (2\pi)^{3} \delta^{(3)}(\mathbf{k}-\mathbf{K}).
\end{equation}
Thereafter, eigenequations from Eq.~(\ref{eq: R eigeneqs.}) can be used with Eq.~\eqref{eq: R pm relations} to substitute values for terms with $R_{\pm}$ and $\overline{R}_{\pm}$ and this ultimately yields Eq.~(\ref{eq:NumDen H_D operators}).
\section{Deriving Relation between $E$, $F$ and $\Omega_{k}$}\label{AppB}
We will now show how the relation between $E$, $F$ and $\Omega_{k}$ from Eq.~(\ref{eq:NumDen E^2+F^2, with C}) comes about. To do so, we first assume $f(\eta)=g(\eta)\equiv X_{k}(\eta)$ for Eq.~(\ref{eq:NumDen f, g identity}), and we can then write:
\begin{equation}\label{eq: f=g identity}
|g'(\eta)|^{2}+\Omega_{k}^{2}(\eta)|g(\eta)|^{2}+2(m_{\psi}a(\eta)+h\varphi(\eta))\text{Im}(g(\eta)g'^{*}(\eta))=C.
\end{equation}
Eqs.~(\ref{eq:NumDen X_k mode eq_eta}) and (\ref{eq:NumDen E(eta),F(eta)}) can also be written as:
\begin{equation}\label{eq: diff eq., E, F}
\begin{split}
&g''(\eta)+\left[k^{2}+(h\varphi(\eta)+m_{\psi}a(\eta))^{2}-ih\varphi'(\eta)\right]g(\eta) = 0,\\
&E(\eta)=\left[h\varphi(\eta)+m_{\psi}a(\eta)\right]\left[|g'(\eta)|^{2}+\Omega_{k}^{2}(\eta)|g(\eta)|^{2}\right]+2\Omega_{k}^{2}(\eta)\text{Im}\left(g(\eta)g'^{*}(\eta)\right),\\
&F(\eta)=k\left(g'^{*}(\eta)\right)^{2}+k\Omega_{k}^{2}(\eta)\left(g^{*}(\eta)\right)^{2}.
\end{split}
\end{equation}
Using these, we can show that:
\begin{equation}\label{eq: E^2, F^2}
\begin{split}
E(\eta)^{2}=&\left[h\varphi(\eta)+m_{\psi}a(\eta)\right]^{2}\left[|g'(\eta)|^{2}+\Omega_{k}^{2}(\eta)|g(\eta)|^{2}\right]^{2}+4\Omega_{k}^{4}(\eta)\text{Im}\left(g(\eta)g'^{*}(\eta)\right)^{2}+\\
&4\Omega_{k}^{2}(\eta)\text{Im}\left(g(\eta)g'^{*}(\eta)\right)\left[h\varphi(\eta)+m_{\psi}a(\eta)\right]\left[|g'(\eta)|^{2}+\Omega_{k}^{2}(\eta)|g(\eta)|^{2}\right],\\
|F(\eta)|^{2}=&k^{2}\left[|g'(\eta)|^{4}+\Omega_{k}^{4}(\eta)|g(\eta)|^{4}\right]+k^{2}\Omega_{k}^{2}\left[2|g(\eta)|^{2}|g'(\eta)|^{2}-4\text{Im}\left(g(\eta)g'^{*}(\eta)\right)^{2}\right].
\end{split}
\end{equation}
Then, we can add these up and use Eq.~\eqref{eq: f=g identity} to show:
\begin{equation}\label{eq: E^2 + F^2 proof}
E(\eta)^{2}+|F(\eta)|^{2}=
\Omega_{k}^{2}
\left[|f'(\eta)|^{2}+\Omega_{k}^{2}(\eta)|f(\eta)|^{2}+2(m_{\psi}a+h\varphi)\text{Im}(f(\eta)f'^{*}(\eta))\right]^2=\Omega_{k}^{2}C^{2}.
\end{equation}
\section{Derivation of $|\beta|^{2}$}\label{AppC}
We will show how we can calculate the fermion (or anti-fermion) number density per mode $n_{k}(\eta)=|\beta|^{2}$. There are two equivalent ways to show that this relation holds (in either the Schrödinger or the Heisenberg picture). One is to consider time-dependent vacuum and use $\ket{0_{\eta}}$ to represent the ground state of the Hamiltonian at time $\eta$, which is annihilated by new operators $\hat{c}_{k,s}$, $\hat{d}_{k,s}$ defined in Eq.~\eqref{eq:NumDen Bogoliubov transf. properties}. Then $n_{k}(\eta)$ is given by the expectation value of the time-independent number density operator $\hat{a}_{k,s}^{\dagger}\hat{a}_{k,s}$ (or $\hat{b}_{k,s}^{\dagger}\hat{b}_{k,s}$) as:  $n_{k}(\eta)=\langle{0_{\eta}}|\hat{a}_{k,s}^{\dagger}\hat{a}_{k,s}|0_{\eta}\rangle=\langle{0_{\eta}}|\hat{b}_{k,s}^{\dagger}\hat{b}_{k,s}|0_{\eta}\rangle$.  Alternatively, we consider the time-independent vacuum $\ket{0}$ that is annihilated by $\hat{a}_{k,s}$, $\hat{b}_{k,s}$, to represent the ground state of the Hamiltonian at time $\eta_{0}$. Then $n_{k}(\eta)$ is given by the expectation value of the number density operator at time $\eta$, $\hat{c}_{k,s}^{\dagger}\hat{c}_{k,s}$ (or $\hat{d}_{k,s}^{\dagger}\hat{d}_{k,s}$), as: $n_{k}(\eta)=\langle{0}|\hat{c}_{k,s}^{\dagger}\hat{c}_{k,s}|0\rangle=\langle{0}|\hat{d}_{k,s}^{\dagger}\hat{d}_{k,s}|0\rangle$, where the time dependence is carried in the transformation from the $\hat{a},\hat{b}$ operators to the $\hat{c},\hat{d}$ operators. We can use the first option, and use the Bogoliubov transformations described by Eqs.~\eqref{eq:NumDen Bogoliubov transf.} and \eqref{eq:NumDen Bogoliubov transf. properties}, to see that:
\begin{equation}\label{eq: nk = |beta|^2}
n_{k}(\eta)
=\langle{0_{\eta}|\hat{a}_{k,s}^{\dagger}\hat{a}_{k,s}|0_{\eta}\rangle}
=\langle{0_{\eta}|\left(\alpha\hat{c}_{k,s}^{\dagger}-\beta^{*} \hat{d}_{k,s}\right)\left(\alpha^{*}\hat{c}_{k,s}-\beta \hat{d}_{k,s}^{\dagger}\right)|0_{\eta}\rangle}=|\beta|^{2}.
\end{equation}
As the Bogoliubov transformation diagonalises the Hamiltonian $H_{D}(\eta)$, we can set the non-diagonal terms of Eq.~(\ref{eq:NumDen H_D Bogoliubov coeff.}) to $0$ and this implies:
\begin{equation}\label{eq: diag. coeffs. to 0}
\begin{split}
F\alpha^{2}-F^{*}\beta^{2}-2E\alpha\beta=0.
\end{split}
\end{equation}
Due to their property in Eq.~\eqref{eq:NumDen Bogoliubov transf. properties}, $\alpha$ and $\beta$ can then be written as:
\begin{equation}\label{eq: alpha, beta}
\alpha=\cos \theta e^{i\phi_{\alpha}}, 
\quad
\beta=\sin \theta e^{i\phi_{\beta}}.
\end{equation}
Substituting these into Eq.~(\ref{eq: diag. coeffs. to 0}) gives (using $\phi=\phi_{\alpha}-\phi_{\beta}$):
\begin{equation}\label{eq: alpha, beta substitution 1}
F \cos^{2}\theta e^{i\phi}-F^{*}\sin^{2}\theta e^{-i\phi}=2E\cos\theta\sin\theta,
\end{equation}
As the right-hand side of this equation is purely real, the left-hand side of the equation is its own conjugate and we can write:
\begin{equation}\label{eq: F*/F}
F \cos^{2}\theta e^{i\phi}-F^{*}\sin^{2}\theta e^{-i\phi}=F^{*} \cos^{2}\theta e^{-i\phi}-F\sin^{2}\theta e^{i\phi}\\
\implies e^{2i\phi}=F^{*}/{F}.
\end{equation}
Eq.~(\ref{eq: alpha, beta substitution 1}) can then be solved as a quadratic equation in $\sin^{2}\theta$, and after making the substitution from Eq.~(\ref{eq: F*/F}), the following equation is obtained:
\begin{equation}\label{eq: beta^2}
\begin{split}
&\sin^{2}\theta=|\beta|^{2}=\frac{(F^{*}/F)(4(|F|^{2}+E^{2})\pm 4E\sqrt{E^{2}+|F|^{2}})}{8(F^{*}/F)(E^{2}+|F|^{2})} = 
\frac{1}{2}\pm \frac{E}{2\sqrt{E^{2}+|F|^{2}}} .
\end{split}
\end{equation}
We need to choose the negative solution to ensure that $n_{k}(\eta_{0})=|\beta(\eta_{0})|^{2}=0$.
\section{Comparing Different $n_{k}(\eta)$ Expressions}\label{AppD}
We will show here that the expression for $n_{k}(\eta)$ that we have derived is equivalent to the number density equations present in \cite{Greene:1998nh}, \cite{Greene:2002uot}, and \cite{Garcia-Bellido:2000woy}. The relevant equations from existing literature also assume the same initial conditions as in Eq.~(\ref{eq:NumDen X_k intial cond.}) that imply $C=1$. Together with the additional assumption of light fermions i.e.~$m_{\psi}\approx0$, the number density from Eq.~(\ref{eq:NumDen n_k(eta), with C}) can then be written as:
\begin{equation}\label{eq: n_k full}
n_{k}(\eta)=\frac{\Omega_{k}(\eta)- h\varphi(\eta)[|X'_{k}(\eta)|^{2}+\Omega_{k}^{2}(\eta)|X_{k}(\eta)|^{2}]-2\Omega_{k}^{2}(\eta)\text{Im}(X_{k}(\eta)X'^{*}_{k}(\eta))}{2\Omega_{k}(\eta)}
\end{equation}
Then, Eq.~(\ref{eq: f=g identity}) can be used with $C=1$ and $m_{\psi}=0$ to write:
\begin{equation}\label{eq: using identity}
\begin{split}
&|X'_{k}(\eta)|^{2}+\Omega_{k}^{2}(\eta)|X_{k}(\eta)|^{2}+2h\varphi(\eta)\text{Im}(X_{k}(\eta)X'^{*}_{k}(\eta))=1\\
\implies &|X'_{k}(\eta)|^{2}+\Omega_{k}^{2}(\eta)|X_{k}(\eta)|^{2}=1-2h\varphi(\eta)\text{Im}(X_{k}(\eta)X'^{*}_{k}(\eta))
\end{split}
\end{equation}
Substituting this in Eq.~(\ref{eq: n_k full}) gives:
\begin{equation}\label{eq: n_k form. Greene}
n_{k}(\eta)=\frac{1}{2}-\frac{k^2}{\Omega_{k}(\eta)}\text{Im}(X_{k}(\eta)X'^{*}_{k}(\eta))-\frac{h\varphi(\eta)}{2\Omega_{k}(\eta)}
\end{equation}
This is of the same form as the number density equation written in \cite{Greene:1998nh} and \cite{Garcia-Bellido:2000woy}.

Alternatively, Eq.~(\ref{eq: using identity}) can be used to write:
\begin{equation}\label{eq: substitution for Omega_k}
\Omega_{k}(\eta)=\Omega_{k}(\eta)\times 1=\Omega_{k}(\eta)[|X'_{k}(\eta)|^{2}+\Omega_{k}^{2}(\eta)|X_{k}(\eta)|^{2}+2h\varphi(\eta)\text{Im}(X_{k}(\eta)X'^{*}_{k}(\eta))]
\end{equation}
Substituting this in Eq.~(\ref{eq: n_k full}) gives:
\begin{equation}\label{eq: using substitution for Omega_k}
n_{k}(\eta)=\left(\frac{\Omega_{k}(\eta)-h\varphi(\eta)}{2\Omega_{k}(\eta)}\right)[|X'_{k}(\eta)|^{2}+\Omega_{k}^{2}(\eta)|X_{k}(\eta)|^{2}-2\Omega_{k}\text{Im}(X_{k}(\eta)X'^{*}_{k}(\eta))]
\end{equation}
This is of the same form as the number density equation written in \cite{Greene:2002uot}.
\section{Derivation of $\overline{n}_{\kappa}(\tau)$}\label{AppE}
We will now show the derivation for $\overline{n}_{\kappa}(\tau)$ from Eq.~\eqref{eq:NumDen bar n_kappa(tau)}, following the procedure of \cite{Mostepanenko:1974im}, using the example and notation we have used for $\lambda\phi^{4}$ inflation. Although all equations are written for the particular case of $\lambda\phi^{4}$ inflation, we can write the relevant equations for any other model (such as $m^2 \phi^2$ inflation) in this form as well, by using different transformations to write all the equations in dimensionless variables, and the same ideas will be valid.

In \cite{Mostepanenko:1974im}, expressions are derived which describe the production of particles in vacuum by a spatially uniform periodic electric field. These were then applied to a particular case of particle production from optical laser light. The general ideas of the derivation are also applicable for particle production from the inflaton, and we make the approximation that the inflaton oscillations start at some point of time and occur for $N$ periods, during which particle production takes place and then stops.

To start the derivation, we first know that $f(\tau)=cn\left(\tau,1/2\right)$ from Eq.~\eqref{eq:Inf. f=cn} is periodic with a period $T$ i.e.~$f(\tau + T)=f(\tau)$. As mentioned below Eq.~\eqref{eq:Inf. f=cn}, we set $\tau_{1}=0$ so that $f(\tau)$ is at its maximum at $\tau = 0$. This implies $f(-\tau)=f(\tau)\implies\dot{f}(-\tau)=-\dot{f}(\tau)$. Thus, applying the transformation ($f(\tau)\rightarrow f^{*}(-\tau)$, $X_{\kappa}(\tau)\rightarrow X_{\kappa}^{*}(-\tau)$) to Eq.~\eqref{eq:NumDen X_kappa mode eq.} yields the exact same differential equation, using the fact that $f(\tau)$ is real. Then, let us introduce two solutions, $X^{1}_{\kappa}$ and $X^{2}_{\kappa}$, of the differential equation Eq.~\eqref{eq:NumDen X_kappa mode eq.} with the solutions having the following initial conditions:
\begin{equation}\label{eq: X1, X2 initial cond.}
X^{1}_{\kappa}(0)=1, 
\quad
\dot{X}^{1}_{\kappa}(0)=0, 
\quad
X^{2}_{\kappa}(0)=0, 
\quad
\dot{X}^{2}_{\kappa}(0)=1,
\end{equation}
where $\dot{}$ implies derivatives with respect to $\tau$.
Then we can write the Wronskian for $X^{1}_{\kappa}$ and $X^{2}_{\kappa}$, and use Abel's identity for Eq.~\eqref{eq:NumDen X_kappa mode eq.} with the above initial conditions to get the following relation for all $\tau$:
\begin{equation}\label{eq: X1, X2 relation in tau}
X^{1}_{\kappa}(\tau)\dot{X}^{2}_{\kappa}(\tau)-\dot{X}^{1}_{\kappa}(\tau)X^{2}_{\kappa}(\tau)=1.
\end{equation}
Due to the symmetry of the differential equation under a transformation involving a complex conjugate and $\tau \rightarrow -\tau$, we can say that one of the solutions $X^{1}_{\kappa}$ or $X^{2}_{\kappa}$ is even under the transformation, while the other is odd under the transformation. Then, let us assume the following: 
\begin{equation}\label{eq: X1, X2 transf.}
\begin{split}
&X^{1*}_{\kappa}(-\tau)=X^{1}_{\kappa}(\tau),
\quad
X^{2*}_{\kappa}(-\tau)=-X^{2}_{\kappa}(\tau).\\
\therefore 
&\dot{X}^{1*}_{\kappa}(-\tau)=-\dot{X}^{1}_{\kappa}(\tau),
\quad
\dot{X}^{2*}_{\kappa}(-\tau)=\dot{X}^{2}_{\kappa}(\tau).
\end{split}
\end{equation}
Due to the periodicity of $f(\tau)$, we can write solutions $X^{1}_{\kappa}(\tau +T)$ and $X^{2}_{\kappa}(\tau +T)$ as linear combinations of $X^{1}_{\kappa}(\tau)$ and $X^{2}_{\kappa}(\tau)$. This can be done in the following manner:
\begin{equation}\label{eq: X1, X2 tau+T linear combination eqs}
\begin{split}
&X^{1}_{\kappa}(\tau +T)=a_{1}X^{1}_{\kappa}(\tau)+b_{1}X^{2}_{\kappa}(\tau),\\
&X^{2}_{\kappa}(\tau +T)=a_{2}X^{1}_{\kappa}(\tau)+b_{2}X^{2}_{\kappa}(\tau).
\end{split}
\end{equation}
The coefficients $a_{i},b_{i}$ $(i=1,2)$ can be found by looking at the above equations at $\tau=0$ and the time derivative of the equations at $\tau=0$, and using Eq.~\eqref{eq: X1, X2 initial cond.}. After finding the coefficients, we can write:
\begin{equation}\label{eq: X1, X2 tau+T linear combination}
\begin{split}
&X^{1}_{\kappa}(\tau +T)=X^{1}_{\kappa}(T)X^{1}_{\kappa}(\tau)+\dot{X}^{1}_{\kappa}(T)X^{2}_{\kappa}(\tau),\\
&X^{2}_{\kappa}(\tau +T)=X^{2}_{\kappa}(T)X^{1}_{\kappa}(\tau)+\dot{X}^{2}_{\kappa}(T)X^{2}_{\kappa}(\tau).
\end{split}
\end{equation}
Now, let us define a matrix $\mathbf{M}(\tau)$ using $X^{1}_{\kappa}(\tau)$ and $X^{2}_{\kappa}(\tau)$:
\begin{equation}\label{eq: matrix M(tau)}
\mathbf{M}(\tau)=
\begin{pmatrix}
X^{1}_{\kappa}(\tau) & \dot{X}^{1}_{\kappa}(\tau)\\
X^{2}_{\kappa}(\tau) & \dot{X}^{2}_{\kappa}(\tau)
\end{pmatrix}.
\end{equation}
We have: $\mathbf{M}(0)=\mathbf{I}_{2 \times 2}$, from Eq.~\eqref{eq: X1, X2 initial cond.}. Using the relation from Eq.~\eqref{eq: X1, X2 relation in tau}, we have: $|\mathbf{M}(\tau)|=1$ for all $\tau$. Then, by Eq.~\eqref{eq: X1, X2 tau+T linear combination}, we find:
\begin{equation}\label{eq: matrix M(tau+T)}
\mathbf{M}(\tau + T)=
\begin{pmatrix}
X^{1}_{\kappa}(T) & \dot{X}^{1}_{\kappa}(T)\\
X^{2}_{\kappa}(T) & \dot{X}^{2}_{\kappa}(T)
\end{pmatrix}
\begin{pmatrix}
X^{1}_{\kappa}(\tau) & \dot{X}^{1}_{\kappa}(\tau)\\
X^{2}_{\kappa}(\tau) & \dot{X}^{2}_{\kappa}(\tau)
\end{pmatrix}
=\mathbf{M}(T)\mathbf{M}(\tau).
\end{equation}
If we take $\tau=T$, Eq.~\eqref{eq: matrix M(tau+T)} implies $\mathbf{M}(2T)=(\mathbf{M}(T))^{2}$; taking $\tau=2T$ implies $\mathbf{M}(3T)=(\mathbf{M}(T))^{3}$, and so on. In general, we can write:
\begin{equation}\label{eq: M(nT)}
\mathbf{M}(nT)=(\mathbf{M}(T))^{n},
\end{equation}
where $n=1,2,3\dots$. Now, let us introduce an matrix operation (represented by $ ^{\Delta}$) which gives another $2\times 2$ matrix with the complex conjugate of all terms and with switched diagonal terms. For example, applying it on $\mathbf{M}(\tau)$ implies:
\begin{equation}\label{eq: matrix M(tau) Delta}
\left(\mathbf{M}(\tau)\right)^{\Delta}=
\begin{pmatrix}
X^{1}_{\kappa}(\tau) & \dot{X}^{1}_{\kappa}(\tau)\\
X^{2}_{\kappa}(\tau) & \dot{X}^{2}_{\kappa}(\tau)
\end{pmatrix}
^{\Delta}=
\begin{pmatrix}
\dot{X}^{2*}_{\kappa}(\tau) & \dot{X}^{1*}_{\kappa}(\tau)\\
X^{2*}_{\kappa}(\tau) & X^{1*}_{\kappa}(\tau)
\end{pmatrix}.
\end{equation}
Similarly to taking complex conjugates, if we apply this operation on two arbitrary matrices $W_1$ and $W_2$, then it can be proven that:
\begin{equation}\label{eq: Delta conj. prop.}
\left(W_{1}W_{2}\right)^{\Delta}=W_{2}^{\Delta}W_{1}^{\Delta}.
\end{equation}
Applying $^{\Delta}$ twice on a matrix will yield the same matrix. Now, we apply $^{\Delta}$ to $\mathbf{M}(-\tau)$ and multiply the result with $\mathbf{M}(\tau)$. Then using Eq.~\eqref{eq: X1, X2 transf.} and Eq.~\eqref{eq: X1, X2 relation in tau}, we find:
\begin{equation}\label{eq: M Delta = M inv.}
\mathbf{M}^{\Delta}(-\tau)\mathbf{M}(\tau)=\mathbf{I}_{2 \times 2}
\implies
\mathbf{M}^{\Delta}(-\tau)=\mathbf{M}^{-1}(\tau).
\end{equation}
Then let us apply Eq.~\eqref{eq: matrix M(tau+T)} with $\tau=-T/2$ and use Eq.~\eqref{eq: M Delta = M inv.} to get:
\begin{equation}\label{eq: M(T) = M(Tby2)M(Tby2)Delta}
\mathbf{M}(T)=\mathbf{M}\left(\frac{T}{2}\right)\mathbf{M}^{-1}\left(-\frac{T}{2}\right)=\mathbf{M}\left(\frac{T}{2}\right)\mathbf{M}^{\Delta}\left(\frac{T}{2}\right).
\end{equation}
Applying $^{\Delta}$ on both sides of this equation implies:
\begin{equation}\label{eq: M(T) Delta = M(T)}
\mathbf{M}^{\Delta}(T)=\left(\mathbf{M}^{\Delta}\left(\frac{T}{2}\right)\right)^{\Delta}\mathbf{M}^{\Delta}\left(\frac{T}{2}\right)=\mathbf{M}\left(\frac{T}{2}\right)\mathbf{M}^{\Delta}\left(\frac{T}{2}\right)=\mathbf{M}(T).
\end{equation}
We can now use this with Eq.~\eqref{eq: M(nT)} and find:
\begin{equation}\label{eq: M(nT) Delta}
\mathbf{M}^{\Delta}(nT)=\left((\mathbf{M}(T))^{n}\right)^{\Delta}=\left(\mathbf{M}^{\Delta}(T)\right)^{n}=\left(\mathbf{M}(T)\right)^{n}=\mathbf{M}(nT).
\end{equation}
If we compare the terms of the matrices on both sides of the above equation, we find that the following relations are valid:
\begin{equation}\label{eq: X1, X2 relations at nT}
\dot{X}^{1*}_{\kappa}(nT)=\dot{X}^{1}_{\kappa}(nT),
\quad
X^{2*}_{\kappa}(nT)=X^{2}_{\kappa}(nT), 
\quad
X^{1*}_{\kappa}(nT)=\dot{X}^{2}_{\kappa}(nT).
\end{equation}
Then, for an arbitrary unimodular $2\times 2$ matrix $\mathbf{C}$ (i.e.~$|\mathbf{C}|=1$), it can be proven that the following relation holds for $n=1, 2, 3, \dots$:
\begin{equation}\label{eq: Cn relation}
\mathbf{C}^{n}=\left(\frac{\sinh(nD)}{\sinh(D)}\right)\mathbf{C} - \left(\frac{\sinh\left((n-1)D\right)}{\sinh(D)}\right)\mathbf{I},
\end{equation}
where $D$ is obtained from the trace of $\mathbf{C}$ as :
\begin{equation}\label{eq: cosh D}
\cosh(D)=Tr(\mathbf{C})/2.
\end{equation}
We can see that Eq.~\eqref{eq: Cn relation} is trivially true for $n=1$. For $n=2$, it can be simplified as follows:
\begin{equation}\label{eq: Cn relation n = 2}
\mathbf{C}^{2}=\left(Tr(\mathbf{C})\right)\mathbf{C}-\mathbf{I}.
\end{equation}
This equation will be true if $|\mathbf{C}|=1$. Thereafter, we can prove by induction that Eq.~\eqref{eq: Cn relation} will be true for all $n=1, 2, 3, \dots$. We can then use Eq.~\eqref{eq: Cn relation} for the matrix $\mathbf{M}(\tau)$, defined in Eq.~\eqref{eq: matrix M(tau)}, at $\tau=nT$. Then, using Eq.~\eqref{eq: M(nT)}, this implies:
\begin{equation}\label{eq: M(nT) Cn relation}
\mathbf{M}(nT)=\left(\frac{\sinh(nD)}{\sinh(D)}\right)\mathbf{M}(T) - \left(\frac{\sinh\left((n-1)D\right)}{\sinh(D)}\right)\mathbf{I},
\end{equation}
where $D$ is obtained from the trace of $\mathbf{M}(T)$ as (using Eq.~\eqref{eq: X1, X2 relations at nT}):
\begin{equation}\label{eq: cosh D M(T)}
\cosh(D)=\frac{Tr(\mathbf{M}(T))}{2}=\frac{X^{1}_{\kappa}(T)+\dot{X}^{2}_{\kappa}(T)}{2}
=\text{Re}(X^{1}_{\kappa}(T)),
\end{equation}
where the the last equality comes from Eq.~\eqref{eq: X1, X2 relations at nT}.
This implies that $\cosh(D)$ is purely real, which means that $D$ needs to be purely real or purely imaginary. Now, we can rewrite the Eq.~\eqref{eq: using identity} identity in dimensionless variables as:
\begin{equation}\label{eq: Xk substitution in n_kappa}
|\dot{X}_{\kappa}(\tau)|^{2}+\Omega_{\kappa}^{2}(\tau)|X_{\kappa}(\tau)|^{2}=1-2\sqrt{q}f(\tau)\text{Im}\left(X_{\kappa}(\tau)\dot{X}^{*}_{\kappa}(\tau)\right).
\end{equation}
Substituting this into Eq.~\eqref{eq:NumDen n_kappa(tau), C=1} and simplifying gives the equation:
\begin{equation}\label{eq: n_kappa form. Greene}
n_{\kappa}(\tau)=\frac{1}{2}-\frac{\sqrt{q}f(\tau)}{2\Omega_{\kappa}(\tau)}-\frac{\kappa^{2}}{\Omega_{\kappa}(\tau)}\text{Im}\left(X_{\kappa}(\tau)\dot{X}^{*}_{\kappa}(\tau)\right),
\end{equation}
which is the same as Eq.~\eqref{eq: n_k form. Greene}, written in dimensionless variables.
We can also use the identity of Eq.~\eqref{eq:NumDen f, g identity} with $X^{1}_{\kappa}$ and $X^{2}_{\kappa}$, at $\tau=0$, and the initial conditions from Eq.~\eqref{eq: X1, X2 initial cond.} to get:
\begin{equation}\label{eq: C using X1, X2}
C=-i\sqrt{q}f(0).
\end{equation}
Now we use Eq.~\eqref{eq:NumDen f, g identity} again with $X^{1}_{\kappa}(NT)$ and $X^{2}_{\kappa}(NT)$ and Eq.~\eqref{eq: X1, X2 relations at nT}, where $N$ is the number of oscillation periods that we are considering (starting from $\tau=0$), and plug in the $C$ above (with $f(0)=f(NT)$) to get:
\begin{equation}\label{eq: X1, X2 relation using f, g}
X^{1}_{\kappa}(nT)\left(
\Omega_{\kappa}^{2}(NT)X^{2}_{\kappa}(NT)+\dot{X}^{1}_{\kappa}(NT)+2\sqrt{q}f(NT)
\text{Im}\left(X^{1}_{\kappa}(NT)\right)\right)=0.
\end{equation}
Then we have that the oscillations $f(\tau)$ for Eq.~\eqref{eq:NumDen X_kappa mode eq.}, which drive the mode equation, are `turned off' at times $\tau > NT$ or $\tau < 0$. This would imply that $f$ will be constant for these times and $\dot{f}$ will be $0$. Hence, Eq.~\eqref{eq:NumDen X_kappa mode eq.} takes the form of a simple harmonic oscillator equation for these times as:
\begin{equation}\label{eq: X_kappa mode eq. for lt 0 and gt nT}
\ddot{X_{\kappa}}(\tau)+\Omega^{2}_{\kappa\pm}X_{\kappa}(\tau)=0,
\end{equation}
where: $\Omega_{\kappa+}=\Omega_{\kappa}(NT)=\sqrt{\kappa^{2}+qf_{+}^{2}}$ and $\Omega_{\kappa-}=\Omega_{\kappa}(0)=\sqrt{\kappa^{2}+qf_{-}^{2}}$ represent the fixed values of $\Omega_{\kappa}(\tau)$ for $\tau\geq NT$ and $\tau\leq 0$ respectively. We need $N$ to be large for the $\tau\geq NT$ approximation to be valid for our mode equation. The general solution for Eq.~\eqref{eq: X_kappa mode eq. for lt 0 and gt nT} for the $\tau\geq NT$ case is given by:  
\begin{equation}\label{eq: Xk approximation A, B}
X_{\kappa}^{+}(\tau)=A e^{i\Omega_{\kappa+}\tau}+B e^{-i\Omega_{\kappa+}\tau},
\end{equation}
where $A$ and $B$ are some complex coefficients. Then, using this general solution with the initial conditions from Eq.~\eqref{eq:NumDen X_kappa initial cond.}, for a time $\tau\leq 0$, we can write the solution as follows:
\begin{equation}\label{eq: Xk approximation leq 0}
X_{\kappa}^{-}(\tau)=X_{\kappa}(0) e^{i\Omega_{\kappa-}\tau}.
\end{equation}
For $0 \leq \tau \leq NT$, we can use the initial conditions from Eq.~\eqref{eq:NumDen X_kappa initial cond.} and the solutions $X^{1}_{\kappa}$ and $X^{2}_{\kappa}$ defined by the initial conditions from Eq.~\eqref{eq: X1, X2 initial cond.} to construct the following approximate solution for Eq.~\eqref{eq:NumDen X_kappa mode eq.}:
\begin{equation}\label{eq: Xk approximation bet. 0 and nT}
X_{\kappa}(\tau)=X_{\kappa}(0)\left(X^{1}_{\kappa}(\tau)+i\Omega_{\kappa}(\tau)X^{2}_{\kappa}(\tau)\right).
\end{equation}
Clearly, this $X_{\kappa}(\tau)$ smoothly joins to $X_{\kappa}^{-}(\tau)$ at $\tau = 0$ (as $X_{\kappa}(0)=X_{\kappa}^{-}(0)$ and $\dot{X}_{\kappa}(0)=\dot{X}_{\kappa}^{-}(0)$ for the $X_{\kappa}(\tau)$ above). We also expect it to smoothly join $X_{\kappa}^{+}(\tau)$ at $\tau = NT$, which implies the boundary conditions at $\tau=NT$. This is possible if the following relations are satisfied:
\begin{equation}\label{eq: smooth join at nT 1}
X_{\kappa}(0)\left(X^{1}_{\kappa}(NT)+i\Omega_{\kappa}(NT)X^{2}_{\kappa}(NT)\right)
=A e^{i\Omega_{\kappa}(NT)NT}+B e^{-i\Omega_{\kappa}(NT)NT},
\end{equation}
\begin{equation}\label{eq: smooth join at nT 2}
\frac{X_{\kappa}(0)}{i\Omega_{\kappa}(NT)}
\left(\dot{X}^{1}_{\kappa}(NT)+i\Omega_{\kappa}(NT)\dot{X}^{2}_{\kappa}(NT)\right)
=A e^{i\Omega_{\kappa}(NT)NT}-B e^{-i\Omega_{\kappa}(NT)NT}.
\end{equation}
Then, let us consider Eq.~\eqref{eq: Xk approximation A, B} at $\tau=NT$ and plug it into Eq.~\eqref{eq: n_kappa form. Greene} and this gives:
\begin{equation}\label{eq: plug in Xk approximation A, B at nT}
\overline{n}_{\kappa}(NT)=\frac{1}{2}-\frac{\sqrt{q}f(NT)}{2\Omega_{\kappa}(NT)}-\kappa^{2}\left(|B|^{2}-|A|^{2}\right).
\end{equation}
We can use Eqs.~\eqref{eq: smooth join at nT 1} and \eqref{eq: smooth join at nT 2} to get an expression for $\left(|B|^{2}-|A|^{2}\right)$ in terms of $X^{1}_{\kappa}(NT)$ and $X^{2}_{\kappa}(NT)$. We can then substitute the expression into Eq.~\eqref{eq: plug in Xk approximation A, B at nT}. Finally, we use Eqs.~\eqref{eq:NumDen X_kappa initial cond.}, \eqref{eq: X1, X2 relations at nT}, \eqref{eq: X1, X2 relation using f, g}, and the periodicity of $f(\tau)$ to simplify the expression for $\overline{n}_{\kappa}(NT)$ and we get: 
\begin{equation}\label{eq: n kappa X1(nT)}
\overline{n}_{\kappa}(NT)=\frac{\kappa^{2}}{\Omega_{\kappa}^{2}}\left(\text{Im}\left(X^{1}_{\kappa}(NT)\right)\right)^{2}.
\end{equation} 
Now, we can compare the $(1,1)$ entries of the matrices on both sides of Eq.~\eqref{eq: M(nT) Cn relation} and use Eq.~\eqref{eq: matrix M(tau)} to get the following equation for $X^{1}_{\kappa}(NT)$: 
\begin{equation}\label{eq: X1(nT) Cn relation}
X^{1}_{\kappa}(NT)=\left(\frac{\sinh(ND)}{\sinh(D)}\right)X^{1}_{\kappa}(T) - \left(\frac{\sinh\left((N-1)D\right)}{\sinh(D)}\right),
\end{equation}
where $D$ is given by Eq.~\eqref{eq: cosh D M(T)}. For the above equation, we find that the terms of form $\left(\sinh(ND)/\sinh(D)\right)$ will be purely real. This is a consequence of the fact that $D$ is purely real or purely imaginary. Hence, we can plug the value for $X^{1}_{\kappa}(NT)$ into Eq.~\eqref{eq: n kappa X1(nT)} and we get the following:
\begin{equation}\label{eq: n kappa(nT) with D}
\overline{n}_{\kappa}(NT)=\frac{\kappa^{2}}{\Omega_{\kappa}^{2}}
\frac{\sinh^{2}(ND)}{\sinh^{2}(D)}\left(\text{Im}\left(X^{1}_{\kappa}(T)\right)\right)^{2}.
\end{equation}
Eq.~\eqref{eq: cosh D M(T)} implies that $D$ will be purely real or purely imaginary. However, as we consider fermions, we also need to ensure that $\overline{n}_{\kappa}(NT)$ respects the Pauli principle by ensuring that $\overline{n}_{\kappa}(NT)\leq 1$. If $D$ were purely real, $\sinh^{2}(ND)/\sinh^{2}(D)$ grows exponentially with $N$ and will not be bounded.\footnote{This case corresponds to considering bosons.} Hence, the only possibility for us is that $D$ is purely imaginary and $-1<\text{Re}(X^{1}_{\kappa}(T))<1$. Hence, we can write: $D=id$, where $0 \leq d \leq \pi$. This leads to the following relations:
\begin{equation}\label{eq: D=id relations}
\cosh(D)=\cos(d), 
\quad
\sinh(ND)=i\sin(Nd).
\end{equation}
We use these relations in Eq.~\eqref{eq: n kappa(nT) with D} and get the expression for $\overline{n}_{\kappa}(nT)$ as:
\begin{equation}\label{eq: n kappa(nT) final}
\overline{n}_{\kappa}(NT)=\frac{\kappa^{2}}{\Omega_{\kappa}^{2}}
\frac{\sin^{2}(Nd)}{\sin^{2}(d)}\left(\text{Im}\left(X^{1}_{\kappa}(T)\right)\right)^{2},
\end{equation}
\begin{equation}\label{eq: cos d}
\cos(d)=\text{Re}\left(X^{1}_{\kappa}(T)\right).
\end{equation}
This is of the same form as Equation (30) of \cite{Mostepanenko:1974im}. This expression for $\overline{n}_{\kappa}(NT)$ matches $n_{\kappa}(\tau)$ from Eq.~\eqref{eq:NumDen n_kappa(tau), C=1} at integer multiples of $T$. As shown in \cite{Greene:2002uot}, we want to convert $\overline{n}_{\kappa}(NT)$ to the continuous function $\overline{n}_{\kappa}(\tau)$ in Eq.~\eqref{eq:NumDen bar n_kappa(tau)}. This is because we want to focus only on the long period of $n_{\kappa}(\tau)$ to study the filling of $\kappa$ modes, and $\overline{n}_{\kappa}$ provides a numerically stable approximation for $n_{\kappa}(\tau)$, without any of the `spiky' short period behaviour of $n_{\kappa}(\tau)$. We can then do this by considering $\tau=NT$ and comparing $\sin^{2}(Nd)$ with $\sin^{2}(\nu_{\kappa}\tau)$. There are infinitely many solutions which will satisfy this, but we shall choose the following (instead of $\cos\left(\nu_{\kappa}T\right)=-\text{Re}\left(X^{1}_{\kappa}(T)\right)$, which was the case in \cite{Greene:2002uot}): 
\begin{equation}\label{eq: cos nu kappa*T combined}
\cos\left(\nu_{\kappa}T\right)=|\text{Re}\left(X^{1}_{\kappa}(T)\right)|.
\end{equation}
This is done to ensure that $\nu_{\kappa}$ is always the lowest frequency, in order to match the low-frequency pattern followed by $n_{\kappa}(\tau)$, irrespective of the sign of $\text{Re}\left(X^{1}_{\kappa}(T)\right)$. This gives us the formula for $\nu_{\kappa}$ from Eq.~\eqref{eq:NumDen nu_kappa and F_kappa}.
\section{Back-reaction}\label{AppF}
In Section~\ref{sec:Lambda phi4}, we mentioned that we neglect back-reaction for our analysis. Here, we show that for the range of $q$ values that we consider, the energy transferred into the fermions is a very small fraction of the energy of the inflaton field, justifying neglecting back-reaction. We can write an integral similar to the one in Eq.~\eqref{eq:NumDen total n, dimless} to describe the total comoving energy density accumulated in the fermions produced \cite{Greene:2002uot}. We approximate the energy density per mode of the fermions by $\overline{\Omega}_{\kappa}$ from Eq.~\eqref{eq:ResPeakPos Omkapavg} and approximate the occupancy of the $\kappa$ modes by $F_{\kappa}/2$, with $F_{\kappa}$ described in Eq.~\eqref{eq:NumDen nu_kappa and F_kappa}. Then, the ratio of the total comoving energy density in the fermions to that of the inflaton field is given by:
\begin{equation}\label{eq: FermInfEnergyRatio}
\frac{\rho_{\psi}}{\rho_{\varphi}}=\frac{4\lambda}{\pi^{2}}
\int d\kappa \kappa^{2}\overline{\Omega}_{\kappa}F_{\kappa},
\end{equation}
where $\rho_{\varphi}=\lambda\varphi_{0}^{4}/4$. We have used the time average of $\Omega_{\kappa}$ and $\overline{n}_{\kappa}$ and hence, the ratio in Eq.~\eqref{eq: FermInfEnergyRatio} is also independent of time. Guided by CMB observations \cite{Greene:2002uot, Planck:2018vyg} of the amplitude of the power spectrum of scalar density perturbations $A_{S}$, we set $\lambda = 10^{-13}$. We then calculate this ratio for three different $q$ values and obtain the corresponding fractions for the comoving energy density of the inflaton field transferred into the produced fermions, as shown in Table \ref{tab: FermInfRatios}.
\begin{table}[H]
    \centering
    \begin{tabular}{|p{3cm}||p{3cm}|} 
    \hline
    $q$ & $\rho_{\psi}/\rho_{\varphi}$ \\ [0.5ex] 
    \hline\hline
    $10^{2}$ & $1.2\cdot 10^{-12}$ \\
    \hline
    $10^{3}$ & $1.8\cdot 10^{-11}$ \\
    \hline
    $10^{5}$ & $5.9\cdot 10^{-9}$ \\ [1ex]
    \hline
    \end{tabular}
    \caption{Three $q$ values and the $\rho_{\psi}/\rho_{\varphi}$ ratios they correspond to, for $\lambda=10^{-13}$.}
    \label{tab: FermInfRatios}
\end{table}
We numerically fit these values and obtain an approximate parametric estimate for the ratio. The estimate will be different for different values of $\lambda$, and we also account for this in the following equation:
\begin{equation}\label{eq: FermInfEnergyRatioEstimate}
\frac{\rho_{\psi}}{\rho_{\varphi}}\approx\left(4.0\cdot10^{-15}\right)
\times\left(\frac{\lambda}{10^{-13}}\right)
\times q^{1.2}.
\end{equation}
Based on this, we find that less than $1\%$ of the energy of the inflaton is transferred into the fermions for $q\lesssim10^{10}$, assuming $\lambda=10^{-13}$. We focus on much smaller $q$ values in our analysis, justifying neglecting back-reaction. 

In \cite{Giudice:1999fb}, back-reaction was analyzed for fermionic preheating with a $m^{2}\phi^{2}$ inflaton potential and it was found, from the fraction of the energy density in the produced fermions, that back-reaction effects are not relevant for $q\lesssim 10^{8}$.
\section{Resonance Peaks for fermions with a $\phi\phi \overline{\psi}\psi$ Coupling}\label{AppG}
As further justification for our energy conservation arguments from Section~\ref{sec: 3} and the relation Eq.~\eqref{eq:ResPeakPos peak position Omkapavg}, we apply our analysis to the case of production of light fermions ($m_{\psi}\approx 0$) with the same $\lambda\phi^{4}$ inflaton potential but with a $h_{1}\phi\phi \overline{\psi}\psi$ coupling between the inflaton and fermions, where $h_{1}$ is now a dimensionful coupling. The description of fermion production for this case and the equations representing the number density will be exactly the same as the ones in Section~\ref{sec: 2}, with the only differences being how the mode equation and the coupling parameter are defined. Similar to Eq.~\eqref{eq:NumDen X_k mode eq_eta}, the mode equation for the case with a $h_{1}\phi\phi \overline{\psi}\psi$ coupling can first be written in conformal variables as:
\begin{equation}\label{eq:PhiPhiPsiPsibar X_k mode eq_eta}
X_{k}''(\eta)+\left[
k^{2}+
\left(h_{1}\varphi(\eta)\varphi(\eta)\right)^{2}
-i\left(2h_{1}\varphi(\eta)\varphi'(\eta)\right)
\right]
X_{k}(\eta) = 0.
\end{equation}
We can write this mode equation in dimensionless variables by using the transformations from Eq.~\eqref{eq:NumDen eta to tau transf.} for most of the variables, with the exception of the coupling parameter. Instead of $q$ from Eq.~\eqref{eq:NumDen eta to tau transf.}, we define $q_1$ as:
\begin{equation}\label{eq:PhiPhiPsiPsibar q}
q_{1}=\frac{h_{1}^{2}\varphi_{0}^{2}}{\lambda}.
\end{equation}
We can then write the mode equation for this scenario in dimensionless variables as:
\begin{equation}\label{eq:PhiPhiPsiPsibar X_kappa mode eq.}
\ddot{X_{\kappa}}(\tau)+\left(\kappa^{2}+q_{1} f^{4}(\tau)-2i\sqrt{q_{1}}f(\tau)\dot{f}(\tau)\right)X_{\kappa}(\tau)=0.
\end{equation}
And we can write its `frequency' as:
\begin{equation}\label{eq:PhiPhiPsiPsibar Omega_k(tau)}
\Omega_{\kappa}(\tau)=\sqrt{\kappa^{2}+q_{1} f^{4}(\tau)}.
\end{equation}
Here $f(\tau)$ is the same Jacobi Elliptic function from Eq.~\eqref{eq:Inf. f=cn}. $F_{\kappa}$ is defined the same way as in Eq.~\eqref{eq:NumDen nu_kappa and F_kappa} but it is now computed by numerically solving the mode equation of Eq.~\eqref{eq:PhiPhiPsiPsibar X_kappa mode eq.}.

We can apply a similar energy conservation argument based on the perturbative description of the process $n\varphi\rightarrow\Psi\overline{\Psi}$ as the one described in Section~\ref{sec: 3}. We can then find a semi-analytic relation that can predict the locations of resonance peaks for this case, using the time average of $\Omega_{\kappa}(\tau)$ from Eq.~\eqref{eq:PhiPhiPsiPsibar Omega_k(tau)}, represented by $\overline{\Omega}_{\kappa}(q_{1})$. As described in Eqs.~\eqref{eq:ResPeakPos integral amp} and \eqref{eq:ResPeakPos Omkapavg with n varphi}, the integrals over internal spacetime points that appear in the amplitude for the process $n\varphi\rightarrow\Psi\overline{\Psi}$ yield a term of the form:
\begin{equation}\label{eq:PhiPhiPsiPsibar integral amp}
\delta^{(3)}\left(\bm{\kappa}_{\Psi}+\bm{\kappa}_{\overline{\Psi}}\right)
\int d\tau f(\tau)^{n}e^{2i\overline{\Omega}_{\kappa}(q_{1}) \tau},
\end{equation}
where we have used the large $\tau$ approximation for the accumulated phase of the fermion-antifermion pair. We can provide a simple qualitative explanation of the relevant physics by looking at the lowest order term in the expansion of $f(\tau)$ from Eq.~\eqref{eq:Inf. cn(u,1/2)}:
\begin{equation}\label{eq: JacobiCN lowest order}
\frac{4\pi\sqrt{2}}{T}\cos\left(\omega_{\varphi}\tau\right)
=\frac{2\pi\sqrt{2}}{T}\left(e^{i\omega_{\varphi}\tau}+e^{-i\omega_{\varphi}\tau}\right),
\end{equation}
where $\omega_{\varphi}=2\pi/T$. Plugging this term into the amplitude of Eq.~\eqref{eq:PhiPhiPsiPsibar integral amp} yields a term that gives the relation from Eq.~\eqref{eq:ResPeakPos Omkapavg with n varphi}. As done before, we can then qualitatively explain the resonance structure for this case using Feynman diagrams.

\begin{figure}[H]
    \subfloat
    {%
    \includegraphics[width=0.5\textwidth,height=0.2\textheight]{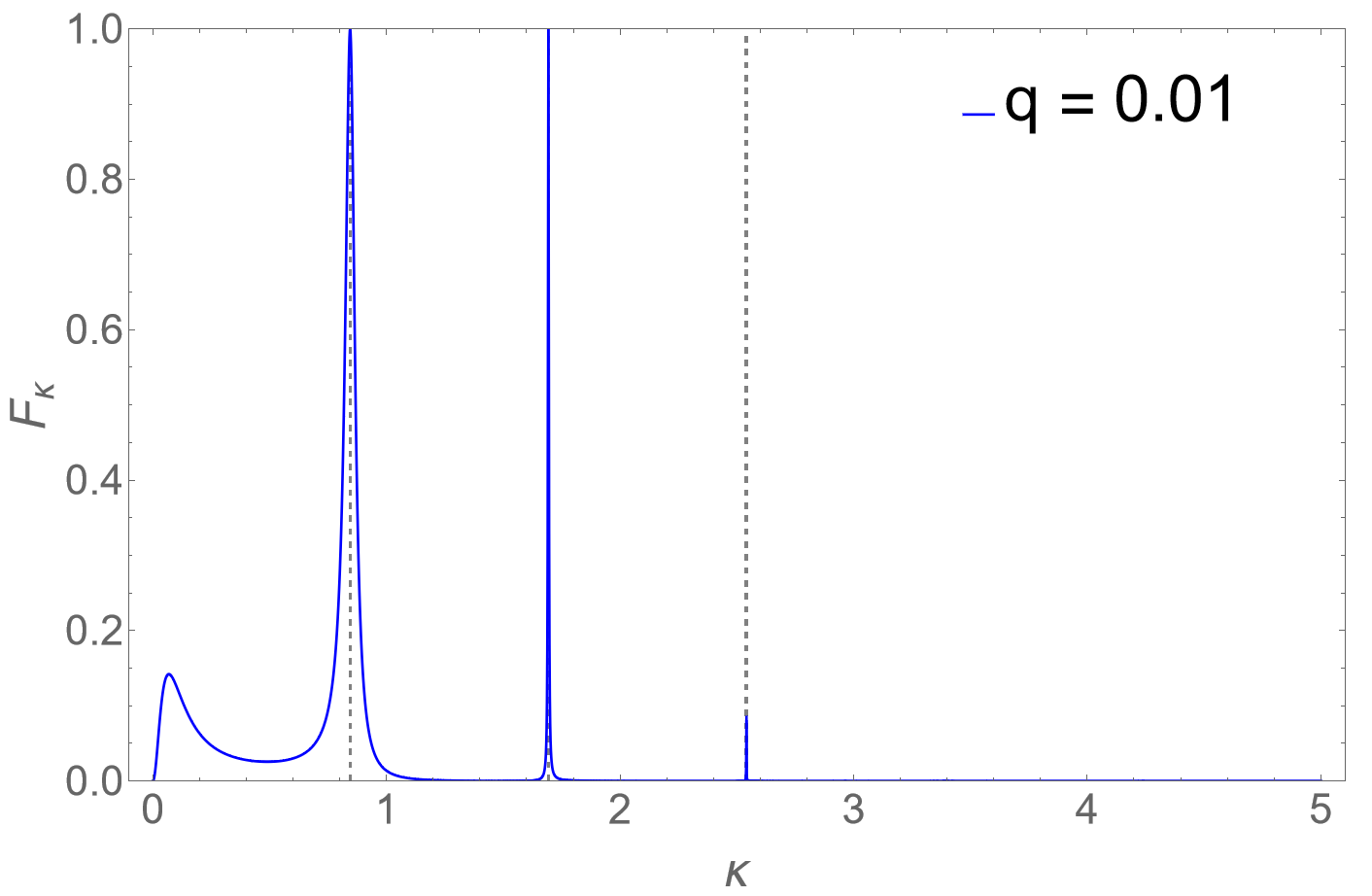}
    }
    \hfill
    \subfloat
    {%
    \includegraphics[width=0.5\textwidth,height=0.2\textheight]{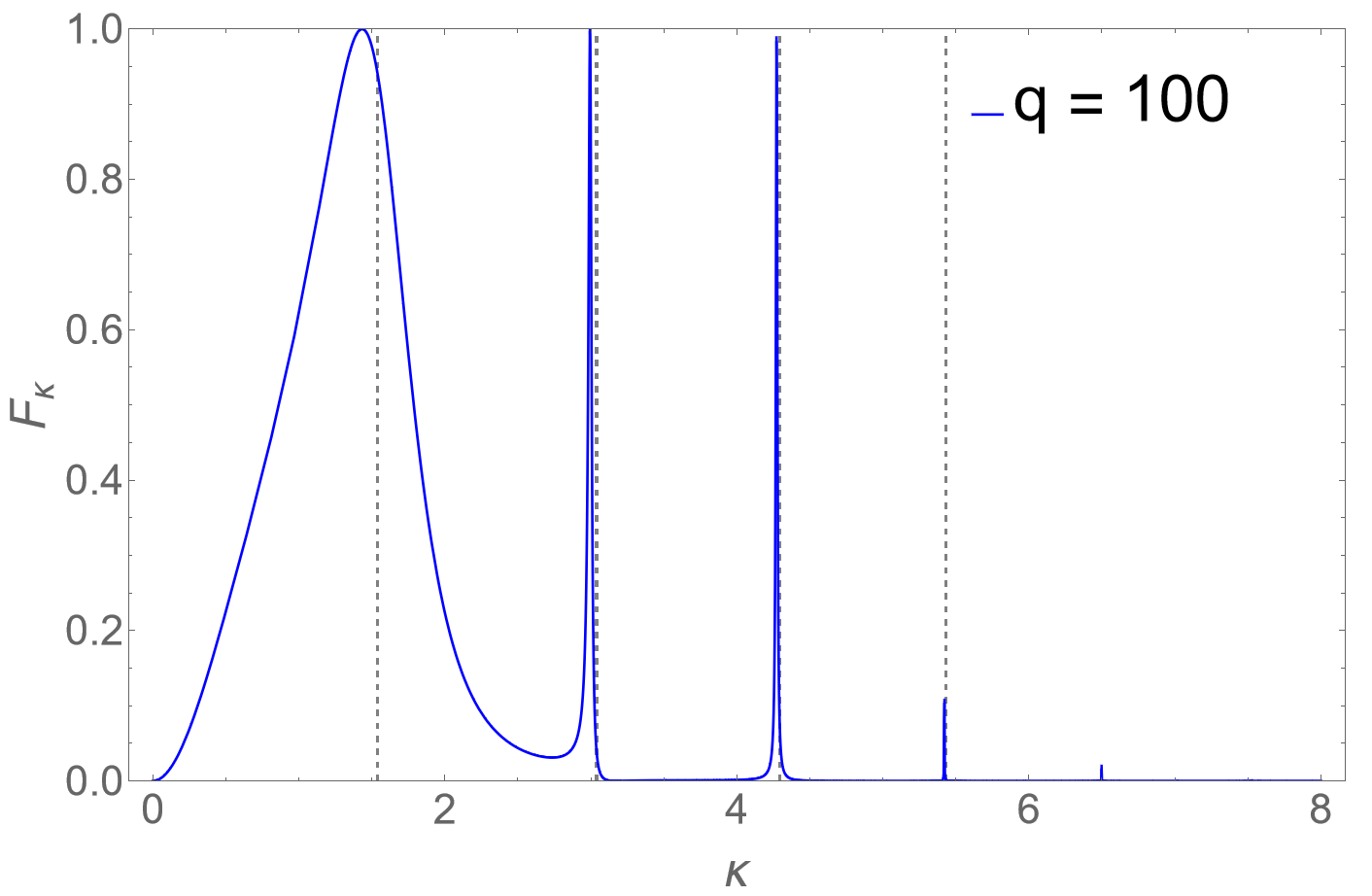}
    }
    \caption{$F_{\kappa}$ (in blue) for $q=0.01$ (left) and $q=100$ (right). Dashed vertical lines represent the locations of first few resonance peaks, found using Eq.~\eqref{eq:PhiPhiPsiPsibar peak position Omkapavg}.}
    \label{fig:PhiPhiPsiPsibar Fkappa plots}
\end{figure}

For small values of $q_{1}$, $\overline{\Omega}_{\kappa}(q_{1})\approx\kappa$ and relevant diagrams are of the form of the left and middle ones in Fig.~\ref{fig:PhiPhiPsiPsibar feyndiags all q}. Clearly, for each subsequent addition of a $\varphi^{4}$ vertex to the $n=2$ case (left diagram of Fig.~\ref{fig:PhiPhiPsiPsibar feyndiags all q}), the number of external $\varphi$ increases by $2$. This implies that $n$ must be even, i.e.~$n=2l$, where $l=1,2,3\dots$. Hence, for small values of $q_{1}$, resonance peaks of $F_{\kappa}$ should be present whenever $\kappa$ is an even multiple of $\pi/T$. This is exactly what we observe for $q_{1}=0.01$ in the left plot of Fig.~\ref{fig:PhiPhiPsiPsibar Fkappa plots}, where we show the locations of the resonance peaks corresponding to $l=1,2,3$.
\begin{figure}[H]
    \centering
    \begin{tikzpicture}[baseline=(b.base)]
    \begin{feynman}
        \vertex (c);
        \vertex [above left=of c] (a) {$\varphi$};
        \vertex [below left=of c] (b) {$\varphi$};
        \vertex [above right=of c] (f1) {$\Psi$};
        \vertex [below right=of c] (f2) {$\overline{\Psi}$};
        \diagram*{
        (a) -- [scalar] (c) -- [fermion] (f1),
        (b) -- [scalar] (c) -- [anti fermion] (f2)
        };
    \end{feynman}    
    \end{tikzpicture}\quad
    \begin{tikzpicture}[baseline=(b.base)]
    \begin{feynman}
        \vertex (c);
        \vertex [above left=of c] (a);
        \vertex [below left=of c] (b) {$\varphi$};
        \vertex [above left=of a] (a1) {$\varphi$};
        \vertex [left=of a] (a2) {$\varphi$};
        \vertex [above=of a] (a3) {$\varphi$};
        \vertex [above right=of c] (f1) {$\Psi$};
        \vertex [below right=of c] (f2) {$\overline{\Psi}$};
        \diagram*{
        (a1) -- [scalar] (a) -- [scalar] (c) -- [fermion] (f1),
        (c) -- [anti fermion] (f2),
        (a2) -- [scalar] (a), (a3) -- [scalar] (a),
        (c) -- [scalar] (b)
        };
    \end{feynman}    
    \end{tikzpicture}
    \begin{tikzpicture}[baseline=(b.base)]
    \begin{feynman}
        \vertex (phi1) at (-2,2.5) {$\varphi$};
        \vertex (phi2) at (-2,1.5) {$\varphi$};
        \vertex (phi3) at (-2,0.5) {$\varphi$};
        \vertex (phi4) at (-2,-0.5) {$\varphi$};
        \vertex (phi5) at (-2,-1.5) {$\varphi$};
        \vertex (phi6) at (-2,-2.5) {$\varphi$};
        
        \vertex (v1) at (0,2);
        \vertex (v2) at (0,0);
        \vertex (v3) at (0,-2);

        \vertex (f1) at (2,2) {$\Psi$};
        \vertex (f2) at (2,-2) {$\overline{\Psi}$};
    
        \diagram*{
            (v3) -- [anti fermion] (f2),
            (v3) -- [fermion] (v2) -- [fermion] (v1) -- [fermion] (f1),
            (phi1) -- [scalar] (v1), (phi2) -- [scalar] (v1),
            (phi3) -- [scalar] (v2), (phi4) -- [scalar] (v2),
            (phi5) -- [scalar] (v3), (phi6) -- [scalar] (v3)
        };
    \end{feynman}    
    \end{tikzpicture}
    \caption{Feynman diagrams describing the perturbative process $n\varphi\rightarrow\Psi\overline{\Psi}$. Shown first is the diagram for $n=2$ (left), which is the same for both small and large $q_{1}$ values. For small $q_{1}$, shown then is a representative diagram for $n=4$ (middle). For large $q_{1}$, shown is the diagram for $n=6$ (right).}
    \label{fig:PhiPhiPsiPsibar feyndiags all q}
\end{figure}
For large values of $q_{1}$, the relevant diagrams are of the form of the left and right ones in Fig.~\ref{fig:PhiPhiPsiPsibar feyndiags all q}. Diagrams with an odd number of fermion propagators will have zero amplitude based on the arguments in Section~\ref{sec: 3}. Thus, at large $q_1$ naively we would expect resonance peaks only for $\overline{\Omega}_{\kappa}(q_{1})=(4l+2)\pi/T$. However, peaks at all even multiples of $\pi/T$ again appear; e.g.\ for $\overline{\Omega}_{\kappa}(q_{1})=4\pi/T$ the diagram corresponding to $n=6$ is relevant, as if we plug Eq.~\eqref{eq: JacobiCN lowest order} into Eq.~\eqref{eq:PhiPhiPsiPsibar integral amp} for $n=6$, we still obtain three energy conservation relations corresponding to resonance peaks:
\begin{equation}\label{eq:PhiPhiPsiPsibar EnergyConservRels}
\overline{\Omega}_{\kappa}(q_{1})=\frac{2\pi}{T},
\quad
\overline{\Omega}_{\kappa}(q_{1})=\frac{4\pi}{T},
\quad
\overline{\Omega}_{\kappa}(q_{1})=\frac{6\pi}{T}.
\end{equation}
The first two of these relations correspond respectively to moving either one or two of the inflatons in the right diagram of Fig.~\ref{fig:PhiPhiPsiPsibar feyndiags all q} from the initial to the final state. 
For the case involving a $\lambda\phi^{4}$ potential, an additional contribution to these peaks comes from the higher order terms in the series expansion of the Jacobi Elliptic function described in Eq.~\eqref{eq:Inf. cn(u,1/2)}.

Hence, resonance peaks are present whenever $\overline{\Omega}_{\kappa}(q_{1})$ is an even multiple of $\pi/T$ for all $q_1$ in this model and we arrive at a semi-analytic relation that predicts the locations of resonance peaks for all $q_{1}$ values:
\begin{equation}\label{eq:PhiPhiPsiPsibar peak position Omkapavg}
\overline{\Omega}_{\kappa}(q_{1})=\frac{\pi}{T}(2l)\text{, where }l=1,2,3,\dots
\end{equation}
Here, $\overline{\Omega}_{0}(q_{1})$ sets a bound for the smallest $l$ value that is possible for a particular $q_{1}$. Similar to Eq.~\eqref{eq:ResPeakPos peak position l0}, this smallest accessible $l$ value can be obtained as follows:
\begin{equation}\label{eq:PhiPhiPsiPsibar peak position l0}
l_{0}(q_{1})=\left\lceil\frac{1}{2}\left(\overline{\Omega}_{0}(q_{1})\frac{T}{\pi}\right)\right\rceil.
\end{equation}
Using these, we find the locations of the first four resonance peaks with $q_{1}=100$ and they are approximately consistent with what we observe for the numerically evaluated $F_{\kappa}$, as shown in the right plot of Fig.~\ref{fig:PhiPhiPsiPsibar Fkappa plots}. In Fig.~\ref{fig:PhiPhiPsiPsibar Fkappa plots}, it also appears that the bulk region vanishes for large $q$, but exploring why this happens is beyond the scope of this work.

Finally, we also analyze the scenario with a $h_{1}\phi\phi\overline{\psi}\psi$ coupling for a $m^{2}\phi^{2}$ inflaton potential and find that the relation in Eq.~\eqref{eq:PhiPhiPsiPsibar peak position Omkapavg} works for that case as well. For small $q_{1}$ values however, the higher resonance peaks (with $l\geq 2$) vanish as only the left diagram of Fig.~\ref{fig:PhiPhiPsiPsibar feyndiags all q} contributes. The higher resonance peaks appear again for large $q_{1}$ values since the same arguments we gave above for the case with a $\lambda\phi^{4}$ potential are still applicable for the $m^{2}\phi^{2}$ scenario. 

\bibliographystyle{apsrev-title}
\bibliography{bib}

@article{Bhusal:2025oqg,
    author = "Bhusal, Nabeen and Ch{\'a}vez M., M. Ernesto and Garcia, Marcos A. G. and Menkara, Adriana G. and Pierre, Mathias",
    title = "{Fermion reheating with a quartic inflaton potential}",
    eprint = "2512.16203",
    archivePrefix = "arXiv",
    primaryClass = "hep-ph",
    reportNumber = "DESY-25-192",
    month = "12",
    year = "2025"
}

@article{Traschen:1990sw,
    author = "Traschen, Jennie H. and Brandenberger, Robert H.",
    title = "{Particle Production During Out-of-equilibrium Phase Transitions}",
    reportNumber = "BROWN-HET-731",
    doi = "10.1103/PhysRevD.42.2491",
    journal = "Phys. Rev. D",
    volume = "42",
    pages = "2491--2504",
    year = "1990"
}

@article{Shtanov:1994ce,
    author = "Shtanov, Y. and Traschen, Jennie H. and Brandenberger, Robert H.",
    title = "{Universe reheating after inflation}",
    eprint = "hep-ph/9407247",
    archivePrefix = "arXiv",
    reportNumber = "BROWN-HET-957",
    doi = "10.1103/PhysRevD.51.5438",
    journal = "Phys. Rev. D",
    volume = "51",
    pages = "5438--5455",
    year = "1995"
}

@article{Carena:2021bqm,
    author = "Carena, Marcela and Coyle, Nina M. and Li, Ying-Ying and McDermott, Samuel D. and Tsai, Yuhsin",
    title = "{Cosmologically degenerate fermions}",
    eprint = "2108.02785",
    archivePrefix = "arXiv",
    primaryClass = "hep-ph",
    reportNumber = "FERMILAB-PUB-21-325-T",
    doi = "10.1103/PhysRevD.106.083016",
    journal = "Phys. Rev. D",
    volume = "106",
    number = "8",
    pages = "083016",
    year = "2022"
}

@article{Greene:1998nh,
    author = "Greene, Patrick B. and Kofman, Lev",
    title = "{Preheating of fermions}",
    eprint = "hep-ph/9807339",
    archivePrefix = "arXiv",
    reportNumber = "UH-IFA-98-44",
    doi = "10.1016/S0370-2693(99)00020-9",
    journal = "Phys. Lett. B",
    volume = "448",
    pages = "6--12",
    year = "1999"
}

@article{Giudice:1999fb,
    author = "Giudice, G. F. and Peloso, M. and Riotto, A. and Tkachev, I.",
    title = "{Production of massive fermions at preheating and leptogenesis}",
    eprint = "hep-ph/9905242",
    archivePrefix = "arXiv",
    reportNumber = "CERN-TH-99-117",
    doi = "10.1088/1126-6708/1999/08/014",
    journal = "JHEP",
    volume = "08",
    pages = "014",
    year = "1999"
}

@article{Greene:2000ew,
    author = "Greene, Patrick B. and Kofman, Lev",
    title = "{On the theory of fermionic preheating}",
    eprint = "hep-ph/0003018",
    archivePrefix = "arXiv",
    reportNumber = "CITA-2000-05",
    doi = "10.1103/PhysRevD.62.123516",
    journal = "Phys. Rev. D",
    volume = "62",
    pages = "123516",
    year = "2000"
}

@phdthesis{Greene:2002uot,
    author = "Greene, Patrick Bradley",
    title = "{Aspects of Preheating After Inflation}",
    school = "Toronto U.",
    year = "2002",
    url = "http://hdl.handle.net/1807/121431"
}

@book{Kolb:1990vq,
    author = "Kolb, Edward W. and Turner, Michael S.",
    title = "{The Early Universe}",
    reportNumber = "FERMILAB-BOOK-1990-01",
    doi = "10.1201/9780429492860",
    isbn = "978-0-429-49286-0, 978-0-201-62674-2",
    publisher = "Taylor and Francis",
    volume = "69",
    month = "5",
    year = "2019"
}

@book{Thomson:2013zua,
    author = "Thomson, Mark",
    title = "{Modern particle physics}",
    doi = "10.1017/CBO9781139525367",
    isbn = "978-1-107-03426-6, 978-1-139-52536-7",
    publisher = "Cambridge University Press",
    address = "New York",
    month = "10",
    year = "2013"
}

@article{Garcia-Bellido:2000woy,
    author = "Garcia-Bellido, Juan and Mollerach, Silvia and Roulet, Esteban",
    title = "{Fermion production during preheating after hybrid inflation}",
    eprint = "hep-ph/0002076",
    archivePrefix = "arXiv",
    reportNumber = "FT-UAM-00-05, IFT-UAM-CSIC-00-05",
    doi = "10.1088/1126-6708/2000/02/034",
    journal = "JHEP",
    volume = "02",
    pages = "034",
    year = "2000"
}

@article{Dolgov:1989us,
    author = "Dolgov, A. D. and Kirilova, D. P.",
    title = "{ON PARTICLE CREATION BY A TIME DEPENDENT SCALAR FIELD}",
    reportNumber = "JINR-E2-89-321",
    journal = "Sov. J. Nucl. Phys.",
    volume = "51",
    pages = "172--177",
    year = "1990"
}

@article{Mostepanenko:1974im,
    author = "Mostepanenko, V. M. and Frolov, V. M.",
    title = "{Particle creation from vacuum by homogeneous electric field with a periodical time dependence}",
    journal = "Yad. Fiz.",
    volume = "19",
    pages = "885--896",
    year = "1974"
}

@Book{abramowitz+stegun,
 author    = "Milton {Abramowitz} and Irene A. {Stegun}",
 title     = "Handbook of Mathematical Functions with
              Formulas, Graphs, and Mathematical Tables",
 publisher = "Dover",
 year      =  1964,
 address   = "New York City",
 edition   = "{ninth Dover printing, tenth GPO printing}"
}

@article{Planck:2018vyg,
    author = "Aghanim, N. and others",
    collaboration = "Planck",
    title = "{Planck 2018 results. VI. Cosmological parameters}",
    eprint = "1807.06209",
    archivePrefix = "arXiv",
    primaryClass = "astro-ph.CO",
    doi = "10.1051/0004-6361/201833910",
    journal = "Astron. Astrophys.",
    volume = "641",
    pages = "A6",
    year = "2020",
    note = "[Erratum: Astron.Astrophys. 652, C4 (2021)]"
}

@article{Planck:2018jri,
    author = "Akrami, Y. and others",
    collaboration = "Planck",
    title = "{Planck 2018 results. X. Constraints on inflation}",
    eprint = "1807.06211",
    archivePrefix = "arXiv",
    primaryClass = "astro-ph.CO",
    doi = "10.1051/0004-6361/201833887",
    journal = "Astron. Astrophys.",
    volume = "641",
    pages = "A10",
    year = "2020"
}

@misc{reference.wolfram_2025_jacobicn, author="{Wolfram Research}", title="{JacobiCN}", year="1988", howpublished="\url{https://reference.wolfram.com/language/ref/JacobiCN.html}"
}

@article{Bogolyubov:1958kj,
    author = "Bogolyubov, N. N. and Tolmachev, V. V. and Shirkov, D. V.",
    title = "{A New method in the theory of superconductivity}",
    reportNumber = "JINR-R-139",
    doi = "10.1002/prop.19580061102",
    journal = "Fortsch. Phys.",
    volume = "6",
    pages = "605--682",
    year = "1958"
}

@article{Kofman:1997yn,
    author = "Kofman, Lev and Linde, Andrei D. and Starobinsky, Alexei A.",
    title = "{Towards the theory of reheating after inflation}",
    eprint = "hep-ph/9704452",
    archivePrefix = "arXiv",
    reportNumber = "IFA-97-28, SU-ITP-97-18",
    doi = "10.1103/PhysRevD.56.3258",
    journal = "Phys. Rev. D",
    volume = "56",
    pages = "3258--3295",
    year = "1997"
}

@article{Kofman:1994rk,
    author = "Kofman, Lev and Linde, Andrei D. and Starobinsky, Alexei A.",
    title = "{Reheating after inflation}",
    eprint = "hep-th/9405187",
    archivePrefix = "arXiv",
    reportNumber = "UH-IFA-94-35, SU-ITP-94-13, YITP-U-94-15",
    doi = "10.1103/PhysRevLett.73.3195",
    journal = "Phys. Rev. Lett.",
    volume = "73",
    pages = "3195--3198",
    year = "1994"
}

@article{Kolb:2023ydq,
    author = "Kolb, Edward W. and Long, Andrew J.",
    title = "{Cosmological gravitational particle production and its implications for cosmological relics}",
    eprint = "2312.09042",
    archivePrefix = "arXiv",
    primaryClass = "astro-ph.CO",
    doi = "10.1103/RevModPhys.96.045005",
    journal = "Rev. Mod. Phys.",
    volume = "96",
    number = "4",
    pages = "045005",
    year = "2024"
}

@article{Adshead:2015kza,
    author = "Adshead, Peter and Sfakianakis, Evangelos I.",
    title = "{Fermion production during and after axion inflation}",
    eprint = "1508.00891",
    archivePrefix = "arXiv",
    primaryClass = "hep-ph",
    doi = "10.1088/1475-7516/2015/11/021",
    journal = "JCAP",
    volume = "11",
    pages = "021",
    year = "2015"
}

@article{Adshead:2018oaa,
    author = "Adshead, Peter and Pearce, Lauren and Peloso, Marco and Roberts, Michael A. and Sorbo, Lorenzo",
    title = "{Phenomenology of fermion production during axion inflation}",
    eprint = "1803.04501",
    archivePrefix = "arXiv",
    primaryClass = "astro-ph.CO",
    doi = "10.1088/1475-7516/2018/06/020",
    journal = "JCAP",
    volume = "06",
    pages = "020",
    year = "2018"
}

@article{Linde:1983gd,
    author = "Linde, Andrei D.",
    title = "{Chaotic Inflation}",
    doi = "10.1016/0370-2693(83)90837-7",
    journal = "Phys. Lett. B",
    volume = "129",
    pages = "177--181",
    year = "1983"
}

@article{Dolgov:1982th,
    author = "Dolgov, A. D. and Linde, Andrei D.",
    title = "{Baryon Asymmetry in Inflationary Universe}",
    reportNumber = "ITEP-78-1982",
    doi = "10.1016/0370-2693(82)90292-1",
    journal = "Phys. Lett. B",
    volume = "116",
    pages = "329",
    year = "1982"
}

@article{Preskill:1982cy,
    author = "Preskill, John and Wise, Mark B. and Wilczek, Frank",
    editor = "Srednicki, M. A.",
    title = "{Cosmology of the Invisible Axion}",
    reportNumber = "HUTP-82-A048, NSF-ITP-82-103",
    doi = "10.1016/0370-2693(83)90637-8",
    journal = "Phys. Lett. B",
    volume = "120",
    pages = "127--132",
    year = "1983"
}

@article{Abbott:1982af,
    author = "Abbott, L. F. and Sikivie, P.",
    editor = "Srednicki, M. A.",
    title = "{A Cosmological Bound on the Invisible Axion}",
    reportNumber = "PRINT-82-0695 (BRANDEIS)",
    doi = "10.1016/0370-2693(83)90638-X",
    journal = "Phys. Lett. B",
    volume = "120",
    pages = "133--136",
    year = "1983"
}

@article{Caprini:2018mtu,
    author = "Caprini, Chiara and Figueroa, Daniel G.",
    title = "{Cosmological Backgrounds of Gravitational Waves}",
    eprint = "1801.04268",
    archivePrefix = "arXiv",
    primaryClass = "astro-ph.CO",
    doi = "10.1088/1361-6382/aac608",
    journal = "Class. Quant. Grav.",
    volume = "35",
    number = "16",
    pages = "163001",
    year = "2018"
}

@article{Linde:1993cn,
    author = "Linde, Andrei D.",
    title = "{Hybrid inflation}",
    eprint = "astro-ph/9307002",
    archivePrefix = "arXiv",
    reportNumber = "SU-ITP-93-17",
    doi = "10.1103/PhysRevD.49.748",
    journal = "Phys. Rev. D",
    volume = "49",
    pages = "748--754",
    year = "1994"
}

@article{Bassett:2005xm,
    author = "Bassett, Bruce A. and Tsujikawa, Shinji and Wands, David",
    title = "{Inflation dynamics and reheating}",
    eprint = "astro-ph/0507632",
    archivePrefix = "arXiv",
    doi = "10.1103/RevModPhys.78.537",
    journal = "Rev. Mod. Phys.",
    volume = "78",
    pages = "537--589",
    year = "2006"
}

@article{Kolb:1996jt,
    author = "Kolb, Edward W. and Linde, Andrei D. and Riotto, Antonio",
    title = "{GUT baryogenesis after preheating}",
    eprint = "hep-ph/9606260",
    archivePrefix = "arXiv",
    reportNumber = "FERMILAB-PUB-96-133-A, SU-ITP-96-22",
    doi = "10.1103/PhysRevLett.77.4290",
    journal = "Phys. Rev. Lett.",
    volume = "77",
    pages = "4290--4293",
    year = "1996"
}

@article{Chung:1998ua,
    author = "Chung, Daniel J. H. and Kolb, Edward W. and Riotto, Antonio",
    title = "{Nonthermal supermassive dark matter}",
    eprint = "hep-ph/9805473",
    archivePrefix = "arXiv",
    reportNumber = "FERMILAB-PUB-98-154-A, CERN-TH-98-160, OUTP-98-40-P",
    doi = "10.1103/PhysRevLett.81.4048",
    journal = "Phys. Rev. Lett.",
    volume = "81",
    pages = "4048--4051",
    year = "1998"
}

@article{Allahverdi:2010xz,
    author = "Allahverdi, Rouzbeh and Brandenberger, Robert and Cyr-Racine, Francis-Yan and Mazumdar, Anupam",
    title = "{Reheating in Inflationary Cosmology: Theory and Applications}",
    eprint = "1001.2600",
    archivePrefix = "arXiv",
    primaryClass = "hep-th",
    doi = "10.1146/annurev.nucl.012809.104511",
    journal = "Ann. Rev. Nucl. Part. Sci.",
    volume = "60",
    pages = "27--51",
    year = "2010"
}

@article{Amin:2014eta,
    author = "Amin, Mustafa A. and Hertzberg, Mark P. and Kaiser, David I. and Karouby, Johanna",
    title = "{Nonperturbative Dynamics Of Reheating After Inflation: A Review}",
    eprint = "1410.3808",
    archivePrefix = "arXiv",
    primaryClass = "hep-ph",
    doi = "10.1142/S0218271815300037",
    journal = "Int. J. Mod. Phys. D",
    volume = "24",
    pages = "1530003",
    year = "2014"
}

@article{Bjaelde:2010vt,
    author = "Bjaelde, Ole Eggers and Das, Subinoy",
    title = "{Dark Matter Decaying into a Fermi Sea of Neutrinos}",
    eprint = "1002.1306",
    archivePrefix = "arXiv",
    primaryClass = "astro-ph.CO",
    doi = "10.1103/PhysRevD.82.043504",
    journal = "Phys. Rev. D",
    volume = "82",
    pages = "043504",
    year = "2010"
}

@article{Choi:2020tqp,
    author = "Choi, Gongjun and Suzuki, Motoo and Yanagida, Tsutomu T.",
    title = "{Degenerate Sub-keV Fermion Dark Matter from a Solution to the Hubble Tension}",
    eprint = "2002.00036",
    archivePrefix = "arXiv",
    primaryClass = "hep-ph",
    doi = "10.1103/PhysRevD.101.075031",
    journal = "Phys. Rev. D",
    volume = "101",
    number = "7",
    pages = "075031",
    year = "2020"
}

@article{Choi:2020nan,
    author = "Choi, Gongjun and Suzuki, Motoo and Yanagida, Tsutomu. T.",
    title = "{Degenerate fermion dark matter from a broken $U(1)_{\rm B-L}$  gauge symmetry}",
    eprint = "2004.07863",
    archivePrefix = "arXiv",
    primaryClass = "hep-ph",
    doi = "10.1103/PhysRevD.102.035022",
    journal = "Phys. Rev. D",
    volume = "102",
    number = "3",
    pages = "035022",
    year = "2020"
}

@article{Kolb:1998ki,
    author = "Kolb, Edward W. and Chung, Daniel J. H. and Riotto, Antonio",
    editor = "Falomir, H. and Gamboa Saravi, R. E. and Schaposnik, F. A.",
    title = "{WIMPzillas!}",
    eprint = "hep-ph/9810361",
    archivePrefix = "arXiv",
    reportNumber = "FERMILAB-CONF-98-325-A",
    doi = "10.1063/1.59655",
    journal = "AIP Conf. Proc.",
    volume = "484",
    number = "1",
    pages = "91--105",
    year = "1999"
}

@article{Bezrukov:2008ut,
    author = "Bezrukov, F. and Gorbunov, D. and Shaposhnikov, M.",
    title = "{On initial conditions for the Hot Big Bang}",
    eprint = "0812.3622",
    archivePrefix = "arXiv",
    primaryClass = "hep-ph",
    doi = "10.1088/1475-7516/2009/06/029",
    journal = "JCAP",
    volume = "06",
    pages = "029",
    year = "2009"
}

@article{Garcia:2021iag,
    author = "Garcia, Marcos A. G. and Kaneta, Kunio and Mambrini, Yann and Olive, Keith A. and Verner, Sarunas",
    title = "{Freeze-in from preheating}",
    eprint = "2109.13280",
    archivePrefix = "arXiv",
    primaryClass = "hep-ph",
    reportNumber = "UMN-TH-4101/21, FTPI-MINN-21/19, CERN-TH-2021-121",
    doi = "10.1088/1475-7516/2022/03/016",
    journal = "JCAP",
    volume = "03",
    number = "03",
    pages = "016",
    year = "2022"
}

@article{Klaric:2022qly,
    author = "Klaric, Juraj and Shkerin, Andrey and Vacalis, Georgios",
    title = "{Non-perturbative production of fermionic dark matter from fast preheating}",
    eprint = "2209.02668",
    archivePrefix = "arXiv",
    primaryClass = "gr-qc",
    reportNumber = "FTPI-MINN-22-24, UMN-TH-4133/22, CP3-22-43",
    doi = "10.1088/1475-7516/2023/02/034",
    journal = "JCAP",
    volume = "02",
    pages = "034",
    year = "2023"
}

\end{document}